\documentclass[manuscript]{acmart}

\AtBeginDocument{%
  }

\setcopyright{acmlicensed}
\copyrightyear{2024}
\acmYear{2024}
\acmDOI{10.1145/1122445.1122456}

\acmJournal{TORS}
\acmVolume{1}
\acmNumber{1}
\acmArticle{1}
\acmMonth{12}

\usepackage{pdflscape}
\usepackage{threeparttable}
\usepackage{xspace}
\usepackage{adjustbox}
\usepackage{subcaption}
\usepackage{relsize}
\usepackage{array}
\usepackage{amsfonts}
\usepackage{tabularx}
\usepackage{siunitx}
\usepackage[inline]{enumitem}
\usepackage{pifont}
\usepackage[super]{nth}
\usepackage{multirow}
\usepackage[normalem]{ulem}
\usepackage[utf8]{inputenc}
\usepackage{makecell}
\usepackage[acronym,nomain]{glossaries}
\setacronymstyle{long-short}
\glsdisablehyper
\newacronym[
longplural={User-Impressions Matrices},
description={%
    A binary matrix (\ensuremath{UIM \in \{0, 1\}^{m,n}}) where an entry \ensuremath{UIM_{u,i} = 1} if the user \ensuremath{u} has been impressed with item \ensuremath{i}. Otherwise \ensuremath{UIM_{u,i} = 0}
}]{uim}{UIM}{User-Impressions Matrix}

\newacronym[
longplural={Counting User-Impressions Matrices},
description={%
    A matrix (\ensuremath{UIM \in \{0, 1\}^{m,n}}) where an entry \ensuremath{UIM_{u,i} = 1} if the user \ensuremath{u} has been impressed with item \ensuremath{i}. Otherwise \ensuremath{UIM_{u,i} = 0}
}]{cuim}{C-UIM}{Counting User-Impressions Matrix}

\newacronym[
longplural={User-Rating Matrices},
description={%
    A binary matrix (\ensuremath{URM \in \{0, 1\}^{m,n}}) where an entry \ensuremath{URM_{u,i} = 1} if the user \ensuremath{u} has interacted with item \ensuremath{i}. Otherwise \ensuremath{UIM_{u,i} = 0}
}]{urm}{URM}{User-Rating Matrix}

\newacronym{knn}{KNN}{K-Nearest Neighbors}
\newacronym{fm}{FM}{Factorization Machine}
\newacronym{mf}{MF}{Matrix Factorization}
\newacronym{pmf}{PMF}{Probabilistic Matrix Factorization}

\newacronym{ir}{IR}{Information Retrieval}
\newacronym{rs}{RS}{Recommender System}

\newacronym{sgd}{SGD}{Stochastic Gradient Descent}
\newacronym{opsgd}{OP-SGD}{One-Pass Stochastic Gradient Descent}

\newacronym{bpr}{BPR}{Bayesian Personalized Ranking}
\newacronym{warp}{WARP}{Weighted Approximate-Rank Pairwise}
\newacronym{sslim}{S-SLIM}{Sparse Linear Methods with Side Information}

\newacronym{pctr}{pCTR}{predicted click-through rate}
\newacronym{ctr}{CTR}{click-through rate}
\newacronym{cvr}{CVR}{Conversion Rate}
\newacronym{ctcvr}{CTCVR}{Click-Through \& Conversion Rate}

\newacronym{gbdt}{GBDT}{Gradient Boosting Decision Trees}
\newacronym{em}{EM}{Expectation-Maximization}

\newacronym{ml}{ML}{Machine Learning}
\newacronym{dl}{DL}{Deep Learning}
\newacronym{rl}{RL}{reinforcement learning}
\newacronym{mdp}{MDP}{Markov Decision Process}

\newacronym{mab}{MAB}{multi-armed bandits}
\newacronym{ipw}{IPW}{inverse propensity weighting}
\newacronym{ips}{IPS}{Inverse Propensity Score}
\newacronym{acf}{AC}{actor-critic}
\newacronym{sac}{SAC}{soft actor-critic}
\newacronym{es}{ES}{evolution strategies}

\newacronym{gan}{GAN}{generative adversarial networks}
\newacronym{dnn}{DNN}{deep neural network}
\newacronym{gnn}{GNN}{graph neural network}

\newacronym{resnet}{ResNet}{residual networks}
\newacronym{dcn}{DCN}{deep \& cross network}
\newacronym{mmoe}{MMoE}{multi-gate mixture-of-experts}
\newacronym{dqn}{DQN}{deep Q-network}
\newacronym{mlp}{MLP}{multilayer perceptrons}
\newacronym{rnn}{RNN}{recurrent neural networks}
\newacronym{vae}{VAE}{variational auto-encoder}
\newacronym{kd}{KD}{knowledge distillation}
\newacronym{plm}{PLM}{pre-trained language model}

\newacronym{ott}{OTT}{Over-The-Top}
\newacronym{vod}{VoD}{Video-on-Demand}
\newacronym{utc}{UTC}{Coordinated Universal Time}
\newacronym{vcpu}{vCPU}{virtual Central Processing Unit}

\newacronym{map}{MAP}{Mean Average Precision}
\newacronym{roc}{ROC}{Receiver Operating Characteristic curve}
\newacronym{rocauc}{ROC-AUC}{Receiver Operating Characteristic Area Under the Curve}
\newacronym{sauc}{S-AUC}{Stratified Area Under the Curve}
\newacronym{ndcg}{NDCG}{Normalized Discounted Cumulative Gain}
\newacronym{mrr}{MRR}{Mean Reciprocal Rank}

\newacronym{nce}{NCE}{Negative Cross-Entropy}

\newacronym{msid}{MSI}{models sampling impressions}
\newacronym{mli}{MLI}{models learning from impressions}
\newacronym{mlim}{MLI-M}{models learning from impressions}
\newacronym{mlif}{MLI-F}{models learning from impressions' features}

\newacronym{cf}{CF}{collaborative filtering}
\newacronym{cbcf}{CBCF}{content-based collaborative filtering}
\newacronym{cfside}{CF-SI}{collaborative filtering with side information}
\newacronym{impressionsbased}{IARS}{impression-aware recommender systems}
\newacronym{sessionbased}{SBRS}{session-based recommender systems}
\newacronym{sessionaware}{SBRS}{session-based recommender systems}
\newacronym{contextaware}{CARS}{context-aware recommender systems}
\newacronym{sequenceaware}{SARS}{sequence-aware recommender systems}
\newacronym{reinforcementbased}{RLRS}{reinforcement learning based recommender systems}

\newacronym{idf}{IDF}{impressions discounting framework}
\newacronym{hfc}{HFC}{hard frequency capping}
\newacronym{sfc}{SFC}{soft frequency capping}
\newacronym{far}{FAR}{fatigue-aware recommendation}

\newacronym{rsc16}{RSC16}{ACM RecSys Challenge 2016}
\newacronym{rsc19}{RSC19}{ACM RecSys Challenge 2019}
\newacronym{mindnrc}{MIND-NRC}{{MIND} News Recommendation Competition}

\newacronym{slr}{SLR}{systematic literature review}
\newacronym{core}{CORE}{Computing Research and Education Association of Australasia}

\newacronym{acmdl}{ACM DL}{ACM Digital Library}
\newacronym{ieeexplore}{IEEE}{IEEE Xplore}
\newacronym{sciencedirect}{SD}{Science Direct}
\newacronym{springerlink}{SL}{Springer Link}

\newacronym{rns}{RNS}{Reinforced Negative Sampler}
\newacronym{mind}{MIND}{Microsoft News Dataset}

\newacronym{linkedin-pymk}{PYMK}{LinkedIn People You May Know}

\newacronym{uma2}{UMA$^{2}$}{Unbiased Model Agnostic Matching Approach}
\newacronym{nsdn}{NSDN}{Negative Samples Debias Network}

\newacronym{drl-rec}{DRL-Rec}{Distilled Reinforcement Learning Framework for Recommendation}

\newacronym{wg4rec}{WG4Rec}{}

\newacronym{lijar}{LiJAR}{LinkedIn's Job Applications Forecasting and Redistribution}

\newacronym{tir}{TIR}{Trigger-Induced Recommendations}
\usepackage[edges,external]{forest}

\usepackage{pgfplots}
\DeclareUnicodeCharacter{2212}{−}
\usepgfplotslibrary{groupplots}
\usepgfplotslibrary{dateplot}
\usetikzlibrary{external}
\usetikzlibrary{fadings}
\usetikzlibrary{shapes}
\usetikzlibrary{shadows}
\usetikzlibrary{patterns}
\usetikzlibrary{shapes.arrows}
\usetikzlibrary{shapes.geometric}
\usetikzlibrary{backgrounds}
\usetikzlibrary{decorations.pathmorphing}
\usetikzlibrary{fit}
\usetikzlibrary{backgrounds}
\usetikzlibrary{arrows}
\pgfplotsset{
    compat=newest,
    /tikz/prefix=figures/data/,
}
\pgfplotsset{
  siunitxlabels/.style={
    /pgfplots/typeset ticklabel/.code={\pgfmathparse{\tick}$\num[zero-decimal-to-integer]{\pgfmathresult}$},
  },
}
\pgfplotsset{
    every axis/.append style={
        font={\small},
        label style={font=\small},
        tick label style={font=\small}  
    }
}
\usepackage{soul} 
\soulregister\acrlong7 
\soulregister\gls7
\soulregister\autoref7

\newif\ifusetikz
\newif\ifusepdf
\newif\ifusepng

\usetikztrue 
\usepdffalse 
\usepngfalse 

\begin{document}
\fancyhead{}

\newcommand{\numuniqueworks}{\num{1351}\xspace}
\newcommand{\numincludedworks}{\num{43}\xspace}

\newcommand{\idest}{i.e.,\xspace}
\newcommand{\eg}{e.g.,\xspace}

\newcommand{\setpositiveintegers}{\ensuremath{\mathbb{Z^{+}}}\xspace}
\newcommand{\setreals}{\ensuremath{\mathbb{R}}\xspace}
\newcommand{\functionindicator}{\ensuremath{\mathbb{I}}\xspace}

\newcommand{\useru}{\ensuremath{u}\xspace}
\newcommand{\itemi}{\ensuremath{i}\xspace}
\newcommand{\sessions}{\ensuremath{s}\xspace}

\newcommand{\functionrecommender}{\ensuremath{f}\xspace}

\newcommand{\setusers}{\ensuremath{\mathcal{U}}\xspace}
\newcommand{\setitems}{\ensuremath{\mathcal{I}}\xspace}
\newcommand{\setuserprofile}{\ensuremath{\mathcal{H}}\xspace}

\newcommand{\vectorinteraction}{\ensuremath{\vec{p}}\xspace}
\newcommand{\vectorimpression}{\ensuremath{\vec{e}}\xspace}

\newcommand{\listsideinfo}{\ensuremath{\vec{s}_{\useru, \itemi}}\xspace}
\newcommand{\listrecommendations}{\ensuremath{\vec{e}_{\useru, \itemi}}\xspace}
\newcommand{\listcontext}{\ensuremath{\vec{c}_{\useru, \itemi}}\xspace}
\newcommand{\listcontexttwo}{\ensuremath{\vec{c}_{\useru, \itemi}^{*}}\xspace}

\newcommand{\relevancescore}{\ensuremath{\Tilde{r}_{\useru, \itemi}}\xspace}
\newcommand{\recommenderrelevance}{\ensuremath{\Hat{r}_{\useru, \itemi}}\xspace}

\newcommand{\vectorrelevancescoreuser}{\ensuremath{\vec{r}_{\useru}}\xspace}

\newcommand{\numinteractions}[1]{\ensuremath{np_{#1}}\xspace}
\newcommand{\numimpressions}[1]{\ensuremath{ne_{#1}}\xspace}

\newcommand{\mfuserfactor}{\ensuremath{p^{T}_{\useru}}\xspace}
\newcommand{\mfitemfactor}{\ensuremath{q_{\itemi}}\xspace}
\newcommand{\mfbias}{\ensuremath{b}\xspace}

\newcommand{\rlstate}{\ensuremath{s}\xspace}
\newcommand{\rlaction}{\ensuremath{a}\xspace}
\newcommand{\rlreward}{\ensuremath{r}\xspace}
\newcommand{\rlpolicy}{\ensuremath{\pi}\xspace}

\newcommand{\queryq}{\ensuremath{q}\xspace}  
\newcommand{\presentationscore}{\ensuremath{\Tilde{p}_{\useru,\itemi}}\xspace}  
\newcommand{\discountingfactor}{\ensuremath{d_{u,i}}\xspace} 
\newcommand{\greedyv}{\ensuremath{GREEDY_{V}}\xspace}  
\newcommand{\greedyd}{\ensuremath{GREEDY_{D}}\xspace}  
\newcommand{\spatioposition}{\ensuremath{p}\xspace}  
\newcommand{\spatiopositioncoeff}{\ensuremath{\theta_{\itemi}}\xspace}  
\newcommand{\sfcfrequencybias}{\ensuremath{\mathbf{w}}\xspace} 
\newcommand{\rllongtail}[1]{\ensuremath{LT_{#1}}\xspace}

\newcommand{\datasetyahoorsixa}{\textsc{Yahoo! - R6A}\xspace}
\newcommand{\datasetyahoorsixb}{\textsc{Yahoo! - R6B}\xspace}
\newcommand{\datasetcw}{\textsc{ContentWise Impressions}\xspace}
\newcommand{\datasetmind}{\textsc{MIND}\xspace}
\newcommand{\datasetfinn}{\textsc{FINN.no Slates}\xspace}
\newcommand{\datasetslrs}{\textsc{SL4RS}\xspace}
\newcommand{\datasetsearchads}{\textsc{Search Ads}\xspace}
\newcommand{\datasetpandor}{\textsc{PANDOR}\xspace}
\newcommand{\datasetaliccp}{\textsc{Ali-CCP}\xspace}
\newcommand{\datasetalimama}{\textsc{Alimama}\xspace}
\newcommand{\datasetinshop}{\textsc{In-Shop Combo}\xspace}
\newcommand{\datasetcrossshop}{\textsc{Cross-Shop Combo}\xspace}
\newcommand{\datasetkwaisystem}{\textsc{Kwai\_FAIR System}\xspace}
\newcommand{\datasetkwairandom}{\textsc{Kwai\_FAIR Experiment}\xspace}

\newcommand{\datasetlinkedinpymk}{\textsc{LinkedIn PYMK}\xspace}
\newcommand{\datasetlinkedinendorsement}{\textsc{LinkedIn Endorsement}\xspace}
\newcommand{\datasettaobaoindustrial}{\textsc{Taobao Industrial}\xspace}
\newcommand{\datasetlreconeb}{\textsc{LRec1-B}\xspace}
\newcommand{\datasetwtsoneb}{\textsc{WTS-1B}\xspace}
\newcommand{\datasetarticlethreem}{\textsc{Article-333M}\xspace}
\newcommand{\datasetavazu}{\textsc{Avazu}\xspace}

\newcommand{\datasetxingsixteen}{\textsc{XING 2016}\xspace}
\newcommand{\datasetxingseventeen}{\textsc{XING 2017}\xspace}
\newcommand{\datasettrivagonineteen}{\textsc{Trivago 2019}\xspace}
\newcommand{\datasetsharechattwentythree}{\textsc{ShareChat 2023}\xspace}

\title{Impression-Aware Recommender Systems}

\author{{Fernando B.} {P{\'{e}}rez Maurera}}
\orcid{0000-0001-6578-7404}
\affiliation{%
  \institution{Politecnico di Milano}
  \city{Milan}
  \country{Italy}
}
\affiliation{%
  \institution{ContentWise}
  \city{Milan}
  \country{Italy}
}
\email{fernandobenjamin.perez@polimi.it}

\author{Maurizio {Ferrari Dacrema}}
\orcid{0000-0001-7103-2788}
\affiliation{%
  \institution{Politecnico di Milano}
  \city{Milan}
  \country{Italy}
}
\email{maurizio.ferrari@polimi.it}

\author{Pablo Castells}
\orcid{0000-0003-0668-6317}
\affiliation{%
\institution{Universidad Aut\'{o}noma de Madrid}
\city{Madrid}
\country{Spain}
}
\affiliation{%
  \institution{Amazon}
  \city{Madrid}
  \country{Spain}
}
\email{pablo.castells@uam.es}

\author{Paolo Cremonesi}
\orcid{0000-0002-1253-8081}
\affiliation{%
  \institution{Politecnico di Milano}
  \city{Milan}
  \country{Italy}
}
\email{paolo.cremonesi@polimi.it}

\renewcommand{\shortauthors}{{P{\'{e}}rez Maurera}, et al.}

\begin{abstract}
  Novel data sources bring new opportunities to improve the quality of recommender systems and serve as a catalyst for the creation of new paradigms on personalized recommendations.
  Impressions are a novel data source containing the items shown to users on their screens.
  Past research focused on providing personalized recommendations using interactions, and occasionally using impressions when such a data source was available.
  Interest in impressions has increased due to their potential to provide more accurate recommendations.
  Despite this increased interest, research in recommender systems using impressions is still dispersed.
  Many works have distinct interpretations of impressions and use impressions in recommender systems in numerous different manners.
  To unify those interpretations into a single framework, we present a systematic literature review on recommender systems using impressions, focusing on three fundamental perspectives: \emph{recommendation models}, \emph{datasets}, and \emph{evaluation methodologies}.
  We define a theoretical framework to delimit recommender systems using impressions and a novel paradigm for personalized recommendations, called impression-aware recommender systems.
  We propose a classification system for recommenders in this paradigm, which we use to categorize the recommendation models, datasets, and evaluation methodologies used in past research.
  Lastly, we identify open questions and future directions, highlighting missing aspects in the reviewed literature.
\end{abstract}


\received{05 February 2024}
\received[revised]{08 August 2024}
\received[accepted]{07 December 2024}

\maketitle

\section{Introduction}
\label{sec:introduction}

Collaborative filtering can be seen as a paradigmatic approach to personalized recommendations, where the system tracks the \emph{interactions} with the available options and predicts good choices for individuals by cross-examining the activity of all users.
Examples of interactions are the purchase of products (termed \emph{implicit} interactions) or ratings that users emit to convey their level of satisfaction with products (termed \emph{explicit} interactions).
Despite the success of collaborative filtering models, they present several limitations~\cite{DBLP:conf/recsys/AharonABLABLRS13/non-impressions/offset-algorithm-yahoo,DBLP:conf/sigir/LiCLXZ22/deep-learning/two-tower-context-aware-cold-start,DBLP:journals/tois/WangM00M23/a-survey-on-the-fairness-of-recsys}.
For instance, they tend to recommend popular items, create filter bubbles, or fail to recommend relevant items to cold users.
Previous research works~\cite{DBLP:reference/sp/AdomaviciusBTU22/context-aware-recommender-systems-from-foundations-to-recent-developments,DBLP:journals/kbs/VillegasSDT18/non-impressions/survey-context-aware-characterizing-context-aware-recommender-systems-a-systematic-literature-review} propose partial solutions to those limitations using additional data sources, \eg metadata of products, location of users, social connections between users, among others.
Past research~\cite{DBLP:conf/recsys/NingK12/non-impressions/slim-with-side-information-for-top-n-recommendations} has shown that using additional data sources beyond interactions may improve the quality of collaborative filtering recommenders.
In this work, we survey the type of recommender systems using \emph{impressions} as an additional data source.
An \emph{impression} is a collection of items shown on-screen to a given user at a particular time.
A user is said to be \emph{impressed} to each item inside the impression.
In the literature, impressions are also called \emph{past recommendations}~\cite{DBLP:conf/recsys/SatoSTSZO19/others/uplift-based-evaluation-and-optimization-of-recommenders}, \emph{previous recommendations}~\cite{DBLP:conf/cikm/AzzaliniACDMA22/others/socrate-a-recommendation-system-with-limited-availability-items}, \emph{exposures}~\cite{DBLP:conf/www/LiangCMB16/heuristics/modeling-user-exposure-in-recommendations,DBLP:conf/sigir/ShihHJLLC16/heuristics/integrating-exposure-to-rating-prediction}, or
\emph{slates}~\cite{DBLP:conf/recsys/EideLFRJV21/finn-no-slates-dataset,DBLP:journals/datamine/EideLF22/dynamic-slate-recommendation-with-gated-recurrent-units-and-thompson-sampling,DBLP:journals/tois/0003YW0RLSCR23/impression-aware/on-the-behavior-leakage-from-recommender-system-exposure}.

\begin{figure}[t]
  \centering
  \ifusetikz \tikzsetnextfilename{feedback-levels}
\tikzset{
  pattern size/.store in=\mcSize,
  pattern size = 5pt,
  pattern thickness/.store in=\mcThickness,
  pattern thickness = 0.3pt,
  pattern radius/.store in=\mcRadius,
  pattern radius = 1pt}\makeatletter
\pgfutil@ifundefined{pgf@pattern@name@__matchapatterngrid}{
  \pgfdeclarepatternformonly[\mcThickness,\mcSize]{__matchapatterngrid}
  {\pgfqpoint{-\mcThickness}{-\mcThickness}}
  {\pgfpoint{\mcSize}{\mcSize}}
  {\pgfpoint{\mcSize}{\mcSize}}
  {\pgfsetcolor{\tikz@pattern@color}
    \pgfsetlinewidth{\mcThickness}
    \pgfpathmoveto{\pgfpointorigin}
    \pgfpathlineto{\pgfpoint{\mcSize}{0}}
    \pgfpathmoveto{\pgfpointorigin}
    \pgfpathlineto{\pgfpoint{0}{\mcSize}}
    \pgfusepath{stroke}}}
\makeatother

\tikzset{
  pattern size/.store in=\mcSize,
  pattern size = 5pt,
  pattern thickness/.store in=\mcThickness,
  pattern thickness = 0.3pt,
  pattern radius/.store in=\mcRadius,
  pattern radius = 1pt}
\makeatletter
\pgfutil@ifundefined{pgf@pattern@name@__matchapatterndots}{
  \makeatletter
  \pgfdeclarepatternformonly[\mcRadius,\mcThickness,\mcSize]{__matchapatterndots}
  {\pgfpoint{-0.5*\mcSize}{-0.5*\mcSize}}
  {\pgfpoint{0.5*\mcSize}{0.5*\mcSize}}
  {\pgfpoint{\mcSize}{\mcSize}}
  {
    \pgfsetcolor{\tikz@pattern@color}
    \pgfsetlinewidth{\mcThickness}
    \pgfpathcircle\pgfpointorigin{\mcRadius}
    \pgfusepath{stroke}
  }}
\makeatother

\begin{tikzpicture}[
    x=0.5pt,
    y=0.5pt,
    yscale=-1,
    xscale=1
  ]

  \draw  [fill={rgb, 255:red, 208; green, 223; blue, 227 }  ,fill opacity=1 ] (51,154) .. controls (51,78.34) and (195.39,17) .. (373.5,17) .. controls (551.61,17) and (696,78.34) .. (696,154) .. controls (696,229.66) and (551.61,291) .. (373.5,291) .. controls (195.39,291) and (51,229.66) .. (51,154) -- cycle ;
  \draw  [fill={rgb, 255:red, 217; green, 234; blue, 211 }  ,fill opacity=1 ][dash pattern={on 4.5pt off 4.5pt}] (199,153.5) .. controls (199,91.92) and (308.91,42) .. (444.5,42) .. controls (580.09,42) and (690,91.92) .. (690,153.5) .. controls (690,215.08) and (580.09,265) .. (444.5,265) .. controls (308.91,265) and (199,215.08) .. (199,153.5) -- cycle ;
  \draw  [fill={rgb, 255:red, 255; green, 242; blue, 204 }  ,fill opacity=1 ][dash pattern={on 0.84pt off 2.51pt}] (339.5,150) .. controls (339.5,103.61) and (415.05,66) .. (508.25,66) .. controls (601.45,66) and (677,103.61) .. (677,150) .. controls (677,196.39) and (601.45,234) .. (508.25,234) .. controls (415.05,234) and (339.5,196.39) .. (339.5,150) -- cycle ;
  \draw    (69.97,229.55) -- (131.5,153) ;
  \draw [shift={(131.5,153)}, rotate = 308.79] [color={rgb, 255:red, 0; green, 0; blue, 0 }  ][fill={rgb, 255:red, 0; green, 0; blue, 0 }  ][line width=0.75]      (0, 0) circle [x radius= 3.35, y radius= 3.35]   ;
  \draw    (176.23,273) -- (184.5,218) ;
  \draw [shift={(184.5,218)}, rotate = 278.55] [color={rgb, 255:red, 0; green, 0; blue, 0 }  ][fill={rgb, 255:red, 0; green, 0; blue, 0 }  ][line width=0.75]      (0, 0) circle [x radius= 3.35, y radius= 3.35]   ;
  \draw    (176.23,273) -- (237.5,195) ;
  \draw [shift={(237.5,195)}, rotate = 308.15] [color={rgb, 255:red, 0; green, 0; blue, 0 }  ][fill={rgb, 255:red, 0; green, 0; blue, 0 }  ][line width=0.75]      (0, 0) circle [x radius= 3.35, y radius= 3.35]   ;

  \draw (118,39.4) node [anchor=north west][inner sep=0.75pt]    {\textit{A}};
  \draw (233,70.4) node [anchor=north west][inner sep=0.75pt]    {\textit{B}};
  \draw (339,100.4) node [anchor=north west][inner sep=0.75pt]    {\textit{C}};
  \draw (55,232.4) node [anchor=north west][inner sep=0.75pt]    {\textit{D}};
  \draw (164,275.4) node [anchor=north west][inner sep=0.75pt]    {\textit{E}};

  \draw (21,304.4) node [anchor=north west][inner sep=0.75pt]    {\textit{A:} \textbf{Catalog}};
  \draw (130,305.4) node [anchor=north west][inner sep=0.75pt]    {\textit{B:} \textbf{Impressions}};
  \draw (263,305.4) node [anchor=north west][inner sep=0.75pt]    {\textit{C:} \textbf{Interactions}};
  \draw (396,305.4) node [anchor=north west][inner sep=0.75pt]    {\textit{D:} \textbf{Non-impressions}};
  \draw (572,305.4) node [anchor=north west][inner sep=0.75pt]    {\textit{E:} \textbf{Non-interactions}};

\end{tikzpicture} \fi 
  \ifusepdf \includegraphics{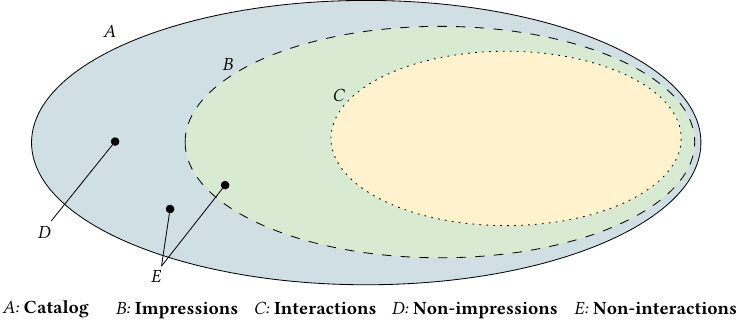} \fi
  \ifusepng \includegraphics{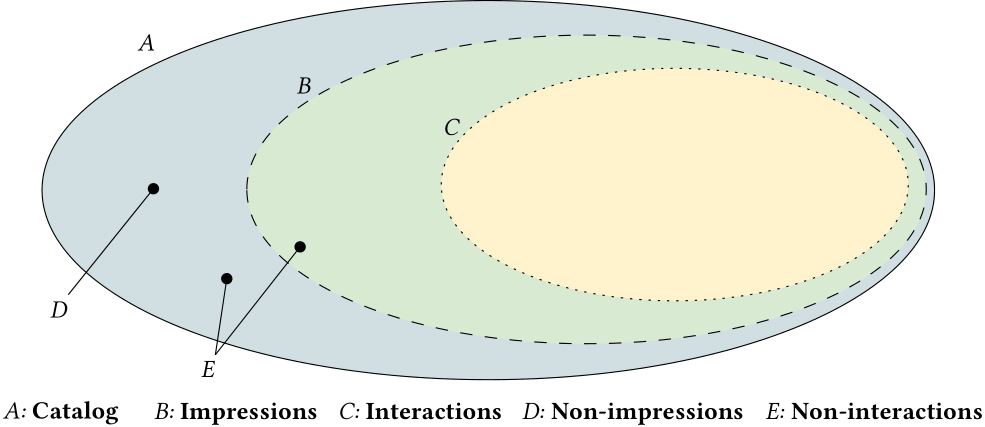} \fi 
  \caption{
    Categorization of items in a recommender system using impressions.
    \textbf{Catalog} are all items (solid line).
    \textbf{Impressions} are shown items (dashed line).
    \textbf{Interactions} are shown and interacted items (dotted line).
    \textbf{Non-impressions} are not shown items ($D = A - B$, \idest between solid and dashed lines).
    \textbf{Non-interactions} are not interacted items ($E = A - C$, \idest between solid and dotted lines).
  }
  \Description[Five user-item feedback levels]{User-item feedback is classified into five groups: Catalog, Impressions, Interactions, Non-impressions, and Non-interactions}
  \label{fig:feedback-levels-diagram}
\end{figure}

Certain data sources may serve as catalysts to accelerate the creation of novel paradigmatic approaches to personalized recommendations.
Impressions are a data source that fosters the creation and research of a novel learning paradigm for personalized recommendations.
We call such a paradigm \gls{impressionsbased}.
It encapsulates those recommender systems that leverage impressions and interactions to learn users' preferences, regardless of the system's complexity or how they process such data sources.
In this learning paradigm, impressions provide many benefits and innovation potential compared to interactions alone in at least three aspects.
First, impressions allow the exploration of characteristics of recommender systems that are often unexplored: the items shown on-screen to users, their arrangement, and how users interact with them.
Second, impressions enable a refined modeling of users' preferences due to the signals impressions carry.
Third, impressions partition the user feedback into further granular levels of users, as shown in \autoref{fig:feedback-levels-diagram}.
The literature contains studies about the new capabilities and opportunities available when using impressions.
For instance, \citet{DBLP:conf/recsys/ZhaoWAHK18/impressions-signals/interpreting-user-inaction-in-recsys} study why users interact with some impressions and not others, while other authors propose methods to learn whether users value impressions positively or negatively~\cite{DBLP:conf/iir/PerezMaureraFDC22/replication-of-impressions,DBLP:conf/recsys/PerezMaureraFDC22/towards-the-evaluation-of-recommender-systems-with-impressions}, and others have researched whether users' perception of impressions change after repeated exposures to the same item~\cite{DBLP:conf/cikm/AharonKLSBESSZ19/heuristics/soft-frequency-cap,DBLP:conf/kdd/LeeLTS14/impression-discounting}.

To the best of our knowledge, this work is the first systematic literature review of impressions in recommender systems.
Other papers~\cite{DBLP:journals/ipm/YangZ22/non-impressions/survey-on-ctr-prediction-in-online-advertising,DBLP:journals/csur/GharibshahZ21/non-impressions/survey-ctr-prediction-user-response-prediction-in-online-advertising} review impressions on non-personalized services focused on click-through rate (ratio between interactions and impressions) prediction, while some reviews study impressions in other types of non-personalized services, \eg online advertisements~\cite{DBLP:conf/www/RichardsonDR07/non-impressions-recsys/overview-of-impressions,DBLP:conf/imc/KrishnanS13/non-impressions/ctr-prediction-video-ads}, search engines~\cite{DBLP:conf/icml/GraepelCBH10/non-impressions/predicting-ctr-in-search-engine-at-web-scale,DBLP:conf/sigir/KonigGW09/non-impressions/related-works/ctr-prediction-for-news-queries}, social media~\cite{DBLP:conf/kdd/LiLMWP15/non-impressions/ctr-prediction-for-social-media}, and others~\cite{DBLP:conf/sigir/LiuXGTZHL20/non-impressions/ctr-prediction-for-downloads,DBLP:journals/datasci/BarbaroGMSWH20/non-impressions/ctr-prediction-mobile-apps}.

Despite existing many research works on impressions in recommender systems in the literature, \eg the papers of \citet{DBLP:conf/kdd/LeeLTS14/impression-discounting}, \citet{DBLP:conf/recsys/WuASB16/user-fatigue/netflix-fatigue-modeling-using-impressions}, or \citet{DBLP:conf/recsys/ZhaoWAHK18/impressions-signals/interpreting-user-inaction-in-recsys}, existing research is dispersed, presents different terminology, and covers distinct and unrelated topics.
Moreover, the effective use of impressions in recommender systems is still in a nascent state, leading to the under-utilization of this data source.
This work proactively introduces and defines the \gls{impressionsbased} learning paradigm, \idest a paradigm that covers all those recommendation models that leverage impressions to produce relevant and personalized recommendations to users.
Such a proactive approach reduces the friction existing nowadays and provides a comprehensive framework to classify and study impressions in recommender systems for the future.
Consequently, this work aims at unifying the existing research into a single document by reviewing and analyzing  \gls{impressionsbased} under three fundamental topics in recommender systems: recommendation models, public datasets, and evaluation methodologies.

\subsection{Organization}
\label{subsec:introduction:organization}

As previously mentioned, we study \emph{recommendation models}, \emph{datasets with impressions}, and \emph{evaluation methodologies} of \gls{impressionsbased}.
We first define a mathematical framework to describe this type of recommenders in \autoref{sec:impressions-based-recsys}.
Then, we state the similarities and differences of \gls{impressionsbased} with related types of recommenders under the framework.
Lastly, we describe our proposed taxonomies to categorize the literature in this area.
Each taxonomy classifies papers based on different aspects.

We categorize the relevant literature according to the taxonomies and describe each work in \autoref{sec:reviewed-papers}.
For this purpose, we discuss each work in six dimensions, each corresponding to one category of one of the proposed taxonomies.
In doing so, we first identify and describe common patterns in papers, \eg whether several papers use the same recommender or technique.
Then, we describe each contribution in more detail regarding those patterns.

We present and classify the datasets with impressions in \autoref{sec:datasets-with-impressions}.
In particular, the section emphasizes the discussion upon \emph{public} datasets: published datasets accessible via the Internet or by request to publishers.
We propose a categorization of public datasets based on their type of impressions.
We briefly present other types of datasets, \eg those used in competitions or never published.

We analyze the challenges, opportunities, and special considerations when evaluating \gls{impressionsbased} in \autoref{sec:evaluation}.
In particular, we focus on the most common research goals and provide guidelines to ensure the evaluation of \gls{impressionsbased} is executed correctly in the published research.
We describe the two most common research goals followed by the reviewed publications.
Our discussion includes guidelines for future works to ensure evaluation methodologies are consistent with the research goals.
We close the section with descriptions of several challenges contended when using impressions.

In \autoref{sec:future-directions}, we present open research questions and future research directions.
We describe how impressions enable us to pursue novel directions in recommender system research.
Notably, with impressions, we have access to items exposed to users, their frequency, and, in some situations, their arrangement on-screen.
This information is crucial in recommender systems, as it alleviates particular roadblocks in the literature.

\subsection{Contributions}
\label{subsec:introduction:contributions}

We undertake a systematic literature review on \gls{impressionsbased}, collecting, discussing, and analyzing relevant work in this area.
We identify recurrent topics in the reviewed literature and discuss future research directions.
Specific sought contributions in this work include:
\begin{itemize}
  \item A theoretical framework of \gls{impressionsbased} aiming to unify diverse conceptual representations existing in prior works.

  \item A comprehensive characterization of \gls{impressionsbased} under different perspectives. Those perspectives include the design of recommenders, handling of impressions, or users' preferences toward impressions.

  \item A review of topics to improve the research quality in future works. Those topics include current public datasets with impressions and evaluation methodologies.

  \item A thorough analysis of current trends and open research questions. Those illustrate the short and long-term topics of interest for future research.
\end{itemize}

\subsection{Paper Selection Criteria}
\label{subsec:introduction:paper-selection-criteria}

\begin{figure*}[t]
  \centering
  \begin{subfigure}[t]{0.49\textwidth}
    \centering
    \ifusetikz \tikzsetnextfilename{papers-by-year-csv}
\pgfkeys{/pgf/number format/.cd,1000 sep={\,}}

\begin{tikzpicture}

  \definecolor{chocolate2267451}{RGB}{226,74,51}
  \definecolor{dimgray85}{RGB}{85,85,85}
  \definecolor{gainsboro229}{RGB}{229,229,229}

  \begin{axis}[
      width=0.9\textwidth,
      height=6cm,
      xbar,
      xmin=0,
      axis background/.style={fill=gainsboro229},
      axis line style={white},
      tick align=outside,
      tick pos=left,
      xlabel=\textcolor{dimgray85}{Number of papers},
      ylabel=\textcolor{dimgray85}{Publication year},
      xmajorgrids,
      ymajorgrids,
      xtick style={color=dimgray85},
      ytick style={color=dimgray85},
      x grid style={white},
      y grid style={white},
      xtick={0,2,4,6,8,10,12,14},
      ytick=data,
      nodes near coords, nodes near coords align={horizontal},
      bar width=6pt,
    ]
    \addplot [xbar, draw=none,fill=chocolate2267451, very thin] table [x=num_papers, y=year, col sep=comma] {figures/data/papers-by-year.csv};
  \end{axis}
\end{tikzpicture} \fi 
    \ifusepdf \includegraphics{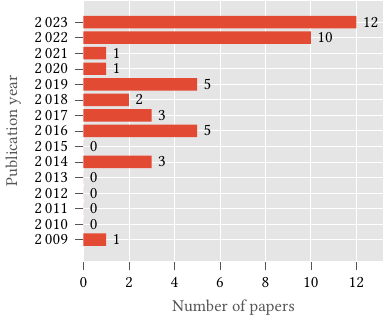} \fi
    \ifusepng \includegraphics{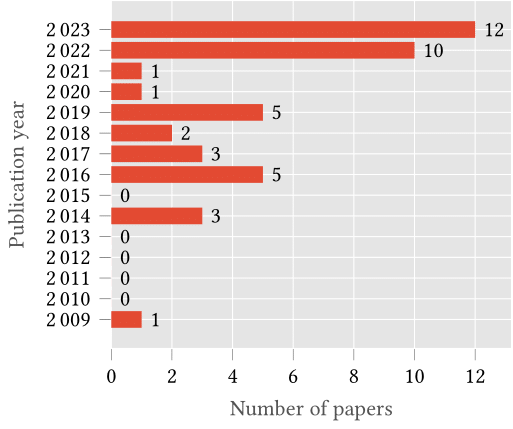} \fi 
  \end{subfigure}
  \hfill
  \begin{subfigure}[t]{0.49\textwidth}
    \centering
    \ifusetikz \tikzsetnextfilename{papers-by-venue-csv}
\pgfkeys{/pgf/number format/.cd,1000 sep={\,}}

\begin{tikzpicture}

  \definecolor{chocolate2267451}{RGB}{226,74,51}
  \definecolor{dimgray85}{RGB}{85,85,85}
  \definecolor{gainsboro229}{RGB}{229,229,229}

  \begin{axis}[
      width=0.9\textwidth,
      height=6cm,
      xbar,
      xmin=0,
      axis background/.style={fill=gainsboro229},
      axis line style={white},
      tick align=outside,
      tick pos=left,
      xlabel=\textcolor{dimgray85}{Number of papers},
      ylabel=\textcolor{dimgray85}{Venue},
      xmajorgrids,
      ymajorgrids,
      xtick style={color=dimgray85},
      ytick style={color=dimgray85},
      x grid style={white},
      y grid style={white},
      xtick={0,2,4,6,8,10,12},
      ytick=data,
      symbolic y coords={TOIS,J. Scheduling,Information Sciences,CSCW,IJCAI,WWW,RecSys,CIKM,SIGIR,WSDM,KDD},
      nodes near coords, nodes near coords align={horizontal},
      bar width=6pt,
    ]
    \addplot [xbar, draw=none,fill=chocolate2267451, very thin] table [x=num_papers, y=venue, col sep=comma] {figures/data/papers-by-venue.csv};
  \end{axis}

\end{tikzpicture} \fi 
    \ifusepdf \includegraphics{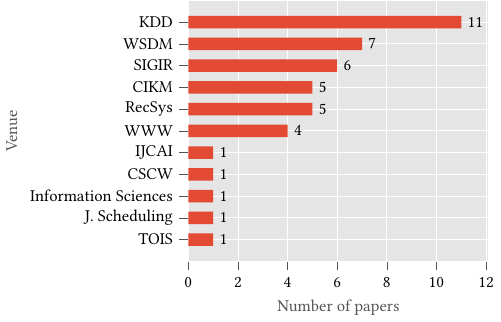} \fi
    \ifusepng \includegraphics{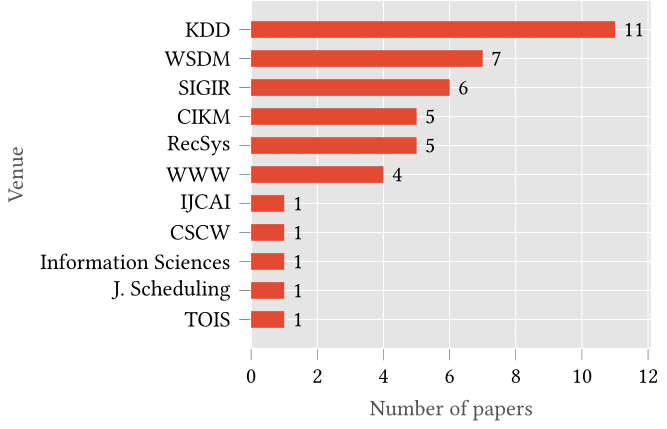} \fi 
  \end{subfigure}
  \caption{Number of papers reviewed in this work by their publication year (left) and venue (right).}
  \Description[Reviewed papers arranged by publication year or venue]{Reviewed papers arranged by publication year: 2023 with 12 papers; 2022 with ten papers; 2021 with one paper; 2020 with one paper; 2019 with five papers; 2018 with two papers; 2017 with three papers; 2016 with five papers; 2015 with 0 papers; 2014 with three papers; 2013 with 0 papers; 2012 with 0 papers; 2011 with 0 papers; 2010 with 0 papers; 2009 with one paper. Reviewed papers arranged by venue: KDD with 11 papers; WSDM with seven papers; SIGIR with six papers; CIKM with five papers; RecSys with five papers; WWW with four papers; IJCAI with one paper; CSCW with one paper; Information Sciences with one paper; J. Scheduling with one paper; TOIS with one paper.}
  \label{fig:tikz-reviewed-papers}
\end{figure*}

The selected \gls{impressionsbased} papers in this literature review conform to the following selection criteria: papers are peer-reviewed, of regular conference or journal types, and published in top-tier venues.
For our purpose, we select top-tier A* and A conferences in the CORE 2021 ranking,\footnote{The \acrfull{core} 2021 ranking is available at: \url{https://portal.core.edu.au/conf-ranks/}} and Q1 journals in the Scimago 2021 ranking in computer science.\footnote{The Scimago 2021 ranking in computer science is available at: \url{https://www.scimagojr.com/journalrank.php?area=1700}}

We queried five popular academic search engines to retrieve candidate papers related to \gls{impressionsbased}.
Specifically, we queried the ACM DL, IEEE Xplore, ScienceDirect, SpringerLink, and Google Scholar.
For each search engine, We built a search query matching the keyword \emph{recommender system} with keywords related to impressions, namely \emph{impression, exposure, slate, past recommendation}, or \emph{previous recommendation}.\footnote{The specific search query is: recommender system AND (impression OR exposure OR slate OR past recommendation OR previous recommendation)}
When possible, we instructed the search engine to match such keywords in the papers' titles, abstracts, or contents.
When retrieving papers, we applied the selection criteria to keep those papers conforming to them.
Lastly, we manually inspected the remaining papers to ensure they were relevant to this work.
The manual inspection discarded most papers as the keywords have several meanings in recommender systems, \idest they are used to represent other topics or concepts.
For instance, one paper~\cite{DBLP:journals/cii/ChangLW16/non-impressions/returned-from-search-but-not-ibrs} has the \emph{exposure} and \emph{recommender system} keywords in its text.
Still, the paper does not publish a dataset with impressions or describe an \gls{impressionsbased}.
Instead, the paper uses the keyword exposure to indicate the number of interactions of items.

Overall, we collected \numuniqueworks unique papers from all search engines.
After applying the selection criteria, we kept \num{352} papers.
Lastly, after manually selecting relevant papers, we kept \numincludedworks papers.
This work reviews and discusses those \numincludedworks papers.
In \autoref{fig:tikz-reviewed-papers}, we show the distribution of selected papers by their publication year and venue.

\section{Impression-Aware Recommender Systems}
\label{sec:impressions-based-recsys}

This section presents \acrfull{impressionsbased}, a novel learning paradigm for personalized recommendations.
The first part of the section focuses on the definition and comparison of \gls{impressionsbased} to other paradigms.
Particularly, we propose a theoretical framework to define \gls{impressionsbased} and unify the different interpretations existing in the literature.
Later, we use the same framework to identify and describe the similarities and differences of this paradigm with others.
Then, we present a unique and novel classification system for recommender systems using impressions.
In such a classification system, we categorize papers describing \gls{impressionsbased} under three properties: the design of recommendation models, how they use impressions data, and their stance on impressions.
We close the section by presenting the classification of reviewed papers according to our classification system.

\subsection{Theoretical Framework}
\label{subsec:impressions-based-recsys:problem-description}

A recommender system is a collection of software tools providing personalized selections of items to users based on their past preferences~\cite{DBLP:reference/sp/RicciRS22/non-impressions/recsys-handbook-chapter-1}, \eg recommendations based on previously listened songs.
In this work, we focus on the task of \emph{top-N recommendations}: a scenario where the goal of the recommender system is to generate a selection of $N$ relevant items to the user.
Such selection of items is referred to as \emph{recommendation}, and having as synonyms in the literature as \emph{impression}~\cite{DBLP:conf/kdd/LeeLTS14/impression-discounting,DBLP:conf/recsys/SatoSTSZO19/others/uplift-based-evaluation-and-optimization-of-recommenders}, \emph{exposure}~\cite{DBLP:conf/www/LiangCMB16/heuristics/modeling-user-exposure-in-recommendations,DBLP:conf/sigir/ShihHJLLC16/heuristics/integrating-exposure-to-rating-prediction}, or \emph{slate}~\cite{DBLP:conf/recsys/EideLFRJV21/finn-no-slates-dataset,DBLP:journals/datamine/EideLF22/dynamic-slate-recommendation-with-gated-recurrent-units-and-thompson-sampling,DBLP:journals/tois/0003YW0RLSCR23/impression-aware/on-the-behavior-leakage-from-recommender-system-exposure}.
The recommender system generates the impression, sends it to the user's device, and the device arranges the impression on its screen.
This work assumes an impression is presented as a list: an ordered selection of items sorted by decreasing user relevance.
However, we identify the existence of other arrangements for impressions, \eg a single item or a grid.

\subsubsection{Relevant Terms}
An impression is a single selection of N items; hence, the term \emph{impressions} refers to several selections of N items.
The term \emph{impressed item} refers to a single item inside an impression.
The term \emph{interaction} is any action users perform on impressed items, \eg playing songs or purchasing products.
As users may decide to interact with none, some, or all impressed items inside an impression, we use the terms \emph{interacted impression} and \emph{non-interacted impression} to denote whether the user has interacted with an item in the impression.\footnote{The term \emph{interacted impression} refers to the same concept as the term \emph{interaction} traditionally used in the literature.}

Due to different logging policies, recommender systems may record their impressions and user interactions using various granularity levels.
We use the term \emph{impressions type} to classify impressions into two groups: \emph{contextual} and \emph{global} impressions.
Both groups contain impressions and interactions, where contextual impressions contain the connections between an impression and the interactions its item receives, and global impressions do not.

We use the term \emph{impressions signals} to categorize users' preferences on non-interacted impressions in two levels: \emph{negative}, \emph{neutral}, or \emph{positive}.
A negative signal indicates the user dislikes a non-interacted impression,
A neutral signal indicates the user has no positive or negative preference for the impressed item.
A positive signal indicates the user likes a non-interacted impression but does not interact with it.
In the literature, interactions (hence, interacted impressions) are already considered positive signals.
For instance, the most common assumption of \emph{missing as negatives} in the literature deems interactions as positive and non-interacted items as negative signals.
The literature does not agree on the signal of non-interacted impressions.
In this work, we do not assume a particular signal for them; instead, we study how reviewed papers deem them.

We use the term \emph{recommender type} to classify recommenders into three groups based on how they generate an impression: \emph{end-to-end}, \emph{plug-in}, and \emph{re-ranking}.
End-to-end recommenders generate an impression themselves.
Plug-in recommenders generate an impression by transforming the relevance of items created by another recommender (or a search engine or an editor).\footnote{Plug-in recommenders are not exclusive of \gls{impressionsbased}. They are commonly used in context-aware recommenders and are called \emph{contextual post-filtering}~\cite{DBLP:reference/sp/AdomaviciusBTU22/context-aware-recommender-systems-from-foundations-to-recent-developments}.}
Re-ranking recommenders generate a permutation of an impression generated by another recommender (or entity).
The difference between the plug-in and re-ranking recommenders is the content and order of items in the impression they generate.

\subsubsection{Mathematical Notation}

Throughout this work, we use several mathematical variables.
We define \numimpressions{} and \numinteractions{} as the total number of impressed items and the total number of interacted items in a system, respectively.
We use subscripts to refine sets, functions, and variables according to a specific user or item.
When referring to a given user, we use the subscript \useru, \eg \numimpressions{\useru} is the number of impressed items for the user \useru.
Similarly, the subscript \itemi refers to a given item.
Lastly, the subscript $\useru, \itemi$ refers to a given user and item, \eg \numimpressions{\useru, \itemi} is the number of times \useru has been impressed with \itemi.

\subsubsection{Formal Definition}

Recommender systems define one set containing all users of the system termed the users' set and denoted as \setusers, another set containing all items in the catalog termed the items' set and denoted as \setitems, and another set containing users' past preferences called users' profiles and denoted as \setuserprofile.
The set of users' profiles contains \emph{events}, denoted as \vectorinteraction, \idest mathematical structures representing behaviors of users with the system, \eg interactions or impressions.
The module of a recommender system generating an impression is called the recommendation model; this module is also known in the literature as the recommendation algorithm, technique, or method.
Each recommendation model defines a function, termed \emph{prediction function} and denoted as $\functionrecommender \colon \setusers \times \setitems \times \setuserprofile \rightarrow \setreals$, mapping users, items, and users' profiles into real values.
For a given user \useru and item \itemi, such real values are called the \emph{predicted relevance}, denoted as \relevancescore, and represent the expected preference of \useru over \itemi.
To generate an impression to a given user, the recommendation model computes the predicted relevance of the user to all items in the catalog and selects those with the highest score.

\gls{impressionsbased} are a recommendation paradigm leveraging \emph{impressions} to learn users' preferences on items.
In other words, and different from traditional recommender systems, \gls{impressionsbased} use impressions and collaborative data as their primary data sources (input) rather than the exclusive use of collaborative data or other data sources.
Still, the product (output) of \gls{impressionsbased}, similar to most recommender systems, are impressions.
In \gls{impressionsbased}, the definitions and notations of the set of users, set of items, set of users' profiles, and prediction function are the same as above (denoted as \setusers, \setitems, \setuserprofile, and \functionrecommender, respectively).

\gls{impressionsbased} define events, denoted as $\vectorinteraction = \left( \useru, \itemi, \relevancescore, \listrecommendations \right)$, as a quadruplet composed of the identifier of a user, the identifier of an item, the user-item predicted relevance and a vector of item identifiers.
The item's identifier represents the interacted item, while the vector of item identifiers represents the impression.
Several constraints exist when defining events depending on the impression type and user feedback.
For a contextual impression, the item must be in the impression; for a global impression, either the item or the impression must be empty.
For an interacted impression, the item and the relevance scores are not empty; for a non-interacted impression, the item and the relevance score are empty.

The definition of events allows for multiple interactions between users and items, \eg with different predicted relevance or impressions.
When duplicated interactions are not allowed or needed, then the definition of an event is a tuple of a predicted relevance and a vector of item identifiers, denoted as $\vectorinteraction_{\useru, \itemi} = \left( \relevancescore, \listrecommendations \right)$.
The same constraints apply to this definition of events.

\begin{figure*}[t]
  \centering
  \ifusetikz \tikzsetnextfilename{recommendation-steps}

\definecolor{myyellow}{rgb}{1, 0.9490196078431372, 0.8}
\definecolor{myred}{rgb}{0.9568627450980393, 0.8, 0.8}
\definecolor{mygreen}{rgb}{0.8509803921568627, 0.9176470588235294, 0.8274509803921568}
\definecolor{myblue}{rgb}{0.8117647058823529, 0.8862745098039215, 0.9529411764705882}

\begin{tikzpicture}[
    auto,
    line/.style={
        draw,
        thick,
        -latex',
        shorten >=2pt
      },
    commonnode/.style={
        rectangle,
        thick,
        text width=9em,
        minimum height=3em,
        align=flush center,
        inner sep=0.5em,
        draw=black,
        font=\small,
      },
    manynode/.style={
        commonnode,
        double copy shadow,
        shadow xshift=2pt,
        shadow yshift=-2pt
      },
    numbernode/.style={
        commonnode,
        rectangle,
        align=flush right,
        text width=8em,
        inner sep=0em,
        draw=white,
        font=\scriptsize,
      },
    processnode/.style={
        commonnode,
        fill=mygreen
      },
    usersprofilesnode/.style={
        manynode,
        fill=myyellow
      },
    tripletnode/.style={
        commonnode,
        fill=myyellow
      },
    singlerelevancenode/.style={
        commonnode,
        fill=myred
      },
    multiplerelevancenode/.style={
        manynode,
        fill=myred
      },
    impressionnode/.style={
        manynode,
        fill=myblue
      },
  ]

  \matrix [column sep=4em,row sep=1em]{
    \node [numbernode] (first) {\textbf{(a) Learning phase}};                                                                            & [-3.5em]
    \node [usersprofilesnode] (usersprofiles) {Users profiles $\left(\setuserprofile\right)$};                                     &
    \node [processnode] (iars) {Impression-aware recommender system};                                                              &
    \node [processnode] (predictionfunction1) {Prediction function $\left(\functionrecommender\right)$};                                      \\

    \node [numbernode] (second) {\textbf{(b) Prediction phase}};                                                                          &
    \node [tripletnode] (tripletuseritem) {User-item-profiles triplet $\left(\useru, \itemi, \setuserprofile\right)$};             &
    \node [processnode] (predictionfunction2) {Prediction function $\left(\functionrecommender\right)$};                           &
    \node [singlerelevancenode] (predictedrelevance) {User-item predicted relevance $\left(\relevancescore\right)$};                          \\

    \node [numbernode] (third) {\textbf{(c) Recommendation phase}};                                                                           &
    \node [multiplerelevancenode] (userrelevances) {User-specific predicted relevances $\left( \vectorrelevancescoreuser\right)$}; &
    \node [processnode] (ranking) {Top-N ranking relevance (highest to lowest)};                                                   &
    \node [impressionnode] (impression) {Impression for user $\left(\vectorimpression_{\useru}\right)$};                                                       \\
  };
  \begin{scope}[every path/.style=line]
    \path (usersprofiles) -- (iars);
    \path (iars) -- (predictionfunction1);

    \path (tripletuseritem) -- (predictionfunction2);
    \path (predictionfunction2) -- (predictedrelevance);

    \path (userrelevances) -- (ranking);
    \path (ranking) -- (impression);
  \end{scope}
\end{tikzpicture} \fi 
  \ifusepdf \includegraphics{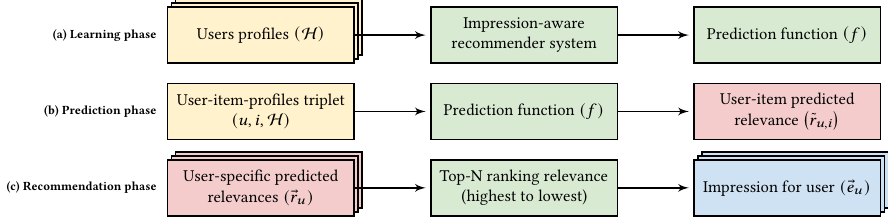} \fi 
  \ifusepng \includegraphics{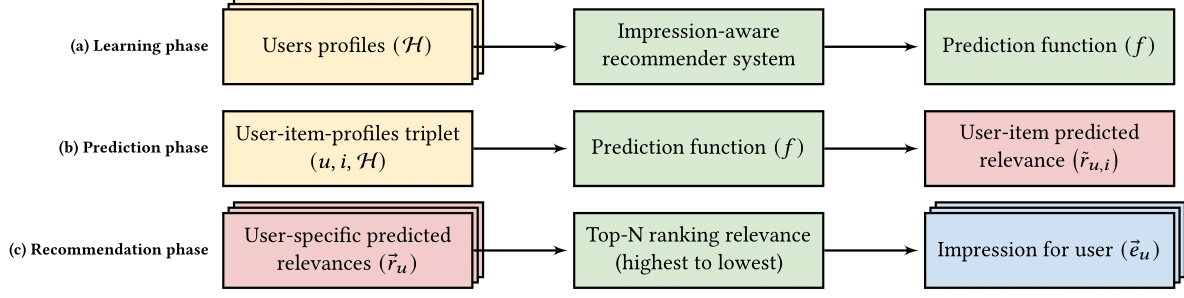} \fi 
  \caption{
    The three phases of any given \gls{impressionsbased} for generating recommendations to a given user.
    \textbf{(a)} illustrates the first phase (\textbf{learning}), where the \gls{impressionsbased} creates a prediction function (\functionrecommender) using the set of users profiles (\setuserprofile).
    \textbf{(b)} illustrates the second phase (\textbf{prediction}), where the \gls{impressionsbased} uses \functionrecommender to predict the relevance score (\relevancescore) of any given user-item-profile triplet.
    \textbf{(c)} illustrates the third phase (\textbf{recommendation}), where the \gls{impressionsbased} generates an impression (recommendation list) to a given user (\useru) by selecting their N-most relevant items based on their predicted relevance scores.
  }
  \Description[Three recommendation phases]{
    The learning Phase starts with a set of users profiles (H), passed as input to an Impression-Aware Recommender System, generating a prediction function (f).
    The prediction phase starts with a user-item-profiles triplet (u, i, H), passed as input to the prediction function (f) (generated in the learning phase), resulting in a predicted user-item relevance score. 
    Lastly, the recommendation phase starts with a set of predicted user-item relevance scores (generated by the prediction phase), passed as input to a Top-N ranker, selecting the Top-N most relevant items; this selection of most relevant items is the resulting impression shown to a given user.}
  \label{fig:tikz-recommendation-steps}
\end{figure*}

\subsection{Recommendation Phases}
\label{subsec:impressions-based-recsys:recommendation-phases}

As \citet{DBLP:reference/sp/RicciRS22/non-impressions/recsys-handbook-chapter-1} states, any kind of recommender system must perform different computations in phases to be able to generate relevant personalized recommendations to a given user.
In particular, \citet{DBLP:reference/sp/RicciRS22/non-impressions/recsys-handbook-chapter-1} identify two stages: \textbf{prediction} and \textbf{recommendation}.
In the prediction stage, the recommender selects all items of the catalog (or a subset of them) and predicts their relevance to any given user \useru.
Formally, as indicated in \autoref{subsec:impressions-based-recsys:problem-description}, for any given user \useru and item \itemi, the recommender predicts \relevancescore.
In the second stage, instead, the recommender takes the computed relevances and selects the top-N items with the highest relevance for any given user.
As mentioned in \autoref{subsec:impressions-based-recsys:problem-description}, this vector of N-most relevant items is called an \emph{impression}.

Based on the theoretical framework described in \autoref{subsec:impressions-based-recsys:problem-description}, recommenders have to go through a further phase prior to the ones identified by \citet{DBLP:reference/sp/RicciRS22/non-impressions/recsys-handbook-chapter-1}.
We call such a stage \textbf{training} where the recommender takes the set of user profiles (\setuserprofile) and creates the prediction function used in the \textbf{prediction} phase.
In other words, this is the stage where the recommender \emph{learns} users' preferences according to the data contained in the users' profiles.

We illustrate the specifics of these three phases for \gls{impressionsbased} in \autoref{fig:tikz-recommendation-steps}.
As seen in the figure, any \gls{impressionsbased} shares the same three phases, differing only in the data sources used in each of them.
As the figure shows, \gls{impressionsbased} leverage \emph{past} impressions inside the set of users' profiles, \idest \setuserprofile, both in the \textbf{learning} and \textbf{prediction} phases.  
However, the figure also shows that \gls{impressionsbased} only use past impressions in all of the three phases.
It is important to stress that impressions are generated as the result of the recommendation phase, which means that those new impressions cannot be available either at learning or prediction. Hence \gls{impressionsbased} can exclusively leverage \emph{previously-generated impressions} to learn and predict users' preferences.
In other words, using an impression outside the set of users' profiles is not possible because the recommender has not generated any new impression and shown it to the user prior to the end of the third phase.

\subsection{Related Recommendation Paradigms}
\label{subsec:impressions-based-recsys:related-domains}

\begin{figure*}[t]
  \centering
  \ifusetikz \tikzsetnextfilename{taxonomy-diagram}
\begin{forest}
  forked edges,
  for tree={draw,align=center,edge={-latex},font=\small}
    [Collaborative filtering\\with side information
        [Impression-aware\\recommender systems
            [Model-centric\\taxonomy]
            [Data-centric\\taxonomy]
            [Signal-centric\\taxonomy]
        ]
        [Context-aware\\recommender systems]
        [Hybrid content-based \\collaborative filtering]
    ]
\end{forest} \fi 
  \ifusepdf \includegraphics{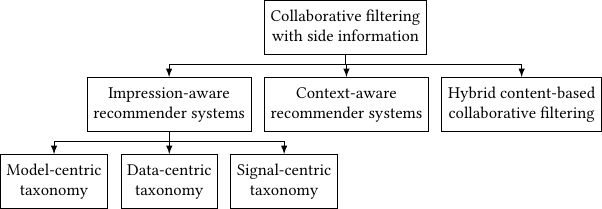} \fi 
  \ifusepng \includegraphics{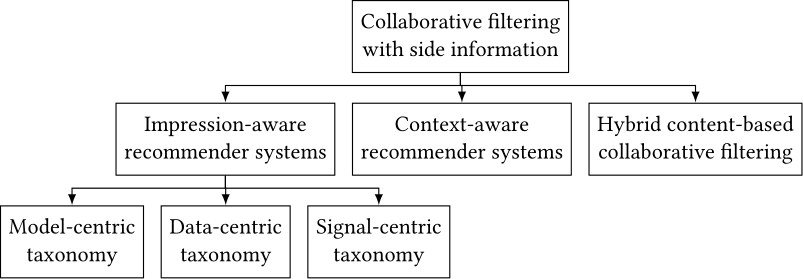} \fi 
  \caption{
    Hierarchy of four learning paradigms, from top to bottom and left to right: \acrfull{cfside}, \acrfull{impressionsbased}, \acrlong{contextaware}, and hybrid content-based collaborative filtering.
    As illustrated, the last three belong to \acrshort{cfside}.
    \gls{impressionsbased} is not equivalent to its sibling paradigms due to the theoretical and practical differences between them, differences that we analyze in \autoref{subsec:impressions-based-recsys:related-domains}.
    Additionally, the diagram places our proposed taxonomies for \acrshort{impressionsbased}, namely model-centric, data-centric, and signal-centric taxonomies; also presented and discussed in \autoref{subsec:impressions-based-recsys:classification}.
  }
  \Description[Hierarchy of learning paradigms]{
    Hierarchy of learning paradigms starting with Collaborative Filtering with side information. 
    That paradigm has three children: Impression-aware recommender systems, context-aware recommender systems, and hybrid content-based collaborative filtering. 
    Impression-aware recommender systems have three children: model-centric taxonomy, data-centric taxonomy, and signal-centric taxonomy. 
    Other learning paradigms are children-less.
  }
  \label{fig:tikz-taxonomy-diagram}
\end{figure*}

\Gls{cf} is the paradigm where recommender systems generate recommendations using interactions.
\Gls{cfside} is an extension of \gls{cf} that uses interactions and additional data sources to generate personalized recommendations~\cite{DBLP:conf/recsys/NingK12/non-impressions/slim-with-side-information-for-top-n-recommendations,DBLP:conf/ijcai/ZhaoXG16/non-impressions/predictive-collaborative-filtering-with-side-information}.
Several paradigms exist within \gls{cfside}, \eg context-aware~\cite{DBLP:reference/sp/AdomaviciusBTU22/context-aware-recommender-systems-from-foundations-to-recent-developments,DBLP:journals/kbs/VillegasSDT18/non-impressions/survey-context-aware-characterizing-context-aware-recommender-systems-a-systematic-literature-review} or hybrid content-based collaborative filtering~\cite{DBLP:conf/iccS/JungPL04/non-impressions/hybrid-cocollaborative-filtering-content-based-filtering}, as illustrated in \autoref{fig:tikz-taxonomy-diagram}.
\gls{impressionsbased} is considered a learning paradigm within \gls{cfside}.
In this section, we analyze why \gls{impressionsbased} is a unique paradigm.
While, at the same time, how it is compatible with other paradigms.

\subsubsection{Comparison With Similar Paradigms}

For a learning paradigm to be equivalent to \gls{impressionsbased}, it is required that such a paradigm share characteristics that can be bijectively projected between \gls{impressionsbased} and such a learning paradigm.
Under our theoretical formulation, this requires that a given learning paradigm shares an equivalent definition of events, the set of user profiles, and the prediction function.
In addition to those, it is also required that the learning paradigm performs equivalent phases when generating recommendations to those performed by \gls{impressionsbased}, as seen in \autoref{subsec:impressions-based-recsys:recommendation-phases}.

After inspecting existing learning paradigms in the recommender systems field, especially those listed by \citet{DBLP:conf/adaptive/Burke07/non-impressions/hybrid-web-recommender-systems} and \citet{DBLP:reference/sp/RicciRS22/non-impressions/recsys-handbook-chapter-1}, we identify one learning paradigm that \emph{is similar but not equivalent} to \gls{impressionsbased}, namely: \Gls{contextaware}.
In particular, \gls{contextaware}~\cite{DBLP:reference/sp/AdomaviciusBTU22/context-aware-recommender-systems-from-foundations-to-recent-developments,DBLP:journals/kbs/VillegasSDT18/non-impressions/survey-context-aware-characterizing-context-aware-recommender-systems-a-systematic-literature-review} are those recommender systems learning from interactions and \emph{contextual} attributes.
Examples of contextual attributes are users' geographical locations or the day and time users access the recommender system~\cite{DBLP:reference/sp/AdomaviciusBTU22/context-aware-recommender-systems-from-foundations-to-recent-developments}.
At a first glance, \gls{impressionsbased} may seem equivalent to \gls{contextaware} where the impressions are the context of the interactions, but this is not the case.
Despite their similarities, both paradigms have theoretical and foundational differences that break a possible equivalence between them, causing \gls{impressionsbased} to be a novel and unique learning paradigm.

In \autoref{tab:comparison-rs-domains}, we list the definitions of events, the set of user profiles, and the prediction function of different recommendation paradigms, including \gls{impressionsbased} and \gls{contextaware}.
From the Table, we can observe the similarities and differences between both recommendation paradigms.
In particular, in terms of their definitions of events, both paradigms share similar but not equivalent definitions, both being quadruplets holding the users and items identifiers, the predicted user-item relevance score, and a vector.
They differ in the contents of the vector, where \gls{impressionsbased} hold an impression, while \gls{contextaware} hold contextual attributes. 
An impression contains a fixed number of item identifiers, while contextual attributes may contain a varying number of features and data types.

In terms of their definitions of prediction function, the differences between their definitions are crucial to establishing the non-equivalence between both paradigms.
As seen in the table, the prediction function of \gls{impressionsbased} predicts the relevance score of any user-item pair by taking a user identifier, an item identifier, and the set of user profiles as input.
Instead, the prediction function of \gls{contextaware} takes an additional argument, a \emph{vector of contextual features} computed on the user-item pair before predicting its relevance.
In other words, this vector is outside the set of user profiles and is \emph{computed} and \emph{used} at the prediction phase of the recommender.
In contrast, as seen in \autoref{fig:tikz-recommendation-steps}, \gls{impressionsbased} compute a new impression \emph{after} predicting the relevance scores of user-item pairs; thus, it is impossible to use such new impressions when predicting the relevance scores associated to them because they are not known yet.\footnote{See \autoref{subsec:impressions-based-recsys:recommendation-phases} for a description of recommender's phases and \autoref{fig:tikz-recommendation-steps} for an illustration of all phases.}

\subsubsection{Compatibility With Other Paradigms}

Despite their uniqueness, \gls{impressionsbased} are compatible with other paradigms, \eg sequence-aware or session-based recommenders.
In this context, compatibility indicates that a recommender system may be from any given paradigm and include impressions in any of their phases to produce relevant recommendations.
Essentially, such a recommender system would retain its learning paradigm and become impression-aware as well.

\begin{table}[t]
  \centering
  \small
  \caption{
    Comparison of the definitions of an event, the users' profile, and the prediction function between \gls{impressionsbased} and similar types of recommenders.
    \useru is a user,
    \itemi is an item,
    \relevancescore is a real number,
    \listsideinfo is side information as a vector of features,
    \listrecommendations is an impression as a vector of item  identifiers,
    and \listcontext and \listcontexttwo are two different vectors of contextual features.
  }
  \label{tab:comparison-rs-domains}
  \begin{minipage}{\linewidth}
    \centering
    \begin{tabular}{lcccc}
      \toprule
      {}
        &  \textbf{\thead{Collaborative\\ filtering}}
        &  \textbf{\thead{Collaborative filtering\\ with side information}}
        &  \textbf{\thead{Impression-aware\\ recommender systems}}
        &  \textbf{\thead{Context-aware\\ recommender systems}}
      \\
      \midrule
      Event (\vectorinteraction)
        &  $\left( \useru, \itemi, \relevancescore  \right)$
        &  $\left( \useru, \itemi, \relevancescore, \listsideinfo  \right)$
        &  $\left( \useru, \itemi, \relevancescore, \listrecommendations \right)$
        &  $\left( \useru, \itemi, \relevancescore, \listcontext \right)$
      \\
      \midrule
      Users profile (\setuserprofile)
        &  \multicolumn{4}{c}{
      $\left\{ \vectorinteraction \right\}$
      }
      \\
      \midrule
      Prediction function (\functionrecommender)
        &  $\functionrecommender \left( \useru, \itemi, \setuserprofile \right)$
        &  $\functionrecommender \left( \useru, \itemi, \setuserprofile, \listsideinfo \right)$
        &  $\functionrecommender \left( \useru, \itemi, \setuserprofile \right)$
        &  $\functionrecommender \left( \useru, \itemi, \setuserprofile, \listcontexttwo \right)$
      \\
      \bottomrule
    \end{tabular}
  \end{minipage}
\end{table}

The compatibility of impressions as a data source with other learning paradigms is possible due to the broad definitions of events in \gls{impressionsbased}: they require impressions, regardless of their representation, \eg whether the recommender uses contextual or global impressions.\footnote{We define contextual and global impressions in \autoref{subsec:impressions-based-recsys:problem-description}.}
In fact, recommenders incorporating impressions may benefit from this additional data source without deviating from their original goal.
An example of these recommenders using impressions while being designed from another paradigm exists in the reviewed literature.
For instance, \citet{DBLP:conf/sigir/GongZ22/deep-learning/session-based-attention-network} describe a session-based recommender using impressions.

\subsection{Classification of Impression-Aware Recommender Systems}
\label{subsec:impressions-based-recsys:classification}

In this section, we present and describe our novel classification system for \gls{impressionsbased}.
This classification system is composed of three taxonomies, each analyzing recommendation models in different dimensions.
We use these three taxonomies of the classification system to analyze recommendation models in the literature, complementing our analyses by also inspecting the previously defined properties of impression-aware recommendation models, \idest the impression's type, impression's signal, and recommender's type.\footnote{See \autoref{subsec:impressions-based-recsys:problem-description} for the definition of these properties.}
In our analyses, we are able to capture finer nuances in proposed recommendation models when using these properties and taxonomies in conjunction.
Consequently, we assemble a comprehensive picture of the recommendation models in the literature in terms of their design and functioning.

\subsubsection{Model-centric Taxonomy}

The model-centric taxonomy classifies papers based on the design of their proposed recommendation model, \idest the module of the recommender system in charge of generating impressions.
In particular, the taxonomy inspects the model's learning technique.
We identify five categories of recommenders from the reviewed papers in the literature.
The definition of each category is:

\begin{itemize}
  \item \textbf{Heuristics} recommenders using ad-hoc rules and techniques.
  \item \textbf{Statistical:} recommenders using probabilistic distributions or statistical properties of users' behavior.
  \item \textbf{Machine learning:} recommenders using machine learning techniques.
  \item \textbf{Deep learning:} recommenders using deep neural networks.
  \item \textbf{Reinforcement learning:} recommenders using a Markov decision process to model users' preferences.
\end{itemize}

\begin{figure*}[t]
  \centering
  \ifusetikz \tikzsetnextfilename{data-centric-taxonomy}

\definecolor{myyellow}{rgb}{
  1, 0.9490196078431372, 0.8
}
\definecolor{myred}{rgb}{0.9568627450980393, 0.8, 0.8}
\definecolor{mygreen}{rgb}{0.8509803921568627, 0.9176470588235294, 0.8274509803921568}
\definecolor{myblue}{rgb}{0.8117647058823529, 0.8862745098039215, 0.9529411764705882}

\begin{tikzpicture}[
    auto,
    line/.style={
        draw,
        thick,
        -latex',
        shorten >=2pt
      },
    commonnode/.style={
        rectangle,
        thick,
        text width=9em,
        minimum height=3em,
        align=flush center,
        inner sep=0.5em,
        draw=black,
        font=\small,
      },
    manynode/.style={
        commonnode,
        double copy shadow,
        shadow xshift=2pt,
        shadow yshift=-2pt
      },
    numbernode/.style={
        commonnode,
        rectangle,
        align=flush right,
        text width=8em,
        inner sep=0em,
        draw=white,
        font=\scriptsize,
      },
    processnode/.style={
        commonnode,
        fill=mygreen
      },
    featuresnode/.style={
        manynode,
        fill=myyellow
      },
    tripletnode/.style={
        commonnode,
        fill=myyellow
      },
    singlerelevancenode/.style={
        commonnode,
        fill=myred
      },
    multiplerelevancenode/.style={
        manynode,
        fill=myred
      },
    impressionnode/.style={
        manynode,
        fill=myblue
      },
  ]

  \matrix [column sep=4em,row sep=1em]{
    \node [numbernode] (first) {\textbf{(a) Category: Features}};    & [-3.5em]
    \node [featuresnode] (features) {Statistical features}; &
    \node [processnode] (iars) {Recommendation model};                 \\

    \node [numbernode] (second) {\textbf{(b) Category: Learn}};   &
    \node [impressionnode] (impression) {Impression};       &
    \node [processnode] (iarstwo) {Recommendation model};              \\

    \node [numbernode] (third) {\textbf{(c) Category: Sample}};    &
    \node [multiplerelevancenode] (sample) {Sampled items}; &
    \node [processnode] (iarsthree) {Recommendation model};            \\
  };
  \begin{scope}[every path/.style=line]
    \path (features) -- (iars);

    \path (impression) -- (iarstwo);

    \path (sample) -- (iarsthree);
  \end{scope}
\end{tikzpicture} \fi 
  \ifusepdf \includegraphics{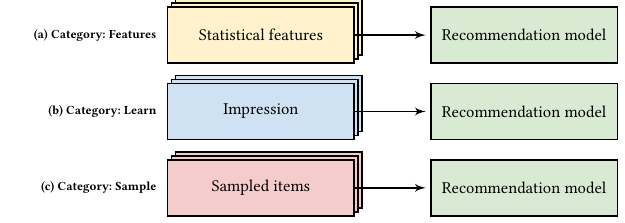} \fi 
  \ifusepng \includegraphics{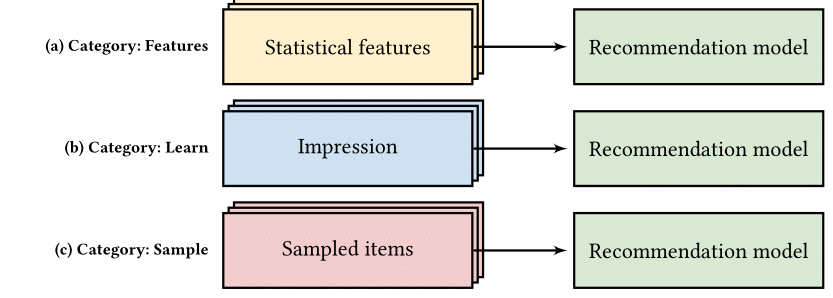} \fi 
  \caption{
    Types of inputs received by recommendation models from the reviewed literature according to the three categories within the data-centric taxonomy.
    \textbf{(a)} illustrates the first category (\textbf{features}), where recommendation models receive statistical features computed from impressions as part of their input.
    \textbf{(b)} illustrates the second category (\textbf{learn}), where recommendation models receive an impression; partially or fully, as part of their input.
    \textbf{(c)} illustrates the third category (\textbf{sample}), where recommendation models receive a vector containing sampled items from the catalog; each item from a possibly different part of the catalog, \idest interacted, solely impressed, or non-impressed.
    The data-centric taxonomy allows for a recommendation model to belong to one or more categories within it.
  }
  \Description[Data categories included in the data-centric taxonomy]{
    Three data categories. 
    The first category is called features. 
    The second category is called learn. 
    The third category is called sampled items. 
    For the first category, a recommendation model receives statistical features as input. 
    For the second, the model receives an impression. 
    For the third, the model receives sampled items.
  }
  \label{fig:tikz-data-centric-taxonomy}
\end{figure*}

\subsubsection{Data-centric Taxonomy}

The data-centric taxonomy classifies papers based on how they process and provide impressions as input for the recommendation model.
One of the main characteristics of the taxonomy, as opposed to the others, is that its categories are non-mutually exclusive, meaning papers may belong to one or more categories.
For instance, six papers are classified into two categories (\textbf{features} and \textbf{learn}) as seen on \autoref{tab:discussion:counts-and-percentages}.
From the reviewed papers, we identify three categories of papers, which we illustrate in \autoref{fig:tikz-data-centric-taxonomy}.
The definition of each category and the number of papers included in the category are:

\begin{itemize}
  \item \textbf{Features:} papers computing features from impressions and receiving such features as input. In this context, a feature is a quantitative property of impressions, \eg the number of times an item has been impressed to a specific user.
  \item \textbf{Learn:} papers handing impressions as input to their recommendation model. This category covers those papers using any impression type and the user feedback on impressed items.
  \item \textbf{Sample:} papers sampling items from the catalog, where at least one sampled item is an impression.
\end{itemize}

\subsubsection{Signal-centric Taxonomy}

The signal-centric taxonomy classifies papers based on how they process users' preferences for non-interacted impressions.
This taxonomy focuses explicitly on non-interacted impressions because the recommender systems literature already assumes the preference of users toward interacted impressions (\idest interactions) as positives; such an assumption is called \emph{missing as negatives}.
From the reviewed papers, we identify two categories of papers.
The definition of each category and the number of papers included in the category are:

\begin{itemize}
  \item \textbf{Assume:} papers assuming users' preference to non-interacted impressions.
  \item \textbf{Learn:} papers learning users' preference to non-interacted impressions.
\end{itemize}

\begin{table}[t]
  \centering
  \small
  \caption{
    Classification of reviewed papers according to the taxonomies and properties defined in this work.
    \textbf{Count} tells the number of papers belonging to a given taxonomy or property.
    \textbf{Percentage} tells the percentage of papers belonging to the classification inside a taxonomy or property.
  }
  \label{tab:discussion:counts-and-percentages}
  \begin{minipage}{\linewidth}
    \centering
    \begin{tabular}{clp{7.5cm}rr}
      \toprule
      \textbf{Classification}
        &  \textbf{Category}
        &  \textbf{Papers References}
        &  \textbf{Count}
        &  \textbf{Percentage}
      \\
      \midrule
      \multirow{6}[0]{*}{\makecell{Model-centric\\taxonomy}}
        &  heuristic
        &  \cite{DBLP:journals/scheduling/BuchbinderFGN14/hard-frequency-cap,DBLP:conf/cscw/ZhaoAHWK17/heuristics/cycling-serpentining,DBLP:conf/kdd/LeeLTS14/impression-discounting,DBLP:conf/kdd/AgarwalCGHHIKMSSZ14/impression-discounting,DBLP:conf/www/LiuRSKMZLJ17/heuristics/memboost-impressions-as-features}
        &  5
        &  11.4\%
      \\
      {}
        &  statistical
        &  \cite{DBLP:conf/www/AgarwalCE09/spatio-temporal-models-for-estimating-ctr,DBLP:conf/recsys/WuASB16/user-fatigue/netflix-fatigue-modeling-using-impressions,DBLP:conf/kdd/ZhangZMCZA16/heuristics/glmix-generalized-linear-mixed-models-for-large-scale-response-prediction,DBLP:conf/kdd/BorisyukZK17/heuristics/lijar-job-boosting-by-impressions,DBLP:conf/kdd/LinCSLLJ23/impression-aware/tree-based-progressive-regression-model-for-watch-time-prediction-in-short-video-recommendation}
        &  5
        &  11.4\%
      \\
      {}
        &  machine learning
        &  \cite{DBLP:conf/www/LiuRSKMZLJ17/heuristics/memboost-impressions-as-features,DBLP:conf/www/MaLS16/user-fatigue/online-news-recommendation-feature-engineering-of-impressions,DBLP:conf/cikm/AharonKLSBESSZ19/heuristics/soft-frequency-cap,DBLP:conf/cikm/PerezMaureraFDSSC20/contentwise-impressions}
        &  4
        &  9.1\%
      \\
      {}
        &  deep learning
        &  \cite{DBLP:conf/recsys/CovingtonAS16/deep-learning/two-stage-reranker/learns-impressions/impressions-features/deep-neural-networks-for-youtube-recommendations,DBLP:conf/kdd/ZhanPSWWMZJG22/heuristics/impressions-to-compute-statistical-features,DBLP:conf/kdd/ZhuD00C22/deep-learning/combo-fashion-clothes-matching,DBLP:conf/cikm/ZhangCXBHDZ22/impressions/keep-an-industrial-pre-training-framework-for-online-recommendation-via-knowledge-extraction-and-plugging,DBLP:conf/ijcai/BiedNPHCCGS23/impression-aware/toward-job-recommendation-for-all,DBLP:conf/wsdm/RenHZZZ23/impressions/slate-aware-ranking-for-recommendation,DBLP:journals/tois/0003YW0RLSCR23/impression-aware/on-the-behavior-leakage-from-recommender-system-exposure,DBLP:conf/sigir/ChenWLC0Z22/deep-learning/two-tower-without-impressions-but-exposure-count,DBLP:conf/sigir/WangCLHCYLC22/counterfactual/three-towers-counterfactual-learning-with-impressions,DBLP:conf/sigir/GongZ22/deep-learning/session-based-attention-network,DBLP:conf/recsys/ZhaoHWCNAKSYC19/deep-learning/two-state-reranker/learns-impressions/impressions-features/recommending-what-video-to-watch-next-a-multitask-ranking-system,DBLP:conf/kdd/MaILYCCNB22/ctr-estimation/an-online-multitask-learning-framework-for-google-feed-ads-auction-models,DBLP:conf/cikm/GongFZQDLJG22/impressions/real-time-short-video-recommendation-on-mobile-devices,DBLP:conf/wsdm/XiLLD0ZT023/impressions/a-birds-eye-view-of-reranking-from-list-level-to-page-level,DBLP:conf/kdd/HuCZWCZ23/impression-aware/boss-a-bilateral-occupational-suitability-aware-recommender-system-for-online-recruitment,DBLP:journals/isci/Lu23/impression-aware/knowledge-distillation-enhanced-multitask-framework-for-recommendation,DBLP:conf/kdd/WangSXDZ23/impression-aware/bert4ctr-an-efficient-framework-to-combine-pre-trained-language-model-with-non-textual-features-for-ctr-prediction}
        &  17
        &  38.6\%
      \\
      {}
        &  reinforcement learning
        &  \cite{DBLP:conf/wsdm/XieZW0L22/adversarial/afe-gan-with-impressions,DBLP:conf/kdd/ChenWWLZZLZ023/impression-aware/controllable-multi-objective-re-ranking-with-policy-hypernetworks,DBLP:conf/wsdm/DeffayetTRR23/impressions/generative-slate-recommendation-with-reinforcement-learning,DBLP:conf/www/PeiYCLSJOZ19/reinforcement-learning/economic-ecommerce-recommender-systems,DBLP:conf/sigir/LiKG16/reinforcement-learning/collaborative-filtering-multi-armed-bandits,DBLP:conf/recsys/McInerneyLHHBGM18/reinforcement-learning/multi-armed-bandit-explore-exploit-and-explain,DBLP:conf/wsdm/GrusonCCMHTC19/reinforcement-learning/impressions-to-stream-for-playlist-recommendation,DBLP:conf/wsdm/ChenBCJBC19/reinforcement-learning/top-k-off-policy-correction-for-a-reinforce-recommender-system,DBLP:conf/cikm/XieZW0L21/reinforcement-learning/teacher-student-rl-network,DBLP:conf/wsdm/GeZYPHHZ22/reinforcement-learning/grouping-items-by-exposure}
        &  10
        &  22.7\%
      \\
      {}
        &  not described
        &  \cite{DBLP:conf/recsys/ZhaoWAHK18/impressions-signals/interpreting-user-inaction-in-recsys,DBLP:conf/sigir/NayakGM23/impression-aware/news-popularity-beyond-the-click-through-rate-for-personalized-recommendations,DBLP:conf/sigir/Sun23/impression-aware/take-a-fresh-look-at-recommender-systems-from-an-evaluation-standpoint}
        &  3
        &  6.8\%
      \\
      \midrule
      \multirow{5}[0]{*}{\makecell{Data-centric\\taxonomy}}
        &  features \& learn
        &  \cite{DBLP:conf/kdd/LeeLTS14/impression-discounting,DBLP:conf/kdd/AgarwalCGHHIKMSSZ14/impression-discounting,DBLP:conf/kdd/LinCSLLJ23/impression-aware/tree-based-progressive-regression-model-for-watch-time-prediction-in-short-video-recommendation,DBLP:conf/www/MaLS16/user-fatigue/online-news-recommendation-feature-engineering-of-impressions,DBLP:conf/cikm/AharonKLSBESSZ19/heuristics/soft-frequency-cap,DBLP:conf/recsys/CovingtonAS16/deep-learning/two-stage-reranker/learns-impressions/impressions-features/deep-neural-networks-for-youtube-recommendations,DBLP:conf/kdd/ZhanPSWWMZJG22/heuristics/impressions-to-compute-statistical-features,DBLP:conf/kdd/ZhuD00C22/deep-learning/combo-fashion-clothes-matching,DBLP:conf/recsys/ZhaoHWCNAKSYC19/deep-learning/two-state-reranker/learns-impressions/impressions-features/recommending-what-video-to-watch-next-a-multitask-ranking-system,DBLP:conf/kdd/HuCZWCZ23/impression-aware/boss-a-bilateral-occupational-suitability-aware-recommender-system-for-online-recruitment,DBLP:conf/kdd/WangSXDZ23/impression-aware/bert4ctr-an-efficient-framework-to-combine-pre-trained-language-model-with-non-textual-features-for-ctr-prediction}
        &  11
        &  25.0\%
      \\
      {}
        &  features
        &  \cite{DBLP:journals/scheduling/BuchbinderFGN14/hard-frequency-cap,DBLP:conf/cscw/ZhaoAHWK17/heuristics/cycling-serpentining,DBLP:conf/www/LiuRSKMZLJ17/heuristics/memboost-impressions-as-features,DBLP:conf/www/AgarwalCE09/spatio-temporal-models-for-estimating-ctr,DBLP:conf/recsys/WuASB16/user-fatigue/netflix-fatigue-modeling-using-impressions,DBLP:conf/kdd/BorisyukZK17/heuristics/lijar-job-boosting-by-impressions,DBLP:conf/sigir/ChenWLC0Z22/deep-learning/two-tower-without-impressions-but-exposure-count,DBLP:conf/cikm/GongFZQDLJG22/impressions/real-time-short-video-recommendation-on-mobile-devices,DBLP:conf/www/PeiYCLSJOZ19/reinforcement-learning/economic-ecommerce-recommender-systems,DBLP:conf/wsdm/GeZYPHHZ22/reinforcement-learning/grouping-items-by-exposure}
        &  10
        &  22.7\%
      \\
      {}
        &  learn
        &  \cite{DBLP:conf/kdd/ZhangZMCZA16/heuristics/glmix-generalized-linear-mixed-models-for-large-scale-response-prediction,DBLP:conf/www/LiuRSKMZLJ17/heuristics/memboost-impressions-as-features,DBLP:conf/cikm/ZhangCXBHDZ22/impressions/keep-an-industrial-pre-training-framework-for-online-recommendation-via-knowledge-extraction-and-plugging,DBLP:conf/ijcai/BiedNPHCCGS23/impression-aware/toward-job-recommendation-for-all,DBLP:conf/wsdm/RenHZZZ23/impressions/slate-aware-ranking-for-recommendation,DBLP:journals/tois/0003YW0RLSCR23/impression-aware/on-the-behavior-leakage-from-recommender-system-exposure,DBLP:conf/sigir/WangCLHCYLC22/counterfactual/three-towers-counterfactual-learning-with-impressions,DBLP:conf/kdd/MaILYCCNB22/ctr-estimation/an-online-multitask-learning-framework-for-google-feed-ads-auction-models,DBLP:conf/wsdm/XiLLD0ZT023/impressions/a-birds-eye-view-of-reranking-from-list-level-to-page-level,DBLP:journals/isci/Lu23/impression-aware/knowledge-distillation-enhanced-multitask-framework-for-recommendation,DBLP:conf/wsdm/XieZW0L22/adversarial/afe-gan-with-impressions,DBLP:conf/kdd/ChenWWLZZLZ023/impression-aware/controllable-multi-objective-re-ranking-with-policy-hypernetworks,DBLP:conf/wsdm/DeffayetTRR23/impressions/generative-slate-recommendation-with-reinforcement-learning,DBLP:conf/sigir/LiKG16/reinforcement-learning/collaborative-filtering-multi-armed-bandits,DBLP:conf/recsys/McInerneyLHHBGM18/reinforcement-learning/multi-armed-bandit-explore-exploit-and-explain,DBLP:conf/wsdm/GrusonCCMHTC19/reinforcement-learning/impressions-to-stream-for-playlist-recommendation,DBLP:conf/wsdm/ChenBCJBC19/reinforcement-learning/top-k-off-policy-correction-for-a-reinforce-recommender-system,DBLP:conf/cikm/XieZW0L21/reinforcement-learning/teacher-student-rl-network}
        &  18
        &  41.0\%
      \\
      {}
        &  sample
        &  \cite{DBLP:conf/cikm/PerezMaureraFDSSC20/contentwise-impressions,DBLP:conf/sigir/GongZ22/deep-learning/session-based-attention-network}
        &  2
        &  4.5\%
      \\
      {}
        &  not described
        &  \cite{DBLP:conf/recsys/ZhaoWAHK18/impressions-signals/interpreting-user-inaction-in-recsys,DBLP:conf/sigir/NayakGM23/impression-aware/news-popularity-beyond-the-click-through-rate-for-personalized-recommendations,DBLP:conf/sigir/Sun23/impression-aware/take-a-fresh-look-at-recommender-systems-from-an-evaluation-standpoint}
        &  3
        &  6.8\%
      \\
      \midrule
      \multirow{4}[0]{*}{\makecell{Signal-centric\\taxonomy}}
        &  assume
        &  \cite{DBLP:journals/scheduling/BuchbinderFGN14/hard-frequency-cap,DBLP:conf/cscw/ZhaoAHWK17/heuristics/cycling-serpentining,DBLP:conf/www/LiuRSKMZLJ17/heuristics/memboost-impressions-as-features,DBLP:conf/kdd/ZhangZMCZA16/heuristics/glmix-generalized-linear-mixed-models-for-large-scale-response-prediction,DBLP:conf/www/LiuRSKMZLJ17/heuristics/memboost-impressions-as-features,DBLP:conf/www/MaLS16/user-fatigue/online-news-recommendation-feature-engineering-of-impressions,DBLP:conf/cikm/AharonKLSBESSZ19/heuristics/soft-frequency-cap,DBLP:conf/cikm/PerezMaureraFDSSC20/contentwise-impressions,DBLP:conf/recsys/CovingtonAS16/deep-learning/two-stage-reranker/learns-impressions/impressions-features/deep-neural-networks-for-youtube-recommendations,DBLP:conf/kdd/ZhuD00C22/deep-learning/combo-fashion-clothes-matching,DBLP:conf/cikm/ZhangCXBHDZ22/impressions/keep-an-industrial-pre-training-framework-for-online-recommendation-via-knowledge-extraction-and-plugging,DBLP:conf/ijcai/BiedNPHCCGS23/impression-aware/toward-job-recommendation-for-all,DBLP:conf/sigir/WangCLHCYLC22/counterfactual/three-towers-counterfactual-learning-with-impressions,DBLP:conf/sigir/GongZ22/deep-learning/session-based-attention-network,DBLP:conf/recsys/ZhaoHWCNAKSYC19/deep-learning/two-state-reranker/learns-impressions/impressions-features/recommending-what-video-to-watch-next-a-multitask-ranking-system,DBLP:conf/kdd/MaILYCCNB22/ctr-estimation/an-online-multitask-learning-framework-for-google-feed-ads-auction-models,DBLP:conf/cikm/GongFZQDLJG22/impressions/real-time-short-video-recommendation-on-mobile-devices,DBLP:conf/wsdm/XiLLD0ZT023/impressions/a-birds-eye-view-of-reranking-from-list-level-to-page-level,DBLP:conf/kdd/HuCZWCZ23/impression-aware/boss-a-bilateral-occupational-suitability-aware-recommender-system-for-online-recruitment,DBLP:journals/isci/Lu23/impression-aware/knowledge-distillation-enhanced-multitask-framework-for-recommendation,DBLP:conf/kdd/WangSXDZ23/impression-aware/bert4ctr-an-efficient-framework-to-combine-pre-trained-language-model-with-non-textual-features-for-ctr-prediction,DBLP:conf/wsdm/XieZW0L22/adversarial/afe-gan-with-impressions,DBLP:conf/www/PeiYCLSJOZ19/reinforcement-learning/economic-ecommerce-recommender-systems,DBLP:conf/sigir/LiKG16/reinforcement-learning/collaborative-filtering-multi-armed-bandits,DBLP:conf/recsys/McInerneyLHHBGM18/reinforcement-learning/multi-armed-bandit-explore-exploit-and-explain,DBLP:conf/wsdm/GrusonCCMHTC19/reinforcement-learning/impressions-to-stream-for-playlist-recommendation,DBLP:conf/wsdm/ChenBCJBC19/reinforcement-learning/top-k-off-policy-correction-for-a-reinforce-recommender-system,DBLP:conf/cikm/XieZW0L21/reinforcement-learning/teacher-student-rl-network}
        &  28
        &  63.6\%
      \\
      {}
        &  learn
        &  \cite{DBLP:conf/kdd/LeeLTS14/impression-discounting,DBLP:conf/kdd/AgarwalCGHHIKMSSZ14/impression-discounting,DBLP:conf/www/AgarwalCE09/spatio-temporal-models-for-estimating-ctr,DBLP:conf/recsys/WuASB16/user-fatigue/netflix-fatigue-modeling-using-impressions,DBLP:conf/kdd/BorisyukZK17/heuristics/lijar-job-boosting-by-impressions,DBLP:conf/kdd/LinCSLLJ23/impression-aware/tree-based-progressive-regression-model-for-watch-time-prediction-in-short-video-recommendation,DBLP:conf/kdd/ZhanPSWWMZJG22/heuristics/impressions-to-compute-statistical-features,DBLP:conf/wsdm/RenHZZZ23/impressions/slate-aware-ranking-for-recommendation,DBLP:conf/kdd/ChenWWLZZLZ023/impression-aware/controllable-multi-objective-re-ranking-with-policy-hypernetworks,DBLP:conf/wsdm/DeffayetTRR23/impressions/generative-slate-recommendation-with-reinforcement-learning,DBLP:conf/recsys/ZhaoWAHK18/impressions-signals/interpreting-user-inaction-in-recsys}
        &  11
        &  25.0\%
      \\
      {}
        &  not described
        &  \cite{DBLP:journals/tois/0003YW0RLSCR23/impression-aware/on-the-behavior-leakage-from-recommender-system-exposure,DBLP:conf/sigir/ChenWLC0Z22/deep-learning/two-tower-without-impressions-but-exposure-count,DBLP:conf/wsdm/GeZYPHHZ22/reinforcement-learning/grouping-items-by-exposure,DBLP:conf/sigir/NayakGM23/impression-aware/news-popularity-beyond-the-click-through-rate-for-personalized-recommendations,DBLP:conf/sigir/Sun23/impression-aware/take-a-fresh-look-at-recommender-systems-from-an-evaluation-standpoint}
        &  5
        &  11.4\%
      \\
      \midrule
      \multirow{4}[0]{*}{\makecell{Impressions\\type}}
        &  contextual
        &  \cite{DBLP:conf/www/MaLS16/user-fatigue/online-news-recommendation-feature-engineering-of-impressions,DBLP:conf/wsdm/RenHZZZ23/impressions/slate-aware-ranking-for-recommendation,DBLP:journals/tois/0003YW0RLSCR23/impression-aware/on-the-behavior-leakage-from-recommender-system-exposure,DBLP:conf/sigir/GongZ22/deep-learning/session-based-attention-network,DBLP:conf/recsys/ZhaoHWCNAKSYC19/deep-learning/two-state-reranker/learns-impressions/impressions-features/recommending-what-video-to-watch-next-a-multitask-ranking-system,DBLP:conf/cikm/GongFZQDLJG22/impressions/real-time-short-video-recommendation-on-mobile-devices,DBLP:conf/wsdm/XiLLD0ZT023/impressions/a-birds-eye-view-of-reranking-from-list-level-to-page-level,DBLP:conf/wsdm/XieZW0L22/adversarial/afe-gan-with-impressions,DBLP:conf/kdd/ChenWWLZZLZ023/impression-aware/controllable-multi-objective-re-ranking-with-policy-hypernetworks,DBLP:conf/wsdm/DeffayetTRR23/impressions/generative-slate-recommendation-with-reinforcement-learning,DBLP:conf/www/PeiYCLSJOZ19/reinforcement-learning/economic-ecommerce-recommender-systems,DBLP:conf/cikm/XieZW0L21/reinforcement-learning/teacher-student-rl-network,DBLP:conf/recsys/ZhaoWAHK18/impressions-signals/interpreting-user-inaction-in-recsys,DBLP:conf/sigir/NayakGM23/impression-aware/news-popularity-beyond-the-click-through-rate-for-personalized-recommendations}
        &  14
        &  31.8\%
      \\
      {}
        &  global
        &  \cite{DBLP:journals/scheduling/BuchbinderFGN14/hard-frequency-cap,DBLP:conf/cscw/ZhaoAHWK17/heuristics/cycling-serpentining,DBLP:conf/kdd/LeeLTS14/impression-discounting,DBLP:conf/kdd/AgarwalCGHHIKMSSZ14/impression-discounting,DBLP:conf/www/LiuRSKMZLJ17/heuristics/memboost-impressions-as-features,DBLP:conf/www/AgarwalCE09/spatio-temporal-models-for-estimating-ctr,DBLP:conf/recsys/WuASB16/user-fatigue/netflix-fatigue-modeling-using-impressions,DBLP:conf/kdd/ZhangZMCZA16/heuristics/glmix-generalized-linear-mixed-models-for-large-scale-response-prediction,DBLP:conf/kdd/BorisyukZK17/heuristics/lijar-job-boosting-by-impressions,DBLP:conf/kdd/LinCSLLJ23/impression-aware/tree-based-progressive-regression-model-for-watch-time-prediction-in-short-video-recommendation,DBLP:conf/www/LiuRSKMZLJ17/heuristics/memboost-impressions-as-features,DBLP:conf/cikm/AharonKLSBESSZ19/heuristics/soft-frequency-cap,DBLP:conf/cikm/PerezMaureraFDSSC20/contentwise-impressions,DBLP:conf/recsys/CovingtonAS16/deep-learning/two-stage-reranker/learns-impressions/impressions-features/deep-neural-networks-for-youtube-recommendations,DBLP:conf/kdd/ZhanPSWWMZJG22/heuristics/impressions-to-compute-statistical-features,DBLP:conf/kdd/ZhuD00C22/deep-learning/combo-fashion-clothes-matching,DBLP:conf/cikm/ZhangCXBHDZ22/impressions/keep-an-industrial-pre-training-framework-for-online-recommendation-via-knowledge-extraction-and-plugging,DBLP:conf/ijcai/BiedNPHCCGS23/impression-aware/toward-job-recommendation-for-all,DBLP:conf/sigir/ChenWLC0Z22/deep-learning/two-tower-without-impressions-but-exposure-count,DBLP:conf/sigir/WangCLHCYLC22/counterfactual/three-towers-counterfactual-learning-with-impressions,DBLP:conf/kdd/MaILYCCNB22/ctr-estimation/an-online-multitask-learning-framework-for-google-feed-ads-auction-models,DBLP:conf/kdd/HuCZWCZ23/impression-aware/boss-a-bilateral-occupational-suitability-aware-recommender-system-for-online-recruitment,DBLP:journals/isci/Lu23/impression-aware/knowledge-distillation-enhanced-multitask-framework-for-recommendation,DBLP:conf/kdd/WangSXDZ23/impression-aware/bert4ctr-an-efficient-framework-to-combine-pre-trained-language-model-with-non-textual-features-for-ctr-prediction,DBLP:conf/sigir/LiKG16/reinforcement-learning/collaborative-filtering-multi-armed-bandits,DBLP:conf/recsys/McInerneyLHHBGM18/reinforcement-learning/multi-armed-bandit-explore-exploit-and-explain,DBLP:conf/wsdm/GrusonCCMHTC19/reinforcement-learning/impressions-to-stream-for-playlist-recommendation,DBLP:conf/wsdm/ChenBCJBC19/reinforcement-learning/top-k-off-policy-correction-for-a-reinforce-recommender-system,DBLP:conf/wsdm/GeZYPHHZ22/reinforcement-learning/grouping-items-by-exposure}
        &  29
        &  65.9\%
      \\
      {}
        &  not described
        &  \cite{DBLP:conf/sigir/Sun23/impression-aware/take-a-fresh-look-at-recommender-systems-from-an-evaluation-standpoint}
        &  1
        &  2.3\%
      \\
      \midrule
      \multirow{5}[0]{*}{\makecell{Impressions\\signal}}
        &  positive
        &  \cite{DBLP:conf/ijcai/BiedNPHCCGS23/impression-aware/toward-job-recommendation-for-all}
        &  1
        &  2.3\%
      \\
      {}
        &  negative
        &  \cite{DBLP:journals/scheduling/BuchbinderFGN14/hard-frequency-cap,DBLP:conf/cscw/ZhaoAHWK17/heuristics/cycling-serpentining,DBLP:conf/kdd/LeeLTS14/impression-discounting,DBLP:conf/kdd/AgarwalCGHHIKMSSZ14/impression-discounting,DBLP:conf/www/LiuRSKMZLJ17/heuristics/memboost-impressions-as-features,DBLP:conf/kdd/ZhangZMCZA16/heuristics/glmix-generalized-linear-mixed-models-for-large-scale-response-prediction,DBLP:conf/www/LiuRSKMZLJ17/heuristics/memboost-impressions-as-features,DBLP:conf/www/MaLS16/user-fatigue/online-news-recommendation-feature-engineering-of-impressions,DBLP:conf/cikm/AharonKLSBESSZ19/heuristics/soft-frequency-cap,DBLP:conf/cikm/PerezMaureraFDSSC20/contentwise-impressions,DBLP:conf/recsys/CovingtonAS16/deep-learning/two-stage-reranker/learns-impressions/impressions-features/deep-neural-networks-for-youtube-recommendations,DBLP:conf/kdd/ZhuD00C22/deep-learning/combo-fashion-clothes-matching,DBLP:conf/cikm/ZhangCXBHDZ22/impressions/keep-an-industrial-pre-training-framework-for-online-recommendation-via-knowledge-extraction-and-plugging,DBLP:conf/sigir/WangCLHCYLC22/counterfactual/three-towers-counterfactual-learning-with-impressions,DBLP:conf/sigir/GongZ22/deep-learning/session-based-attention-network,DBLP:conf/recsys/ZhaoHWCNAKSYC19/deep-learning/two-state-reranker/learns-impressions/impressions-features/recommending-what-video-to-watch-next-a-multitask-ranking-system,DBLP:conf/kdd/MaILYCCNB22/ctr-estimation/an-online-multitask-learning-framework-for-google-feed-ads-auction-models,DBLP:conf/cikm/GongFZQDLJG22/impressions/real-time-short-video-recommendation-on-mobile-devices,DBLP:conf/wsdm/XiLLD0ZT023/impressions/a-birds-eye-view-of-reranking-from-list-level-to-page-level,DBLP:conf/kdd/HuCZWCZ23/impression-aware/boss-a-bilateral-occupational-suitability-aware-recommender-system-for-online-recruitment,DBLP:journals/isci/Lu23/impression-aware/knowledge-distillation-enhanced-multitask-framework-for-recommendation,DBLP:conf/kdd/WangSXDZ23/impression-aware/bert4ctr-an-efficient-framework-to-combine-pre-trained-language-model-with-non-textual-features-for-ctr-prediction,DBLP:conf/wsdm/XieZW0L22/adversarial/afe-gan-with-impressions,DBLP:conf/www/PeiYCLSJOZ19/reinforcement-learning/economic-ecommerce-recommender-systems,DBLP:conf/sigir/LiKG16/reinforcement-learning/collaborative-filtering-multi-armed-bandits,DBLP:conf/recsys/McInerneyLHHBGM18/reinforcement-learning/multi-armed-bandit-explore-exploit-and-explain,DBLP:conf/wsdm/GrusonCCMHTC19/reinforcement-learning/impressions-to-stream-for-playlist-recommendation,DBLP:conf/wsdm/ChenBCJBC19/reinforcement-learning/top-k-off-policy-correction-for-a-reinforce-recommender-system,DBLP:conf/cikm/XieZW0L21/reinforcement-learning/teacher-student-rl-network}
        &  29
        &  65.9\%
      \\
      {}
        &  neutral
        &  \cite{DBLP:conf/www/AgarwalCE09/spatio-temporal-models-for-estimating-ctr,DBLP:conf/recsys/WuASB16/user-fatigue/netflix-fatigue-modeling-using-impressions,DBLP:conf/kdd/BorisyukZK17/heuristics/lijar-job-boosting-by-impressions,DBLP:conf/kdd/LinCSLLJ23/impression-aware/tree-based-progressive-regression-model-for-watch-time-prediction-in-short-video-recommendation,DBLP:conf/kdd/ZhanPSWWMZJG22/heuristics/impressions-to-compute-statistical-features,DBLP:conf/wsdm/RenHZZZ23/impressions/slate-aware-ranking-for-recommendation,DBLP:journals/tois/0003YW0RLSCR23/impression-aware/on-the-behavior-leakage-from-recommender-system-exposure,DBLP:conf/sigir/ChenWLC0Z22/deep-learning/two-tower-without-impressions-but-exposure-count,DBLP:conf/kdd/ChenWWLZZLZ023/impression-aware/controllable-multi-objective-re-ranking-with-policy-hypernetworks,DBLP:conf/wsdm/DeffayetTRR23/impressions/generative-slate-recommendation-with-reinforcement-learning,DBLP:conf/wsdm/GeZYPHHZ22/reinforcement-learning/grouping-items-by-exposure,DBLP:conf/recsys/ZhaoWAHK18/impressions-signals/interpreting-user-inaction-in-recsys,DBLP:conf/sigir/NayakGM23/impression-aware/news-popularity-beyond-the-click-through-rate-for-personalized-recommendations}
        &  13
        &  29.5\%
      \\
      {}
        &  not described
        &  \cite{DBLP:conf/sigir/Sun23/impression-aware/take-a-fresh-look-at-recommender-systems-from-an-evaluation-standpoint}
        &  1
        &  2.3\%
      \\
      \midrule
      \multirow{5}[0]{*}{\makecell{Recommender\\type}}
        &  end-to-end
        &  \cite{DBLP:journals/scheduling/BuchbinderFGN14/hard-frequency-cap,DBLP:conf/www/LiuRSKMZLJ17/heuristics/memboost-impressions-as-features,DBLP:conf/www/AgarwalCE09/spatio-temporal-models-for-estimating-ctr,DBLP:conf/recsys/WuASB16/user-fatigue/netflix-fatigue-modeling-using-impressions,DBLP:conf/kdd/ZhangZMCZA16/heuristics/glmix-generalized-linear-mixed-models-for-large-scale-response-prediction,DBLP:conf/kdd/LinCSLLJ23/impression-aware/tree-based-progressive-regression-model-for-watch-time-prediction-in-short-video-recommendation,DBLP:conf/www/LiuRSKMZLJ17/heuristics/memboost-impressions-as-features,DBLP:conf/www/MaLS16/user-fatigue/online-news-recommendation-feature-engineering-of-impressions,DBLP:conf/cikm/AharonKLSBESSZ19/heuristics/soft-frequency-cap,DBLP:conf/cikm/PerezMaureraFDSSC20/contentwise-impressions,DBLP:conf/kdd/ZhanPSWWMZJG22/heuristics/impressions-to-compute-statistical-features,DBLP:conf/kdd/ZhuD00C22/deep-learning/combo-fashion-clothes-matching,DBLP:conf/cikm/ZhangCXBHDZ22/impressions/keep-an-industrial-pre-training-framework-for-online-recommendation-via-knowledge-extraction-and-plugging,DBLP:conf/ijcai/BiedNPHCCGS23/impression-aware/toward-job-recommendation-for-all,DBLP:journals/tois/0003YW0RLSCR23/impression-aware/on-the-behavior-leakage-from-recommender-system-exposure,DBLP:conf/sigir/ChenWLC0Z22/deep-learning/two-tower-without-impressions-but-exposure-count,DBLP:conf/sigir/WangCLHCYLC22/counterfactual/three-towers-counterfactual-learning-with-impressions,DBLP:conf/sigir/GongZ22/deep-learning/session-based-attention-network,DBLP:conf/kdd/MaILYCCNB22/ctr-estimation/an-online-multitask-learning-framework-for-google-feed-ads-auction-models,DBLP:conf/kdd/HuCZWCZ23/impression-aware/boss-a-bilateral-occupational-suitability-aware-recommender-system-for-online-recruitment,DBLP:journals/isci/Lu23/impression-aware/knowledge-distillation-enhanced-multitask-framework-for-recommendation,DBLP:conf/kdd/WangSXDZ23/impression-aware/bert4ctr-an-efficient-framework-to-combine-pre-trained-language-model-with-non-textual-features-for-ctr-prediction,DBLP:conf/wsdm/XieZW0L22/adversarial/afe-gan-with-impressions,DBLP:conf/wsdm/DeffayetTRR23/impressions/generative-slate-recommendation-with-reinforcement-learning,DBLP:conf/www/PeiYCLSJOZ19/reinforcement-learning/economic-ecommerce-recommender-systems,DBLP:conf/sigir/LiKG16/reinforcement-learning/collaborative-filtering-multi-armed-bandits,DBLP:conf/wsdm/GrusonCCMHTC19/reinforcement-learning/impressions-to-stream-for-playlist-recommendation,DBLP:conf/wsdm/ChenBCJBC19/reinforcement-learning/top-k-off-policy-correction-for-a-reinforce-recommender-system,DBLP:conf/cikm/XieZW0L21/reinforcement-learning/teacher-student-rl-network,DBLP:conf/wsdm/GeZYPHHZ22/reinforcement-learning/grouping-items-by-exposure}
        &  30
        &  68.2\%
      \\
      {}
        &  plug-in
        &  \cite{DBLP:conf/kdd/LeeLTS14/impression-discounting,DBLP:conf/kdd/AgarwalCGHHIKMSSZ14/impression-discounting,DBLP:conf/kdd/BorisyukZK17/heuristics/lijar-job-boosting-by-impressions,DBLP:conf/recsys/McInerneyLHHBGM18/reinforcement-learning/multi-armed-bandit-explore-exploit-and-explain}
        &  4
        &  9.1\%
      \\
      {}
        &  re-ranking
        &  \cite{DBLP:conf/cscw/ZhaoAHWK17/heuristics/cycling-serpentining,DBLP:conf/recsys/CovingtonAS16/deep-learning/two-stage-reranker/learns-impressions/impressions-features/deep-neural-networks-for-youtube-recommendations,DBLP:conf/wsdm/RenHZZZ23/impressions/slate-aware-ranking-for-recommendation,DBLP:conf/recsys/ZhaoHWCNAKSYC19/deep-learning/two-state-reranker/learns-impressions/impressions-features/recommending-what-video-to-watch-next-a-multitask-ranking-system,DBLP:conf/cikm/GongFZQDLJG22/impressions/real-time-short-video-recommendation-on-mobile-devices,DBLP:conf/wsdm/XiLLD0ZT023/impressions/a-birds-eye-view-of-reranking-from-list-level-to-page-level,DBLP:conf/kdd/ChenWWLZZLZ023/impression-aware/controllable-multi-objective-re-ranking-with-policy-hypernetworks}
        &  7
        &  15.9\%
      \\
        &  not described
        &  \cite{DBLP:conf/recsys/ZhaoWAHK18/impressions-signals/interpreting-user-inaction-in-recsys,DBLP:conf/sigir/NayakGM23/impression-aware/news-popularity-beyond-the-click-through-rate-for-personalized-recommendations,DBLP:conf/sigir/Sun23/impression-aware/take-a-fresh-look-at-recommender-systems-from-an-evaluation-standpoint}
        &  3
        &  6.8\%
      \\
      \bottomrule
    \end{tabular}
  \end{minipage}
\end{table}

\subsubsection{Distribution of Papers}

The three taxonomies included in our classification system let us inspect and analyze recommendation models from the literature in a comprehensive manner.
As a reference, in \autoref{tab:discussion:counts-and-percentages}, we show the distribution of papers when grouped according to each of our proposed taxonomies and properties of recommendation models.
The distribution of papers shows both the number and percentage of papers that belong to each group.
From the table, we observe that in most taxonomies and properties, the distribution of papers falls into one or two categories over the others within the same group.

This behavior is mostly observed in the \emph{signal-centric} taxonomy and the three properties we analyze, \idest \emph{impression's type}, \emph{impressions signals}, and \emph{recommender's type}.
These skewed distributions of papers show that the literature leans its efforts toward specific approaches or topics while others remain less explored.
For instance, in the former, most papers (close to \qty{64}{\percent}) simply assume a signal to items impressions based on their received (or lack of) user feedback.
Similarly, in the other groups, around \qty{66}{\percent} of papers use global impressions, give negative signals to non-interacted but impressed items, and use end-to-end recommenders, respectively.

The remaining taxonomies, \idest the \emph{model-centric} and \emph{data-centric} taxonomies, show a less skewed distribution of papers.
In particular, according to the model-centric taxonomy, papers mostly describe recommendation models using deep learning or reinforcement learning; a considerable number of papers use other techniques as well.
According to the data-centric taxonomy, papers extract features from impressions, pass an impression as input to the recommender, or combine both approaches.
In contrast to the previous taxonomy, according to the data-centric taxonomy, a limited number of papers sample items from the catalog or do not describe how they process impressions in recommendation models.

\section{Reviewed Papers}
\label{sec:reviewed-papers}

This section presents the selected \gls{impressionsbased} papers alongside the recommendation models they describe; for a total of \numincludedworks papers.\footnote{We describe the selection criteria of \gls{impressionsbased} papers in \autoref{subsec:introduction:paper-selection-criteria}.}
When describing papers, we present their recommendation models and other aspects, \eg their recommendation domain, complexity, how they use impressions, among others.
We organize this section using the categories of the model-centric taxonomy.
This means that we present papers in the same order as we present the categories inside that taxonomy, \idest heuristics, statistical, machine learning, deep learning, and reinforcement learning.
Due to the varying levels of detail provided by reviewed papers, their different evaluation strategies, and the complexity in fairly evaluating recommendation models, it is beyond the scope of this work to compare the recommendation quality of the recommendation models presented in this section.

\subsection{Heuristics}
\label{subsec:reviewed-papers:heuristics}

In this section, we present five papers describing recommenders using ad-hoc techniques to learn users' preferences.
Four in this category apply \emph{frequency capping}, a technique designed to limit or discourage the selection of an item in a future impression after its number of impressions exceeds a threshold.
Those papers employ two types of frequency capping approaches, namely: hard and soft frequency capping.
Two papers~\cite{DBLP:journals/scheduling/BuchbinderFGN14/hard-frequency-cap,DBLP:conf/cscw/ZhaoAHWK17/heuristics/cycling-serpentining} employ the hard version, meaning their recommenders encode such threshold.
Two papers~\cite{DBLP:conf/kdd/LeeLTS14/impression-discounting,DBLP:conf/kdd/AgarwalCGHHIKMSSZ14/impression-discounting} employ the soft version, meaning their recommenders apply data mining techniques to learn the threshold from user feedback.
The remaining paper~\cite{DBLP:conf/www/LiuRSKMZLJ17/heuristics/memboost-impressions-as-features} does not apply frequency capping.
Instead, it focuses on recommending items with the highest \gls{ctr}, \idest the ratio between the number of interactions and the number of impressions.

\subsubsection{Hard Frequency Capping}

Two papers~\cite{DBLP:journals/scheduling/BuchbinderFGN14/hard-frequency-cap,DBLP:conf/cscw/ZhaoAHWK17/heuristics/cycling-serpentining} use hard frequency capping.
\citet{DBLP:journals/scheduling/BuchbinderFGN14/hard-frequency-cap} describe two recommenders in the domain of online advertisements termed \greedyd and \greedyv.
In the paper, items are advertisements published in an online advertisement system and served by the systems' recommender.
Each item is associated with a \emph{payoff} (the amount of money the advertiser pays for each user-item impression) and two \emph{constraints} (a maximum number of global impressions and a maximum number of user-item impressions).
Both recommenders aim to obtain the highest payoff while respecting the items' constraints.
For a given user, \greedyv recommends items with the highest payoffs.
Instead, \greedyd recommends those items with the highest number of global impressions and the highest number of user-item impressions.
\citet{DBLP:conf/cscw/ZhaoAHWK17/heuristics/cycling-serpentining} describe a session-based re-ranking recommender termed \emph{cycling} in the media domain.
The goal of the recommender is to re-arrange an impression to favor less impressed items over highly impressed items.
The paper defines the relevance score of the cycling recommender as a tuple containing a \emph{presentation score} (denoted as \presentationscore) and the relevance score of another recommender (denoted as \recommenderrelevance):
\begin{equation}
  \label{eq:cycling-presentation-score}
  \begin{split}
    \relevancescore = \left(
    \presentationscore,
    \recommenderrelevance
    \right)
    \,,\quad
    \presentationscore = \left\lfloor
    \frac{\numimpressions{\useru,\itemi}}{3}
    \right\rfloor
  \end{split}
\end{equation}
where \numimpressions{\useru,\itemi} is the number of impressions of a given item \itemi on user \useru, and $\lfloor x \rfloor$ rounds $x$ to the lowest integer.
The paper states the recommender re-ranks an impression by selecting the items with the lowest presentation score and solving ties by selecting the items with the highest recommender's score.

\subsubsection{Soft Frequency Capping}

Two papers~\cite{DBLP:conf/kdd/LeeLTS14/impression-discounting,DBLP:conf/kdd/AgarwalCGHHIKMSSZ14/impression-discounting} use soft frequency capping.
\citet{DBLP:conf/kdd/LeeLTS14/impression-discounting} describe a plug-in recommender termed \gls{idf} in the job and online advertisement domains.
The goal of the recommender is to adjust the relevance of items based on their historical user-item interactions and impressions.
For a given-user item pair, the paper defines the relevance score of its recommender as the product between a discounting factor (denoted as \discountingfactor) and the relevance score of another recommender (denoted as \recommenderrelevance):
\begin{equation}
  \label{eq:impressions-aware-recommender-systems:reviewed-papers:heuristics:impressions-discounting-relevance}
  \relevancescore = \recommenderrelevance \cdot \discountingfactor
\end{equation}
where \discountingfactor is a normalized linear aggregation of several features computed on impressions: the number of days elapsed since the same user-item impression, the number of user-item impressions, and the position on-screen of the last impression.
\citet{DBLP:conf/kdd/AgarwalCGHHIKMSSZ14/impression-discounting} use the \gls{idf} in an industrial recommender in the jobs domain; however, the paper does not provide details of the recommender's implementation, deployment, or other properties.
For instance, the paper does not state which features from impressions the recommender computes.

\subsubsection{CTR Prediction}

\citet{DBLP:conf/www/LiuRSKMZLJ17/heuristics/memboost-impressions-as-features} does not describe a recommender employing a frequency capping technique.
Instead, the paper describes a non-personalized recommender to predict the ratio between interactions and impressions for a given query and item.
The recommender is deployed in an image-sharing service in the social media domain, where items are images, and users search for images by providing a ``search query'' (denoted as \queryq).
The recommender memorizes the best images given a search query based on the number of interactions and impressions each image received.
For a given query-item pair, the recommender's relevance score (denoted as $\Tilde{r}_{\queryq,\itemi}$) is:

\begin{equation}
  \label{eq:memboost-relevance-score}
  \Tilde{r}_{\queryq, \itemi} = \log{\left(
    \frac{
      \numinteractions{\queryq, \itemi} + \alpha
    }{
      \numimpressions{\queryq, \itemi} \cdot CTR + \alpha
    }
    \right)}
\end{equation}
where \numinteractions{\queryq, \itemi} and \numimpressions{\queryq, \itemi} are the number of interactions and impressions of query \queryq on item \itemi, respectively.
$CTR$ is the ratio between the number of interactions and impressions across all queries and items.
$\alpha$ is a hyper-parameter.

\subsection{Statistical}
\label{subsec:reviewed-papers:statistical}

In this section, we present five papers describing recommenders using probabilistic distributions or statistical properties to model users' preferences.
Two papers model user preferences by accounting for user fatigue, \idest modeling the user dissatisfaction with the recommender system upon repeated items in impressions.
Specifically, those papers define user fatigue as a function of the number of impressions.
One paper learns user preferences using logistic regression, while two papers model features from impressions, \eg the future number of interactions and impressions.

\subsubsection{User Fatigue}
Two papers~\cite{DBLP:conf/www/AgarwalCE09/spatio-temporal-models-for-estimating-ctr,DBLP:conf/recsys/WuASB16/user-fatigue/netflix-fatigue-modeling-using-impressions} model user fatigue.
\citet{DBLP:conf/www/AgarwalCE09/spatio-temporal-models-for-estimating-ctr} describe a recommender in the news domain.
The goal of the recommender is to predict the ratio between the number of interactions and the number of impressions (\acrshort{ctr}) by incorporating a factor to account for user fatigue.
The recommender uses impressions, the current time, and the position of items on-screen to build the model of users' preferences.
A simplified version of the recommender's relevance score function is:

\begin{equation}
  \label{eq:spatio-temporal-relevance}
  \begin{split}
    \relevancescore = \spatiopositioncoeff \exp\left(
    g\left(
      R_{\useru}
      \right)
    \right) \,,\quad
    \theta_{\itemi} \sim \Gamma\left(
    \alpha + \numinteractions{\itemi}, \gamma + \numimpressions{\itemi}
    \right)
  \end{split}
\end{equation}

where \spatiopositioncoeff is the \acrshort{ctr} in its first global impression. \numinteractions{\itemi} and \numimpressions{\itemi} are the numbers of interactions and impressions of a given item, respectively.
Lastly, $\alpha$ and $\gamma$ are hyper-parameters.
\citet{DBLP:conf/recsys/WuASB16/user-fatigue/netflix-fatigue-modeling-using-impressions} describe a recommender in the media domain.
The goal of the recommender is to provide personalized recommendations on a two-dimensional layout -- a similar scenario to the one described by \citet{DBLP:journals/fdata/FerrariDacremaFC22/frontiers-offline-evaluation-of-recommender-systems-with-multiple-carousels}.
The paper defines user fatigue as a piece-wise linear function dependent on the number of user-item impressions:

\begin{equation}
  \label{eq:netflix-fatigue-modeling}
  fatigue(\numimpressions{\useru,\itemi}) =
  \begin{cases}
    a_1 \numimpressions{\useru,\itemi} + a_{2}                  & \text{if } \numimpressions{\useru,\itemi} < k \\
    a_{1} k + a_{2} + a_{3}(\numimpressions{\useru,\itemi} - k) & \text{otherwise}
  \end{cases}
\end{equation}
where \numimpressions{\useru,\itemi} is the number of impressions of a given user with a given item; $a_1, a_2, a_3 \in \setreals$ are the control slope, offset, and secondary slope of the users' fatigue function, respectively; and $k \in \setpositiveintegers$ is a threshold parameter.
The paper states the function benefits less-popular items since they are associated, on average, with lower fatigue, while it penalizes popular items as they tend to be recommended often and to cause higher fatigue.
The paper trains the recommender using the expectation-maximization algorithm~\cite{10.2307/2984875/non-impressions/em-expectation-maximization-algorithm}.

\subsubsection{Logistic Regression}
\citet{DBLP:conf/kdd/ZhangZMCZA16/heuristics/glmix-generalized-linear-mixed-models-for-large-scale-response-prediction} learn user preferences using logistic regression.
The paper describes a recommender in the jobs domain, termed \emph{GLMix}.
The goal of the recommender is to classify impressions as interacted or non-interacted impressions using contextual features, \eg time of the day and geographical position of items.
For any user-item pair, the recommender learns the user feedback such pair will receive using logistic regression.
The recommender selects those items with the highest likelihood of becoming interacted impressions.
Consequently, the paper considers interacted impressions as positive signals while non-interacted impressions as negative ones.

\subsubsection{Prediction of Features from Impressions}
\citet{DBLP:conf/kdd/BorisyukZK17/heuristics/lijar-job-boosting-by-impressions} model the future number of interactions and impressions any item will receive during a given period.
The paper describes a non-personalized plug-in recommender in the jobs domain.
In the paper, items are job postings associated with a ``recommendation window'' (a period where they can be recommended) and two constraints (a minimum and a maximum number of interactions they can receive during the recommendation window).
The goal of the recommender is to maximize the number of interactions within each item's recommendation window while respecting the item's constraints.
For a given item, the recommender builds three statistical models:
\begin{itemize}
  \item A confidence interval for the item's number of future interactions.
  \item The item's expected number of impressions.
  \item The item's expected number of interactions.
\end{itemize}
The paper states the expected number of impressions follows a negative binomial distribution.
In contrast, the expected number of interactions is conditioned on the previous but has no closed form.
\citet{DBLP:conf/kdd/LinCSLLJ23/impression-aware/tree-based-progressive-regression-model-for-watch-time-prediction-in-short-video-recommendation} predict the watch time of items in the media domain, where the watch time is the period users spend actively watching videos.
The paper describes a tree-based approach, termed ``tree-based progressive regression,'' to predict a specific range in which the watch time falls.
Then, the paper partitions the watch time into $n$ ranges and assigns each range to a leaf node.
Then, it creates a parent node for every two leaf nodes with consecutive ranges.
Every parent node holds a classifier predicting whether the watch time falls in any of the ranges held by its children.
With the paper's approach, the watch time follows a multinomial distribution dependent on the depth of the tree and the path from the root to any leaf node.

\subsection{Machine Learning}
\label{subsec:reviewed-papers:machine-learning}

In this section, we present four papers describing recommenders using shallow machine learning techniques to model users' preferences.
Two papers describe recommenders using \gls{gbdt}~\cite{DBLP:conf/nips/KeMFWCMYL17/non-impressions/gbdt/lightgbm-a-highly-efficient-gradient-boosting-decision-tree}: an ensemble of decision trees.
Two papers describe modified versions of the traditional matrix factorization technique~\cite{DBLP:reference/sp/KorenRB22/non-impressions/machine-learning/advances-in-collaborative-filtering}.

\subsubsection{Gradient Boosting Decision Trees}
Two papers~\cite{DBLP:conf/www/MaLS16/user-fatigue/online-news-recommendation-feature-engineering-of-impressions,DBLP:conf/www/LiuRSKMZLJ17/heuristics/memboost-impressions-as-features} use \acrshort{gbdt} in their recommenders.
\citet{DBLP:conf/www/LiuRSKMZLJ17/heuristics/memboost-impressions-as-features} describe a \acrshort{gbdt} recommender in the media domain.
The recommender aims to classify user-item pairs as future interacted or non-interacted impressions.
For a given user, the recommender generates an impression by selecting those items it classifies as interacted impressions with higher confidence.
Consequently, the recommender considers non-interacted impressions negative signals while interacted impressions positive ones.
The paper does not provide further details of the recommender.
\citet{DBLP:conf/www/MaLS16/user-fatigue/online-news-recommendation-feature-engineering-of-impressions} describe a LambdaMART recommender in the news domain.
The goal of the recommender is to predict the ratio between the number of interactions and impressions while considering user fatigue (see \autoref{subsec:reviewed-papers:statistical}) on repeated impressions.
The paper models user fatigue by computing four features from impressions and interactions related to user feedback.
Apart from those features, the paper computes eleven more features from impressions.
All the seventeen features computed from impressions by the paper are:
\begin{itemize}
  \item \textbf{CTR}: the ratio between the number of interactions and the number of impressions of a user (one feature).
  \item \textbf{Same item fatigue}: the number of interactions and impressions of a user with the same item (two features).
  \item \textbf{Same category fatigue}: the number of interactions and impressions of a user with a category of item (two features).
  \item \textbf{Positional}: first, average, and last position of an item in impressions for a user (three features).
  \item \textbf{Temporal}: number of user-item impressions in the past \num{3}, \num{10}, \num{30}, \num{120}, and \num{1440} minutes. Also, the elapsed time since the first and last user-item impression. Lastly, the elapsed time since the first and last pair of user and category of item impression (nine features).
\end{itemize}
The paper's results show the recommendation quality increases when including all those features and when compared to the same recommender without such features.
Moreover, the same item fatigue and the temporal features provide the greatest relative improvement.

\subsubsection{Matrix Factorization}
Two papers~\cite{DBLP:conf/cikm/AharonKLSBESSZ19/heuristics/soft-frequency-cap,DBLP:conf/cikm/PerezMaureraFDSSC20/contentwise-impressions} use matrix factorization.
\citet{DBLP:conf/cikm/AharonKLSBESSZ19/heuristics/soft-frequency-cap} describe a recommender in the online advertisements domain.
The goal of the recommender is to predict the ratio between the number of interactions and the number of impressions.
The recommender is a matrix factorization model incorporating a learned bias as a soft frequency capping term (see \autoref{sec:reviewed-papers}) on the number of repeated user-item impressions.
A simplified version of the relevance score of the recommender is:
\begin{equation}
  \label{eq:impressions-aware-recommender-systems:soft-frequency-cap}
  \relevancescore = \mfbias + \mfuserfactor \cdot \mfitemfactor + \sfcfrequencybias
\end{equation}
where \mfuserfactor, \mfitemfactor, and \mfbias are the traditional user latent factors, item latent factors, and bias terms present in traditional matrix factorization recommenders, respectively.
\sfcfrequencybias is the \emph{impressions frequency} bias, the learned personalized bias on the user fatigue.
The recommender learns those factors and biases using SGD, where the expected score of a non-interacted impression is $0$, while the expected score of an interacted impression is $1$.
Hence, the paper considers non-interacted impressions as negative signals, while interacted impressions are positive ones.
\citet{DBLP:conf/cikm/PerezMaureraFDSSC20/contentwise-impressions} describe two recommenders using global impressions in the media domain.
Both recommenders are the traditional matrix factorization model optimized using the \acrshort{bpr} criterion~\cite{DBLP:conf/uai/RendleFGS09/non-impressions/bpr}.
As such, the recommender samples a user, a positive item, and a negative item.
The paper samples positive items from interacted impressions, while negative items are non-interacted impressions or non-impressed items.
Consequently, the paper deems interacted impressions as positive signals, while non-interacted impressions and non-impressed items are considered negative signals.

\subsection{Deep Learning}
\label{subsec:reviewed-papers:deep-learning}

In this section, we present \emph{seventeen} papers describing recommenders using deep neural networks to model users' preferences.
Five papers use the \gls{mlp} architecture: a feed-forward neural network with activation functions in its layers.
Two papers use the \emph{encoder-decoder}~\cite{DBLP:journals/air/RafiqRC23/non-impressions/video-description-a-comprehensive-survey-of-deep-learning-approaches} architecture: a composition of two neural networks called encoder and decoder, where the encoder transforms its input into a latent representation, and the decoder recomposes the input from the latent representation.
Three papers use the \emph{two-tower framework}~\cite{DBLP:conf/recsys/YiYHCHKZWC19/two-tower-framework}: an architecture consisting of two neural networks, where one tower generates user embeddings and the other generates item embeddings.
Five papers use a \gls{mmoe}: a gated ensemble of neural networks for multi-task recommendation.
One paper uses \gls{kd}~\cite{DBLP:journals/corr/HintonVD15/non-impressions/distilling-the-knowledge-in-a-neural-network}: transferring the knowledge of a large recommendation model into a smaller deep learning recommendation model.
Lastly, one paper uses a \gls{plm}~\cite{WANG202351/non-impressions/pre-trained-language-models-and-their-applications,DBLP:journals/csur/MinRSVNSAHR24/non-impressions/recent-advances-in-natural-language-processing-via-large-pre-trained-language-models-a-survey}: a deep neural network that may be fine-tuned to a specific task.\footnote{\gls{mmoe} and \Gls{kd} are architectures that allow the use of various kinds of recommendation models, not only deep learning ones. When a paper uses at least one deep neural network inside \gls{mmoe} or \gls{kd}, we classify the paper as belonging to the deep learning category.}

\subsubsection{Multilayer Perceptron Architecture}
Five papers~\cite{DBLP:conf/recsys/CovingtonAS16/deep-learning/two-stage-reranker/learns-impressions/impressions-features/deep-neural-networks-for-youtube-recommendations,DBLP:conf/kdd/ZhanPSWWMZJG22/heuristics/impressions-to-compute-statistical-features,DBLP:conf/kdd/ZhuD00C22/deep-learning/combo-fashion-clothes-matching,DBLP:conf/cikm/ZhangCXBHDZ22/impressions/keep-an-industrial-pre-training-framework-for-online-recommendation-via-knowledge-extraction-and-plugging,DBLP:conf/ijcai/BiedNPHCCGS23/impression-aware/toward-job-recommendation-for-all} use \gls{mlp}.
\citet{DBLP:conf/recsys/CovingtonAS16/deep-learning/two-stage-reranker/learns-impressions/impressions-features/deep-neural-networks-for-youtube-recommendations} describe a recommender in the media domain.
The recommender is a composition of several dense feed-forward layers with ReLU activations.
The goal of the recommender is to estimate the ``expected watch time per impression'', \idest the time users spend watching an item.
The recommender is a re-ranking one: it receives as input an impression (generated by another recommender) and features of items inside the impressions, while its output is a permutation of the impression.
As per the features from impressions, the recommender computes the number of user-item impressions.
The recommender considers interacted impressions as positive signals, while non-interacted impressions as negative signals.
\citet{DBLP:conf/kdd/ZhanPSWWMZJG22/heuristics/impressions-to-compute-statistical-features} describe a recommender with residual layers in the media domain.
The paper introduces the term ``watch time'' to measure the time users spend watching videos.
The goal of the recommender is to learn user preferences for videos using their watch time.
Unlike other papers, the paper does not distinguish positive or negative signals using user feedback; instead, it uses the watch time.
Hence, the paper introduces a threshold value to classify user-item pairs as positive or negative signals.
Negative signals are user-item pairs with a watch time lower than the threshold, while positive signals are user-item pairs on the opposite side.
For a given item, the recommender computes two features: its number of impressions and its average watch time.
For a given user and item, the recommender encodes both features alongside the identifiers of the user and the item into a ResNet predicting the watch time the user will have on the item.
The recommender generates an impression containing those items with the highest predicted watch time.
\citet{DBLP:conf/kdd/ZhuD00C22/deep-learning/combo-fashion-clothes-matching} describe a recommender with attention layers in the fashion domain.
In the paper, an item combines two fashion garments: top and bottom.
The goal of the recommender is to predict user preferences while accounting for which items receive more interactions than others.
In particular, the paper uses impressions to compute the ratio between the number of interactions and impressions (CTR).
For a given user, the paper creates two sets (termed positive and negative combinations) containing those items with the highest and lowest CTR, respectively.
The input of the recommender is a sextuplet composed of a user, an item, a label, the user's set of positive combinations, the user's set of negative combinations, and a vector of contextual features.
The label indicates whether the user-item pair is an interacted or non-interacted impression.
The paper does not detail which contextual features they compute.
\citet{DBLP:conf/cikm/ZhangCXBHDZ22/impressions/keep-an-industrial-pre-training-framework-for-online-recommendation-via-knowledge-extraction-and-plugging} describe a recommender in the e-commerce domain.
For a given user-item pair, the recommender has three goals: predicting user-item interactions, purchases, and ``add-to-cart''; the latter meaning the user performed a specific type of interaction in the system.
The paper focuses on cross-domain recommendation, \idest the recommender is trained on e-commerce data.
Then, the learned preferences are used by another recommender deployed in the online advertisement domain.
The paper only uses impressions in the former recommender. Furthermore, the paper only uses global impressions to train the recommender in the first task, \idest predicting a user-item pair as an interacted or non-interacted impression.
Hence, the paper deems non-interacted impressions as negative user feedback.
For the remaining tasks, the paper trains the recommender using interactions.
\citet{DBLP:conf/ijcai/BiedNPHCCGS23/impression-aware/toward-job-recommendation-for-all} describe a recommender in the jobs domain.
In the paper, a user is a person registered in an online job-seeking platform, while items are job advertisements.
The paper proposes a \gls{mlp}, called MUSE, consisting of three modules and designed to predict two tasks: hiring and applications.
The first module, called MUSE.0, provides a score for a given user-item pair based on contextual and content features.
The second module, called MUSE.1, uses such a score alongside additional features to predict whether a user-item pair will result in the user being hired.
The third module, called MUSE.2, instead, uses the previous features and predicts whether a user-item pair will result in the user applying for a job.
To train the recommender, the paper uses non-impressed items as negative signals; while it uses impressions as positive signals.

\subsubsection{Encoder-Decoder Architecture}
Two papers~\cite{DBLP:conf/wsdm/RenHZZZ23/impressions/slate-aware-ranking-for-recommendation,DBLP:journals/tois/0003YW0RLSCR23/impression-aware/on-the-behavior-leakage-from-recommender-system-exposure} describe recommenders using the encoder-decoder architecture.
\citet{DBLP:conf/wsdm/RenHZZZ23/impressions/slate-aware-ranking-for-recommendation} describe a recommender in the media and news domains.
When training the recommender, for every user-item pair, it receives as inputs a triplet composed of a vector of features of the item, features of the user, and an impression as a vector of impressed items.
The paper describes the recommender as capable of multi-objective learning.
In the paper's experiments, for a given user-item pair, the recommender is tasked to classify the pair as an interacted or non-interacted impression and to predict the users' watch time.
Hence, the paper considers non-interacted impressions as negative signals.
\citet{DBLP:journals/tois/0003YW0RLSCR23/impression-aware/on-the-behavior-leakage-from-recommender-system-exposure} describe a technique using the encoder-decoder architecture.
The paper does not describe a recommender using impressions but a technique learning from impressions generated by a recommender to derive user feedback.
The paper terms such a model as \emph{attack model}, where the goal of the model is predicting, for a given impression, which items are interacted or non-interacted impressions.
The encoder and decoder networks use a \gls{mlp} architecture with several layers, such as attention or GRU layers.
The paper's results suggest predicting which items in an impression the user will interact with is possible.

\subsubsection{Two-Tower Framework}

Three papers~\cite{DBLP:conf/sigir/ChenWLC0Z22/deep-learning/two-tower-without-impressions-but-exposure-count,DBLP:conf/sigir/WangCLHCYLC22/counterfactual/three-towers-counterfactual-learning-with-impressions,DBLP:conf/sigir/GongZ22/deep-learning/session-based-attention-network} use the two-tower framework.
\citet{DBLP:conf/sigir/ChenWLC0Z22/deep-learning/two-tower-without-impressions-but-exposure-count} describe a recommender in the media and e-commerce domains.
The goal of the recommender is to learn users' preferences while accounting for the popularity of items.
The paper computes the number of impressions of any item.
However, the paper does not specify whether the feature is computed globally, user-wise, item-wise, or pair-wise.
Moreover, the paper only uses the number of impressions in the loss function of the recommender, \idest the recommender does not receive as input a non-interacted impression.
\citet{DBLP:conf/sigir/WangCLHCYLC22/counterfactual/three-towers-counterfactual-learning-with-impressions} describe a recommender in the e-commerce domain.
The goal of the recommender is to estimate users' preferences toward items while accounting for \emph{popularity bias} (overexposure of specific items).
In the paper, an item is an article, and the user may perform two types of interactions: clicking or purchasing items.
The paper defines three levels of user preferences, where the highest level of user preference is a purchased item, then a clicked item, and lastly, a non-interacted item.
The recommender comprises three neural networks, where two networks are trained using all types of impressions.
The input of both networks is a triplet containing the identifier of a user, an item, and a label, where the label indicates whether the user-item pair is a non-interacted or interacted impression.
The other network is trained using only interacted impressions.
\citet{DBLP:conf/sigir/GongZ22/deep-learning/session-based-attention-network} describe a session-based recommender in the news domain.
The goal of the recommender is to recommend the next interacted item for a given sequence of impressions in a session.
In the paper, items are news articles, and sessions are sequences of impressions within a period.
Depending on the user's reading time of an article, the paper considers a user-item pair as an interacted or non-interacted impression.
In particular, interacted impressions are user-item pairs exceeding a certain reading time threshold; analogously, non-interacted impressions are pairs with lower reading time than the threshold.
For a given session-item pair, the recommender is t<rained to distinguish an interacted impression from a non-interacted one; it considers non-interacted impressions as negative signals while interacted ones as positives.
The paper samples items (interacted or non-interacted impressions) from contextual impressions shown in a session.

\subsubsection{Multi-Gate Mixture of Experts}
Five papers~\cite{DBLP:conf/recsys/ZhaoHWCNAKSYC19/deep-learning/two-state-reranker/learns-impressions/impressions-features/recommending-what-video-to-watch-next-a-multitask-ranking-system,DBLP:conf/kdd/MaILYCCNB22/ctr-estimation/an-online-multitask-learning-framework-for-google-feed-ads-auction-models,DBLP:conf/wsdm/XiLLD0ZT023/impressions/a-birds-eye-view-of-reranking-from-list-level-to-page-level,DBLP:conf/cikm/GongFZQDLJG22/impressions/real-time-short-video-recommendation-on-mobile-devices,DBLP:conf/kdd/HuCZWCZ23/impression-aware/boss-a-bilateral-occupational-suitability-aware-recommender-system-for-online-recruitment} use \gls{mmoe}.
\citet{DBLP:conf/recsys/ZhaoHWCNAKSYC19/deep-learning/two-state-reranker/learns-impressions/impressions-features/recommending-what-video-to-watch-next-a-multitask-ranking-system} describe a re-ranker recommender in the media domain.
The paper tasks the recommender to predict different types of user-item feedback, \eg clicks, watch time, likes, and ratings.
For each type of user-item feedback, the recommender outputs a score.
Then, the recommender produces an aggregated score as a weighted linear combination of each feedback score.
To re-rank an impression, the recommender computes the aggregated score of each item in the impression; then, it re-orders the impression by the aggregated score in descending order.
The signals of non-interacted impressions depend on the type of user-item feedback (\eg watching an item or clicking on a dislike button) and the weight associated with such feedback.
As the paper does not disclose the weights, it is not possible to assess how the paper deems non-interacted impressions.
\citet{DBLP:conf/kdd/MaILYCCNB22/ctr-estimation/an-online-multitask-learning-framework-for-google-feed-ads-auction-models} describe a recommender in the online advertisements domain.
In the paper, an item is an advertisement, and the user may perform two types of interactions: clicking or dismissing items.
The paper considers clicking as positive user-item feedback while dismissing it as negative.
The goal of the recommender is to model user preferences to advertisements using those two types of interactions.
The recommender is composed of three networks where only two use impressions.
The first network predicts user preferences by predicting the ratio between the number of interactions and impressions for a given user-item pair.
The second network predicts user preferences by predicting the probability of a user ``dismissing'' an item.
The input of both networks is a triplet containing the identifier of a user, an item, and a label, where the label indicates whether the user-item pair is a non-interacted or interacted impression.
The first network learns from non-interacted and interacted (clicked items) impressions and considers them negative and positive signals, respectively.
Instead, the second network learns from non-interacted and interacted (dismissed items) impressions and considers them positive and negative signals, respectively.
\citet{DBLP:conf/cikm/GongFZQDLJG22/impressions/real-time-short-video-recommendation-on-mobile-devices} describe a re-ranker recommender in the media domain; more specifically, short-videos recommendations.
The paper shows users one video at a time; however, the recommender generates an impression of $N$ items.
The recommender is tasked to re-rank the remaining $N-m$ items, where $m$ is the number of items in the impression the user has already consumed.
The recommender receives as input the impression, the sequence of watched items in the impression, a target video, features computed from the impression, and user feedback on the video (\eg watch time of videos and types of interactions).
For a given item in an impression, the recommender is trained with a \gls{mmoe} to output three probabilities: the user watching another item, watching a certain percentage of the item, and interacting with the item.
\citet{DBLP:conf/wsdm/XiLLD0ZT023/impressions/a-birds-eye-view-of-reranking-from-list-level-to-page-level} describe a re-ranking recommender in the mobile applications domain.
The goal of the recommender is to re-rank the impressions on a page, where a page is a two-dimensional arrangement of several impressions, and each impression is a list.
In the paper, the recommender solely re-ranks the items in the impressions, \idest it generates a permutation of the contents of the impressions and does not change their arrangement on-screen.
The recommender encodes a page as a matrix, where the rows are impressions, and the columns are the items in the impressions.
For a given page, the recommender receives the sequence of user-item interactions and the matrix of impressions as input.
Then, the recommender is trained to predict whether a user-item pair in the matrix is an interaction; hence, non-interacted impressions are deemed negative feedback.
\citet{DBLP:conf/kdd/HuCZWCZ23/impression-aware/boss-a-bilateral-occupational-suitability-aware-recommender-system-for-online-recruitment} describe a recommender in the jobs domain.
The paper proposes a \gls{mmoe} recommender to predict whether a user seeking a job will perform four actions: click, apply, review, and accept.
Thus, the recommender is composed of four experts, one for each user action, designed to correctly predict its corresponding users' actions.
The paper uses impressions to train the recommender.
In particular, on the one hand, non-interacted impressions, \idest those that did not receive any kind of user action, are treated as negative signals across the four experts.
Positive signals, on the other hand, are only those user-item pairs that match a specific user action with the corresponding expert.
Part of the input of the recommender is also the time of impression; however, the paper does not detail such a feature.

\subsubsection{Knowledge Distillation}
\citet{DBLP:journals/isci/Lu23/impression-aware/knowledge-distillation-enhanced-multitask-framework-for-recommendation} proposes a recommender using knowledge distillation in the e-commerce domain.
The recommender uses DeepGBM~\cite{DBLP:conf/kdd/KeXZBL19/non-impressions/deepgbm-a-deep-learning-framework-distilled-by-gbdt-for-online-prediction-tasks}, a framework for knowledge distillation where the teacher and student models are \gls{gbdt} and \gls{mlp}, respectively.
In other words, users' preferences learned by the \gls{gbdt} model are compressed and transferred to the \gls{mlp}.
For a given user-item pair, the recommender estimates the probability of the pair resulting in clicks, purchases, and money spent.
Consequently, the paper uses interacted impressions as positive signals, while non-interacted impressions are negative.

\subsubsection{Pre-trained Language Models}
\citet{DBLP:conf/kdd/WangSXDZ23/impression-aware/bert4ctr-an-efficient-framework-to-combine-pre-trained-language-model-with-non-textual-features-for-ctr-prediction} propose a recommender using a \gls{plm} in online advertisements.
The recommender, called BERT4CTR, is based on the NumBERT\cite{DBLP:conf/blackboxnlp/ZhangRTER20/non-impressions/do-language-embeddings-capture-scales} language model and a novel attention mechanism called Uni-Attention.
The paper optimizes the recommender to predict the \gls{ctr} of a given user-item pair, \idest the ratio between the number of interactions and the number of impressions of the pair.
As such, the paper treats interacted impressions as positive and non-interacted impressions as negative signals.
The paper also computes and includes several features from impressions when training the recommender.
For instance, the user-wise \gls{ctr}, number of impressions for a given item, and the items' position on-screen.

\subsection{Reinforcement Learning}
\label{subsec:reviewed-papers:reinforcement-learning}

In this section, we present \emph{eleven} papers describing recommenders using \gls{rl} to model users' preferences.
Papers in this section model the recommendation task as a Markov decision process~\cite{DBLP:journals/csur/AfsarCF23/non-impressions/reinforcement-learning-based-recommender-systems-a-survey}. The notation and terminology of a recommender using reinforcement learning is:
\begin{itemize}
  \item \textbf{State} (\rlstate): is a tuple containing a user and their impressions and interactions.
  \item \textbf{Action} (\rlaction): is an impression generated by the model for a given user and state.
  \item \textbf{Reward} (\rlreward): indicates the users' preference for the items in an impression. It is a function of the user, state, and action to the set of real numbers.
  \item \textbf{Policy} (\rlpolicy): is the objective function of the recommender.
\end{itemize}
A recommender using reinforcement learning uses the reward to model users' preferences with higher granularity.
Reinforcement learning aims to produce an unbiased estimator maximizing accumulated rewards over time.

One paper uses the policy gradient REINFORCE~\cite{DBLP:journals/ml/Williams92/non-impressions/reinforcement-learning/simple-statistical-gradient-following-algorithms-for-connectionist-reinforcement-learning} algorithm: a method computing a stochastic approximate gradient~\cite{DBLP:conf/aaai/ZhangKOB21/non-impressions/reinforcement-learning/sample-efficient-reinforcement-learning-with-REINFORCE}.
Two papers use the \gls{acf} framework~\cite{DBLP:journals/tsmc/GrondmanBLB12/non-impressions/a-survey-of-actor-critic-reinforcement-learning-standard-and-natural-policy-gradients}: evaluating and updating the policy driving the recommender system during its learning process.
One paper uses \gls{es}~\cite{DBLP:journals/corr/SalimansHCS17/non-impressions/reinforcement-learning/evolution-strategies-as-a-scalable-alternative-to-reinforcement-learning} to model user preferences: optimization algorithms using heuristic emulating the evolution of organisms.
Three papers model the recommendation task as the \gls{mab}~\cite{DBLP:journals/mor/KatehakisV87/non-impressions/reinforcement-learning/the-multi-armed-bandit-problem-decomposition-and-computation,DBLP:journals/eswa/SilvaWSPR22/non-impressions/reinforcement-learning/multi-armed-bandits-in-recommendation-systems-a-survey-of-the-state-of-the-art-and-future-directions} problem: the recommender considers each item as an \emph{arm} and generates impressions by selecting specific arms.
Lastly, four papers use deep neural networks adapted to the reinforcement learning paradigm.

\subsubsection{REINFORCE}
\citet{DBLP:conf/wsdm/XieZW0L22/adversarial/afe-gan-with-impressions} describe a recommender using the REINFORCE algorithm in the e-commerce domain.
The recommender is a \acrlong{gan}, \idest a composition of two neural networks called \emph{generator} and \emph{discriminator}.
The generator network creates user-item pairs representing realistic user feedback, either interacted or non-interacted impressions.
The discriminator is tasked to distinguish whether any user-item pair is real or created by the generator.
Both networks are trained in an adversarial setting: the generator improves by creating realistic user-item pairs deceiving the discriminator; the discriminator improves by classifying the pair's source.

\subsubsection{Actor-Critic Framework}
Two papers~\cite{DBLP:conf/wsdm/DeffayetTRR23/impressions/generative-slate-recommendation-with-reinforcement-learning,DBLP:conf/kdd/ChenWWLZZLZ023/impression-aware/controllable-multi-objective-re-ranking-with-policy-hypernetworks} in the reviewed literature use the \gls{acf} framework.
However, one paper uses the \gls{sac} framework, an off-policy \gls{acf} devised for reinforcement learning using deep neural networks.
\citet{DBLP:conf/kdd/ChenWWLZZLZ023/impression-aware/controllable-multi-objective-re-ranking-with-policy-hypernetworks} describe a re-ranking recommender in the online advertisements domain.
The paper defines the actor as a deep neural network using the encoder-decoder architecture and aiming at re-ranking an impression.
The paper defines uses DeepSet\cite{DBLP:conf/nips/ZaheerKRPSS17/non-impressions/deep-sets} as the encoder network, while it uses PointerNet\cite{DBLP:conf/nips/VinyalsFJ15/non-impressions/pointer-networks} as the decoder network.
Regarding the critic, the paper uses a \gls{mlp} architecture that predicts whether a given impression will receive an interaction from the user, \idest whether any item inside the impression will be interacted with.
As such, the paper considers a positive signal when at least an item in an impression is interacted with; on the contrary, a negative signal is when all items do not receive any interaction.
The paper also describes a business-oriented metric called flow control that limits the number of impressions of certain items.
\citet{DBLP:conf/wsdm/DeffayetTRR23/impressions/generative-slate-recommendation-with-reinforcement-learning} describe a recommender evaluated using simulated user profiles.
The paper uses the \gls{sac}~\cite{DBLP:conf/icml/HaarnojaZAL18/non-impressions/soft-actor-critic} framework to train a recommendation model composed of a \gls{vae}~\cite{DBLP:journals/corr/KingmaW13/non-impressions/auto-encoding-variational-bayes}.
In particular, the paper first trains a \gls{vae} on impressions and their received user feedback, \idest the \gls{vae} learns to generate another impression and to predict the user feedback on such impression.
After training the \gls{vae} model, the paper uses the \gls{vae}'s decoder as the critic of the \gls{sac} framework.

\subsubsection{Evolution Strategies}
\citet{DBLP:conf/www/PeiYCLSJOZ19/reinforcement-learning/economic-ecommerce-recommender-systems} describe a recommender using \gls{es} in the e-commerce domain.
The goal of the recommender is to maximize the financial earnings associated with each type of user-item interaction, \eg the paper considers clicks and purchases to yield different revenues.
The recommender receives as input a tuple containing several features from the user (\eg age) and an impression as a vector of features from impressed items.
The paper considers non-interacted impressions as negative signals, as the reward for this type of user-item feedback is \num{0}.
The reward of other types of user-item feedback varies by the expected economic profit for the action performed by the user.

\subsubsection{Multi-Armed Bandits}
Three papers~\cite{DBLP:conf/sigir/LiKG16/reinforcement-learning/collaborative-filtering-multi-armed-bandits,DBLP:conf/recsys/McInerneyLHHBGM18/reinforcement-learning/multi-armed-bandit-explore-exploit-and-explain,DBLP:conf/wsdm/GrusonCCMHTC19/reinforcement-learning/impressions-to-stream-for-playlist-recommendation} describe \gls{mab} recommenders.
\citet{DBLP:conf/sigir/LiKG16/reinforcement-learning/collaborative-filtering-multi-armed-bandits} describe a recommender in the domain of online advertisements.
The goal of the recommender is to balance exploitation and exploration in their generated recommendations.
In particular, the recommender creates clusters of users and items using global impressions.
The paper defines the reward as a piece-wise function, where non-interacted impressions receive a reward of \num{0}, while interacted impressions receive a reward of \num{1}.
\citet{DBLP:conf/recsys/McInerneyLHHBGM18/reinforcement-learning/multi-armed-bandit-explore-exploit-and-explain} describe a contextual multi-armed bandit recommender in the music domain.
The goal of the recommender is to generate relevant recommendations and ``explanations'', \idest detailing the reasons behind recommendations.
The recommender is tasked to predict whether a given user-item pair is an interacted or non-interacted impression.
Then, the recommender generates impressions containing items with the highest probability of being interactions.
Consequently, the paper deems non-interacted impressions as negative signals.
Moreover, the paper assigns the label ``1'' to interacted impressions and the label ``0'' to non-interacted impressions.
The recommender predicts such labels using a factorization machine~\cite{DBLP:conf/icdm/Rendle10/non-impressions/factorization-machines}.
\citet{DBLP:conf/wsdm/GrusonCCMHTC19/reinforcement-learning/impressions-to-stream-for-playlist-recommendation} describe a recommender and in the music domain.
The goal of the recommender is to maximize the ``impression-to-stream'', \idest the number of interactions per impression on a two-dimensional carousel layout.
The paper defines interacted impressions as playlists with at least one song listened for more than thirty seconds, while non-interacted impressions are playlists without listened songs or with songs reproduced for less than thirty seconds.
Similar to previous papers, the reward is \num{1} when the playlist is an interacted impression, while \num{0} when it is a non-interacted one.

\subsubsection{Deep Neural Networks}
Four papers~\cite{DBLP:conf/wsdm/GeZYPHHZ22/reinforcement-learning/grouping-items-by-exposure,DBLP:conf/cikm/XieZW0L21/reinforcement-learning/teacher-student-rl-network,DBLP:conf/wsdm/ChenBCJBC19/reinforcement-learning/top-k-off-policy-correction-for-a-reinforce-recommender-system,DBLP:conf/wsdm/DeffayetTRR23/impressions/generative-slate-recommendation-with-reinforcement-learning} use deep neural networks adapted to the reinforcement learning paradigm.
\citet{DBLP:conf/wsdm/ChenBCJBC19/reinforcement-learning/top-k-off-policy-correction-for-a-reinforce-recommender-system} describe a recommender in the video domain.
The recommender consists of a \gls{rnn} architecture with chaos-free layers~\cite{DBLP:conf/iclr/0001B17/non-impressions/deep-learning/a-recurrent-neural-network-without-chaos}.
The paper defines the reward function as a non-negative decreasing function of the position of items and whether such items received interactions.
For instance, for an impression with five items, an interaction with the item in the first position yields a reward of \num{5}, while an interaction with the item in the last position yields a reward of \num{1}.
If no item in the impression is interacted with, then the reward is \num{0}.
\citet{DBLP:conf/cikm/XieZW0L21/reinforcement-learning/teacher-student-rl-network} describe a recommender in the product's domain.
The goal of the recommender is to generate relevant recommendations in a resource-constrained scenario.
The recommender is a composition of two \acrshortpl{dqn}~\cite{DBLP:journals/corr/MnihKSGAWR13/non-impressions/reinforcement-learning-deep-q-network}, where each network receives the state as input, and their output is an impression.
The state consists of the set of users' profiles, a vector of the user's interacted impressions, and a set of the user's contextual impressions.
The paper evaluates the recommender in an offline and online setting.
\citet{DBLP:conf/wsdm/GeZYPHHZ22/reinforcement-learning/grouping-items-by-exposure} describe a recommender in the media and e-commerce domains.
The goal of the recommender is to generate accurate and fair recommendations.
The recommender consists of a single \gls{mlp} network with GRU layers.
In the paper, fairness is related to the number of impressions of a given user with the same item.
The paper defines the set of long-tail items (denoted as \rllongtail{\rlstate}) on a given state containing items with the lowest number of impressions, precisely, \rllongtail{\rlstate} encloses \qty{80}{\percent} of the items in the catalog.
The reward is a combination of an accuracy-based and a fairness-based reward.
For a given state and action, the accuracy-based reward (denoted as $\rlreward^{acc}_{\rlstate, \rlaction}$) is the precision~\cite{DBLP:journals/tois/FerrariDacremaBCJ21/a-troubling-analysis} and the fairness-based reward (denoted as $\rlreward^{fair}_{\rlstate, \rlaction}$) is the hinge-loss between the precision of long-tail items and the desired percentage of long-tail items in an impression:
\begin{equation}
  \label{eq:reinforcement-learning:accuracy-and-fairness-reward}
  \begin{split}
    \rlreward^{acc}_{\rlstate, \rlaction}  =
    \frac{
      \sum_{j \in \rlaction}
      {
        \functionindicator\left(j \text{ is interacted in } \rlaction \right)
      }
    }{
      |\rlaction|
    }
    \,,\qquad
    \rlreward^{fair}_{\rlstate, \rlaction}  = \max \left(
    \frac{
      \sum_{j \in \rlaction}
      {
        \functionindicator\left(j \in \rllongtail{\rlstate} \right)
      }
    }{
      |\rlaction|
    },
    \beta
    \right)
  \end{split}
\end{equation}
where $\functionindicator$ returns $1$ if the predicate is true, $0$ otherwise; \rllongtail{\rlstate} is the set of long-tail items for a given user in \rlstate; and $\beta$ is the target percentage of long-tail items in \rlaction.

\subsection{Not Described}
\label{subsec:not-described}

In this section, we describe \emph{three} reviewed papers that do not propose a recommendation model using impressions, but analyze impressions in recommender systems from other angles or contribute to the research and development of \gls{impressionsbased} in other ways.
\citet{DBLP:conf/recsys/ZhaoWAHK18/impressions-signals/interpreting-user-inaction-in-recsys} describe how to extract the signals from non-interacted impressions using user studies on the media domain.
The paper describes participants of the study as registered users of a movie recommender system, who, after a certain number of interactions, were given the option to participate in the study.
Participants were asked about their preference for a non-interacted item in an impression after interacting with it.
Regarding the results of the study, \qty{38.6}{\percent} of the responses were classified as \emph{unaware}, meaning the participant did not know the item was in the impression.
Moreover, only \qty{5.8}{\percent} of the responses were classified as \emph{not enjoy}, meaning the impressed item is a negative signal.
Due to the low number of responses in the last category, the paper states that treating non-interacted impressions as negative signals ``could be problematic''.
\citet{DBLP:conf/sigir/NayakGM23/impression-aware/news-popularity-beyond-the-click-through-rate-for-personalized-recommendations} propose three evaluation metrics for recommender systems on the news domain.
Those metrics inspect all items included in the impressions generated by the recommendation model.
The first metric, called lifespan, measures the period items are deemed as relevant by users.
The second metric, called half-life, measures the period items take to reach half their number of impressions since their first one.
The third metric, called peak-time, measures the period items take to reach their peak number of impressions since their first one; it is computed as a moving average.
\citet{DBLP:conf/sigir/Sun23/impression-aware/take-a-fresh-look-at-recommender-systems-from-an-evaluation-standpoint} proposes several directions to improve users' preferences modeling.
One direction is to exploit contextual impressions in the learning process of recommendation models.
By adding this type of impression, it is expected that the recommendation model can extract more refined information about users' preferences, at least when compared to using interactions alone.
With contextual impressions, it is possible to contextualize the users' decision process, as they contain all the shown items, their position on-screen, and kind of received user feedback.

\subsection{Discussion}

After reviewing the literature of \gls{impressionsbased}, composed by \numincludedworks papers, we identify several trends in the recommendation models present in those papers and how they use impressions.
We observe an increase in the number of \gls{impressionsbased} papers over time, especially in recent years, as \autoref{fig:tikz-reviewed-papers} shows.
A higher number of published papers serves as an indicator of the community's interest in this learning paradigm.
In particular, almost half of the reviewed papers were published between \num{2023} and \num{2022}; furthermore, two-thirds of the reviewed papers were published in the past five years, and all but one were published in the past ten years.

\subsubsection{Model-centric Taxonomy}

\autoref{tab:impressions-recommenders} shows the classification of each reviewed paper.
In the table, each row presents the classification of one paper according to the proposed taxonomies described in \autoref{subsec:impressions-based-recsys:classification} and properties of recommenders and impressions described in \autoref{subsec:impressions-based-recsys:problem-description}.\footnote{One paper~\cite{DBLP:conf/www/LiuRSKMZLJ17/heuristics/memboost-impressions-as-features} appears twice in \autoref{tab:impressions-recommenders} because it describes two recommenders using impressions.}\footnote{The order of papers in the text and the table is the same.}
As seen in the table, the two most popular categories of recommendation models are \emph{deep learning} and \emph{reinforcement learning}, becoming popular in the past five years.
Older papers, instead, describe recommendation models classified as heuristics or statistical.
Between older and recent papers, we find that papers describe recommendation models that use shallow machine learning, particularly in \num{2016}, \num{2017}, \num{2019}, and \num{2020}.

\subsubsection{Data-centric Taxonomy}
The two most popular choices in the literature are extracting features and learning from impressions, as seen in \autoref{tab:impressions-recommenders}.
In the literature, the most popular feature is \emph{the number of user-item impressions}, \idest \numimpressions{\useru, \itemi}.
Particularly, this feature indicates the times a given user has seen a given item inside an impression.
Other features, less observed in the literature, comprise the \gls{ctr}, the on-screen position of impressed items, and the time since the last impression with the same item.
In \autoref{subsec:impressions-based-recsys:classification}, we also define learning from impressions as receiving either a single impressed item or all items in an impression as part of the input of the recommendation model.
In the literature, most papers pass a single impressed item to recommendation models.
This means the input of a recommendation model consists of a $n$-tuple holding a user identifier, an impressed item identifier, the users' action on the item, and a variable number of user-wise and item-wise attributes.

\begin{table}[ht!]
  \centering
  \small
  \caption{
    Categorization of \gls{impressionsbased} papers reviewed in this work, where each row is one paper.
    \textbf{Model-Centric Taxonomy} categorizes papers by their recommender design.
    \textbf{Data-Centric Taxonomy} categorizes papers by how they use impressions.
    \textbf{Signal-Centric Taxonomy} categorizes papers by how they deem non-interacted impressions to user preferences.
    \textbf{Impression Type} categorizes papers using contextual or global impressions.
    \textbf{Impression Signals} categorizes papers using non-interacted impressions as positive, negative, or neutral signals.
    \textbf{Recommender Type} categorizes papers by how the recommender generates impressions.
  }
  \label{tab:impressions-recommenders}
  \begin{minipage}{\linewidth}
    \centering
    \begin{tabular}{clccccc}
      \toprule
      \textbf{\thead{Model-Centric\\Taxonomy}}
        &  \textbf{\thead{Paper\\Reference}}
        &  \textbf{\thead{Data-Centric\\Taxonomy}}
        &  \textbf{\thead{Signal-Centric\\Taxonomy}}
        &  \textbf{\thead{Impressions\\Type}}
        &  \textbf{\thead{Impressions\\Signal}}
        &  \textbf{\thead{Recommender\\Type}}
      \\
      \midrule
      \multirow{5}[0]{*}{Heuristic}
        &  \citet{DBLP:journals/scheduling/BuchbinderFGN14/hard-frequency-cap}
        &  features
        &  assume
        &  global
        &  negative
        &  end-to-end
      \\
      {}
        &  \citet{DBLP:conf/cscw/ZhaoAHWK17/heuristics/cycling-serpentining}
        &  features
        &  assume
        &  global
        &  negative
        &  re-ranking
      \\
      {}
        &  \citet{DBLP:conf/kdd/LeeLTS14/impression-discounting}
        &  features \& learn
        &  learn
        &  global
        &  negative
        &  plug-in
      \\
      {}
        &  \citet{DBLP:conf/kdd/AgarwalCGHHIKMSSZ14/impression-discounting}
        &  features \& learn
        &  learn
        &  global
        &  negative
        &  plug-in
      \\
      {}
        &  \citet{DBLP:conf/www/LiuRSKMZLJ17/heuristics/memboost-impressions-as-features}
        &  features
        &  assume
        &  global
        &  negative
        &  end-to-end
      \\
      \midrule
      \multirow{5}[0]{*}{Statistical}
        &  \citet{DBLP:conf/www/AgarwalCE09/spatio-temporal-models-for-estimating-ctr}
        &  features
        &  learn
        &  global
        &  neutral
        &  end-to-end
      \\
      {}
        &  \citet{DBLP:conf/recsys/WuASB16/user-fatigue/netflix-fatigue-modeling-using-impressions}
        &  features
        &  learn
        &  global
        &  neutral
        &  end-to-end
      \\
      {}
        &  \citet{DBLP:conf/kdd/ZhangZMCZA16/heuristics/glmix-generalized-linear-mixed-models-for-large-scale-response-prediction}
        &  learn
        &  assume
        &  global
        &  negative
        &  end-to-end
      \\
      {}
        &  \citet{DBLP:conf/kdd/BorisyukZK17/heuristics/lijar-job-boosting-by-impressions}
        &  features
        &  learn
        &  global
        &  neutral
        &  plug-in
      \\
      {}
        &  \citet{DBLP:conf/kdd/LinCSLLJ23/impression-aware/tree-based-progressive-regression-model-for-watch-time-prediction-in-short-video-recommendation}
        &  features \& learn
        &  learn
        &  global
        &  neutral
        &  end-to-end
      \\
      \midrule
      \multirow{4}[0]{*}{\makecell{Machine\\learning}}
        &  \citet{DBLP:conf/www/LiuRSKMZLJ17/heuristics/memboost-impressions-as-features}
        &  learn
        &  assume
        &  global
        &  negative
        &  end-to-end
      \\
      {}
        &  \citet{DBLP:conf/www/MaLS16/user-fatigue/online-news-recommendation-feature-engineering-of-impressions}
        &  features \& learn
        &  assume
        &  contextual
        &  negative
        &  end-to-end
      \\
      {}
        &  \citet{DBLP:conf/cikm/AharonKLSBESSZ19/heuristics/soft-frequency-cap}
        &  features \& learn
        &  assume
        &  global
        &  negative
        &  end-to-end
      \\
      {}
        &  \citet{DBLP:conf/cikm/PerezMaureraFDSSC20/contentwise-impressions}
        &  sample
        &  assume
        &  global
        &  negative
        &  end-to-end
      \\
      \midrule
      \multirow{17}[0]{*}{\makecell{Deep\\learning}}
        &  \citet{DBLP:conf/recsys/CovingtonAS16/deep-learning/two-stage-reranker/learns-impressions/impressions-features/deep-neural-networks-for-youtube-recommendations}
        &  features \& learn
        &  assume
        &  global
        &  negative
        &  re-ranking
      \\
      {}
        &  \citet{DBLP:conf/kdd/ZhanPSWWMZJG22/heuristics/impressions-to-compute-statistical-features}
        &  features \& learn
        &  learn
        &  global
        &  neutral
        &  end-to-end
      \\
      {}
        &  \citet{DBLP:conf/kdd/ZhuD00C22/deep-learning/combo-fashion-clothes-matching}
        &  features \& learn
        &  assume
        &  global
        &  negative
        &  end-to-end
      \\
      {}
        &  \citet{DBLP:conf/cikm/ZhangCXBHDZ22/impressions/keep-an-industrial-pre-training-framework-for-online-recommendation-via-knowledge-extraction-and-plugging}
        &  learn
        &  assume
        &  global
        &  negative
        &  end-to-end
      \\
      {}
        &  \citet{DBLP:conf/ijcai/BiedNPHCCGS23/impression-aware/toward-job-recommendation-for-all}
        &  learn
        &  assume
        &  global
        &  positive
        &  end-to-end
      \\
      {}
        &  \citet{DBLP:conf/wsdm/RenHZZZ23/impressions/slate-aware-ranking-for-recommendation}
        &  learn
        &  learn
        &  contextual
        &  neutral
        &  re-ranking
      \\
      {}
        &  \citet{DBLP:journals/tois/0003YW0RLSCR23/impression-aware/on-the-behavior-leakage-from-recommender-system-exposure}
        &  learn
        &  not described
        &  contextual
        &  neutral
        &  end-to-end
      \\
      {}
        &  \citet{DBLP:conf/sigir/ChenWLC0Z22/deep-learning/two-tower-without-impressions-but-exposure-count}
        &  features
        &  not described
        &  global
        &  neutral
        &  end-to-end
      \\
      {}
        &  \citet{DBLP:conf/sigir/WangCLHCYLC22/counterfactual/three-towers-counterfactual-learning-with-impressions}
        &  learn
        &  assume
        &  global
        &  negative
        &  end-to-end
      \\
      {}
        &  \citet{DBLP:conf/sigir/GongZ22/deep-learning/session-based-attention-network}
        &  sample
        &  assume
        &  contextual
        &  negative
        &  end-to-end
      \\
      {}
        &  \citet{DBLP:conf/recsys/ZhaoHWCNAKSYC19/deep-learning/two-state-reranker/learns-impressions/impressions-features/recommending-what-video-to-watch-next-a-multitask-ranking-system}
        &  features \& learn
        &  assume
        &  contextual
        &  negative
        &  re-ranking
      \\
      {}
        &  \citet{DBLP:conf/kdd/MaILYCCNB22/ctr-estimation/an-online-multitask-learning-framework-for-google-feed-ads-auction-models}
        &  learn
        &  assume
        &  global
        &  negative
        &  end-to-end
      \\
      {}
        &  \citet{DBLP:conf/cikm/GongFZQDLJG22/impressions/real-time-short-video-recommendation-on-mobile-devices}
        &  features
        &  assume
        &  contextual
        &  negative
        &  re-ranking
      \\
      {}
        &  \citet{DBLP:conf/wsdm/XiLLD0ZT023/impressions/a-birds-eye-view-of-reranking-from-list-level-to-page-level}
        &  learn
        &  assume
        &  contextual
        &  negative
        &  re-ranking
      \\
      {}
        &  \citet{DBLP:conf/kdd/HuCZWCZ23/impression-aware/boss-a-bilateral-occupational-suitability-aware-recommender-system-for-online-recruitment}
        &  features \& learn
        &  assume
        &  global
        &  negative
        &  end-to-end
      \\
      {}
        &  \citet{DBLP:journals/isci/Lu23/impression-aware/knowledge-distillation-enhanced-multitask-framework-for-recommendation}
        &  learn
        &  assume
        &  global
        &  negative
        &  end-to-end
      \\
      {}
        &  \citet{DBLP:conf/kdd/WangSXDZ23/impression-aware/bert4ctr-an-efficient-framework-to-combine-pre-trained-language-model-with-non-textual-features-for-ctr-prediction}
        &  features \& learn
        &  assume
        &  global
        &  negative
        &  end-to-end
      \\
      \midrule
      \multirow{10}[0]{*}{\makecell{Reinforcement\\learning}}
        &  \citet{DBLP:conf/wsdm/XieZW0L22/adversarial/afe-gan-with-impressions}
        &  learn
        &  assume
        &  contextual
        &  negative
        &  end-to-end
      \\
      {}
        &  \citet{DBLP:conf/kdd/ChenWWLZZLZ023/impression-aware/controllable-multi-objective-re-ranking-with-policy-hypernetworks}
        &  learn
        &  learn
        &  contextual
        &  neutral
        &  re-ranking
      \\
      {}
        &  \citet{DBLP:conf/wsdm/DeffayetTRR23/impressions/generative-slate-recommendation-with-reinforcement-learning}
        &  learn
        &  learn
        &  contextual
        &  neutral
        &  end-to-end
      \\
      {}
        &  \citet{DBLP:conf/www/PeiYCLSJOZ19/reinforcement-learning/economic-ecommerce-recommender-systems}
        &  features
        &  assume
        &  contextual
        &  negative
        &  end-to-end
      \\
      {}
        &  \citet{DBLP:conf/sigir/LiKG16/reinforcement-learning/collaborative-filtering-multi-armed-bandits}
        &  learn
        &  assume
        &  global
        &  negative
        &  end-to-end
      \\
      {}
        &  \citet{DBLP:conf/recsys/McInerneyLHHBGM18/reinforcement-learning/multi-armed-bandit-explore-exploit-and-explain}
        &  learn
        &  assume
        &  global
        &  negative
        &  plug-in
      \\
      {}
        &  \citet{DBLP:conf/wsdm/GrusonCCMHTC19/reinforcement-learning/impressions-to-stream-for-playlist-recommendation}
        &  learn
        &  assume
        &  global
        &  negative
        &  end-to-end
      \\
      {}
        &  \citet{DBLP:conf/wsdm/ChenBCJBC19/reinforcement-learning/top-k-off-policy-correction-for-a-reinforce-recommender-system}
        &  learn
        &  assume
        &  global
        &  negative
        &  end-to-end
      \\
      {}
        &  \citet{DBLP:conf/cikm/XieZW0L21/reinforcement-learning/teacher-student-rl-network}
        &  learn
        &  assume
        &  contextual
        &  negative
        &  end-to-end
      \\
      {}
        &  \citet{DBLP:conf/wsdm/GeZYPHHZ22/reinforcement-learning/grouping-items-by-exposure}
        &  features
        &  not described
        &  global
        &  neutral
        &  end-to-end
      \\
      \midrule
      \multirow{3}[0]{*}{\makecell{Not\\described}}
        &  \citet{DBLP:conf/recsys/ZhaoWAHK18/impressions-signals/interpreting-user-inaction-in-recsys}
        &  not described
        &  learn
        &  contextual
        &  neutral
        &  not described
      \\
      {}
        &  \citet{DBLP:conf/sigir/NayakGM23/impression-aware/news-popularity-beyond-the-click-through-rate-for-personalized-recommendations}
        &  not described
        &  not described
        &  contextual
        &  neutral
        &  not described
      \\
      {}
        &  \citet{DBLP:conf/sigir/Sun23/impression-aware/take-a-fresh-look-at-recommender-systems-from-an-evaluation-standpoint}
        &  not described
        &  not described
        &  not described
        &  not described
        &  not described
      \\
      \bottomrule
    \end{tabular}
  \end{minipage}
\end{table}

\clearpage 

\subsubsection{Signal-centric Taxonomy}
In terms of how papers deem impressions in terms of users' preferences, especially non-interacted impressions, we observe that reviewed papers use several approaches with varying complexity.
On the one hand, most papers treat non-interacted impressions as negative signals implicitly issued by users.
This approach assumes that users' preferences are binary, where an interaction represents strong positive feedback, while a non-interaction represents strong negative feedback.
Moreover, this approach is also the simplest to implement; however, as illustrated in previous research~\cite{DBLP:conf/recsys/ZhaoWAHK18/impressions-signals/interpreting-user-inaction-in-recsys}, it does not take advantage of impressions and their signals.
On the other hand, a handful of papers prefer recommendation models that learn the users' preferences toward impressed items using additional information.
For instance, \citet{DBLP:conf/kdd/LeeLTS14/impression-discounting} computes the number of past impressions for user-item pairs, while \citet{DBLP:conf/kdd/ZhanPSWWMZJG22/heuristics/impressions-to-compute-statistical-features} measure the time spent by a user watching an item.
Approaches that learn users' preferences are more complex in the literature.
Also, they tend to extract more information from impressions.
When combined, both aspects may result in better modeling of users' preferences and, ultimately, better recommendations.

\section{Datasets with Impressions}
\label{sec:datasets-with-impressions}

Researchers and practitioners need access to datasets with impressions to evaluate their proposed \gls{impressionsbased}.
Before 2020, only five datasets were available for research purposes.
The landscape has evolved, as nowadays, researchers can access and use thirteen datasets in their research works.
This section describes datasets with impressions, where we use the definitions of impressions, interactions, and the different types of user-item feedback provided in \autoref{subsec:impressions-based-recsys:problem-description}.
We classify datasets with impressions into three categories: \emph{public}, \emph{expired}, and \emph{private} datasets.
The definition of each category is the same as the one stated by \citet{DBLP:conf/cikm/PerezMaureraFDSSC20/contentwise-impressions}:
\begin{itemize}
  \item \textbf{Public:} datasets accessible via the Internet or by request to publishers. They can be used in future research activities if their license agreements are met.
  \item \textbf{Expired:} datasets available to participants of competitions. They are currently not accessible or cannot be used in research activities.
  \item \textbf{Private:} datasets not published nor publicly available.
\end{itemize}

\autoref{tab:public-datasets:statistics} and \autoref{tab:expired-datasets:statistics} summarize the statistics of the datasets with impressions that we identify in the literature.
\autoref{tab:public-datasets:statistics} reports the statistics of public datasets, where those statistics are computed using our definition of impressions and interactions.
\autoref{tab:expired-datasets:statistics} reports the statistics of expired and private datasets, where those statistics are extracted from the paper using the dataset.
In most cases, papers do not report such statistics or may compute them using different definitions.
This section only describes public and expired datasets, as
papers limit the details of private datasets due to privacy and business property constraints.

\begin{table}[t]
  \centering
  \small
  \caption{Statistics of public datasets with impressions.
    \textbf{Type} refers to the type of impressions in the dataset.
    \textbf{Year} refers to when the dataset is published.
    \textbf{Users} and \textbf{Items} refer to the number of users and items, respectively.
    \textbf{Impressions} refers to the number of impressed user-item pairs.
    \textbf{Interactions} refer to the number of interacted user-item pairs.
    \textbf{Imp x Int} refers to the ratio between the number of impressed and interacted user-item pairs.
    Datasets are grouped by category and by year in ascending order.
  }
  \label{tab:public-datasets:statistics}
  \begin{minipage}{\linewidth}
    \centering
    \begin{tabular}{@{}clcrrrrrrS[table-alignment=right,table-format = 1.2e-1]@{}}
      \toprule
      \textbf{Type}
        &  \textbf{Dataset}
        &  \textbf{Year}
        &  \textbf{Users}
        &  \textbf{Items}
        &  \textbf{Impressions}
        &  \textbf{Interactions}
        &  \textbf{Imp x Int}
      \\
      \midrule
      \multirow{4}[0]{*}{\makecell{Contextual}}
        &  \datasetcw~\cite{DBLP:conf/cikm/PerezMaureraFDSSC20/contentwise-impressions}
        &  2020
        &  42.15 K
        &  28.88 K
        &  253.19 M
        &  10.46 M
        &  24.21
      \\
      {}
        &  \datasetmind~\cite{DBLP:conf/acl/WuQCWQLLXGWZ20/MIND-a-large-scale-dataset-for-news-recommendation}
        &  2020
        &  876.96 K
        &  130.38 K
        &  93.63 M
        &  17.70 M
        &  5.29
      \\
      {}
        &  \datasetfinn~\cite{DBLP:conf/recsys/EideLFRJV21/finn-no-slates-dataset,DBLP:journals/datamine/EideLF22/dynamic-slate-recommendation-with-gated-recurrent-units-and-thompson-sampling}
        &  2021
        &  2.28 M
        &  1.31 M
        &  346.66 M
        &  28.28 M
        &  12.26
      \\
      {}
        &  \datasetslrs~\cite{DBLP:conf/sigir/WangZZDSLWSLF23/rl4rs-a-real-world-dataset-for-reinforcement-learning-based-recommender-system}
        &  2023
        &  -\footnote{The number of users cannot be computed as the dataset does not include unique numerical identifiers for users.}
        &  283
        &  30.95 M
        &  18.69 M
        &  1.66
      \\
      \midrule
      \multirow{10}[0]{*}{\makecell{Global}}
        &  \datasetyahoorsixa~\cite{DBLP:conf/wsdm/LiCLW11/datasets/yahoo-r6a,DBLP:conf/kdd/ChuPBMPCZ09/datasets/yahoo-r6a}
        &  2009
        &  29.85 M
        &  271
        &  44.18 M
        &  1.63 M
        &  27.05
      \\
      {}
        &  \datasetyahoorsixb~\cite{DBLP:conf/icml/GentileLZ14/datasets/yahoo-r6b,DBLP:conf/sigir/LiKG16/reinforcement-learning/collaborative-filtering-multi-armed-bandits}
        &  2011
        &  1.28 M
        &  652
        &  26.75 M
        &  1.03 M
        &  26.03
      \\
      {}
        &  \datasetsearchads\footnote{\url{https://www.kaggle.com/competitions/kddcup2012-track2}}
        &  2012
        &  23.91 M
        &  670.56 K
        &  257.95 M
        &  9.14 M
        &  28.21
      \\
      {}
        &  \datasetpandor~\cite{DBLP:conf/recsys/SidanaLA18/datasets/PANDOR}
        &  2018
        &  5.89 M
        &  14.72 K
        &  48.42 M
        &  337.51 K
        &  143.45
      \\
      {}
        &  \datasetaliccp~\cite{DBLP:conf/sigir/MaZHWHZG18/datasets/ali-ccp-entire-space-multi-task-model}
        &  2018
        &  445 K
        &  4.35 M
        &  82.00 M
        &  3.32 M
        &  24.71
      \\
      {}
        &  \datasetalimama~\cite{DBLP:conf/www/ShenWTZLCL22/datasets/alimama-deep-interest-highlight-network}
        &  2022
        &  1.14 M
        &  846.81 K
        &  25.19 M
        &  1.37 M
        &  18.44
      \\
      {}
        &  \datasetcrossshop~\cite{DBLP:conf/kdd/ZhuD00C22/deep-learning/combo-fashion-clothes-matching}
        &  2022
        &  428.20 K
        &  3.49 M
        &  5.70 M
        &  306.83 K
        &  18.58
      \\
      {}
        &  \datasetinshop~\cite{DBLP:conf/kdd/ZhuD00C22/deep-learning/combo-fashion-clothes-matching}
        &  2022
        &  2.64 M
        &  5.27 M
        &  32.11 M
        &  3.49 M
        &  9.19
      \\
      {}
        &  \datasetkwaisystem~\cite{DBLP:conf/kdd/WangMLLZLLJM22/fairness/fair-item-utility-estimation-and-exposure-redistribution}
        &  2022
        &  5.70 M
        &  12.58 K
        &  5.05 M
        &  2.77 M
        &  1.82
      \\
      {}
        &  \datasetkwairandom~\cite{DBLP:conf/kdd/WangMLLZLLJM22/fairness/fair-item-utility-estimation-and-exposure-redistribution}
        &  2022
        &  10.62 M
        &  12.75 K
        &  10.45 M
        &  1.82 M
        &  5.74
      \\
      \bottomrule
    \end{tabular}
  \end{minipage}
\end{table}

\subsection{Public Datasets}
\label{subsec:datasets-with-impressions:public-datasets}

Public datasets are accessible to researchers and practitioners and can be used in future research activities.
Such datasets are available online or upon request to the publishers.
However, future work must comply with each dataset's license agreements.\footnote{For instance, the \datasetcw dataset cannot be used for commercial purposes.}
\autoref{tab:public-datasets:statistics} summarizes relevant statistics of public datasets, such as the number of users, items, user-item interactions, and user-item impressions.
In this section, we comprehensively describe each public dataset and its attributes, \eg its definition of users, items, and collection period.
We classify public datasets into two categories based on the type of impressions such datasets contain:

\begin{itemize}
  \item \textbf{Contextual:} datasets containing interactions and impressions with their connections, \idest for a user-item interaction, it is known which impression has the interacted item.
  \item \textbf{Global:} datasets containing interactions and impressions without their connections, \idest for a user-item interaction, it is not known which impression has the interacted item.
\end{itemize}

Global datasets have a reduced utility when compared to contextual counterparts.
By their definition, global datasets do not contain the context of interactions, \idest it is not possible to connect interacted items with the impression holding them.
This implies it is not possible to know which other items were impressed when the interaction happened, the position of the interacted item on the impression, or the arrangement of the impression on-screen.
Furthermore, five recommenders in reviewed papers~\cite{DBLP:conf/wsdm/XieZW0L22/adversarial/afe-gan-with-impressions,DBLP:conf/sigir/GongZ22/deep-learning/session-based-attention-network,DBLP:conf/cikm/XieZW0L21/reinforcement-learning/teacher-student-rl-network,DBLP:conf/www/PeiYCLSJOZ19/reinforcement-learning/economic-ecommerce-recommender-systems,DBLP:conf/www/MaLS16/user-fatigue/online-news-recommendation-feature-engineering-of-impressions} use contextual impressions.
Lastly, global datasets cannot be used when sampling interacted and non-interacted items from the same impression, as used in the literature~\cite{DBLP:conf/acl/WuQCWQLLXGWZ20/MIND-a-large-scale-dataset-for-news-recommendation,DBLP:conf/sigir/GongZ22/deep-learning/session-based-attention-network}.

\subsection{Public Datasets with Contextual Impressions}
\label{subsec:datasets-with-impressions:contextual-datasets}

We identify four public datasets containing contextual impressions.
In those datasets, a contextual impression is a tuple containing, at least, a user identifier and an impression as a vector of item identifiers.
In two datasets, the tuple also contains an item identifier representing the interacted item.
When the user did not interact with any item in the impression, the item identifier placed in the dataset is empty or replaced with a unique code.
In the two other datasets, the tuple contains an additional vector of labels with the same number of elements as the impression vector.
In this vector, there is one label for each impressed item, where the label indicates the type of user feedback such item received.

Two datasets contain the item identifier in the contextual impression.
The \datasetcw dataset~\cite{DBLP:conf/cikm/PerezMaureraFDSSC20/contentwise-impressions} contains impressions from an online streaming media service.\footnote{\datasetcw is accessible at \url{https://github.com/ContentWise/contentwise-impressions}}
The data was collected from January to April 2019. Users are anonymized registered accounts with the service, and items are the media content related to TV series and movies.
The dataset also contains on-screen layout information: the position of the impression on a two-dimensional layout. Lastly, most impressions in the dataset do not contain interacted items.
The \datasetfinn~\cite{DBLP:conf/recsys/EideLFRJV21/finn-no-slates-dataset,DBLP:journals/datamine/EideLF22/dynamic-slate-recommendation-with-gated-recurrent-units-and-thompson-sampling} dataset contains contextual impressions from an e-commerce service.\footnote{\datasetfinn is accessible at \url{https://github.com/finn-no/recsys_slates_dataset}} The data was collected over thirty days; however, the dataset does not contain the date and time of impressions or interactions. Users are registered accounts, and items are products and goods. Similar to the \datasetcw dataset, most impressions do not contain interactions. The dataset contains impressions generated by a search engine and a recommender system.

Two datasets contain labels of users' feedback.
The \datasetmind dataset~\cite{DBLP:conf/acl/WuQCWQLLXGWZ20/MIND-a-large-scale-dataset-for-news-recommendation} contains impressions from an online news service.\footnote{\datasetmind is accessible at \url{https://msnews.github.io}.}
The data was collected between October \nth{12} and November \nth{22} 2019, for a total of six weeks, on users with at least five interactions between those dates. Users are anonymized registered accounts, and items are news articles. The dataset does not contain all the impressions generated by the recommender system during the collection period. Instead, it contains the impressions generated between the \nth{5} and \nth{6} weeks. Moreover, impressions in the \nth{6} week are not labeled. This dataset was provided to the participants of the \emph{MIND News Recommendation Competition} where they were tasked to devise a re-ranker recommender (see \autoref{subsec:impressions-based-recsys:problem-description}).\footnote{Details on the MIND News Recommendation Competition are available at \url{https://msnews.github.io/competition.html}}
The \datasetslrs dataset~\cite{DBLP:conf/sigir/WangZZDSLWSLF23/rl4rs-a-real-world-dataset-for-reinforcement-learning-based-recommender-system} contains impressions from an e-commerce service.\footnote{\datasetslrs is accessible at \url{https://github.com/fuxiAIlab/RL4RS}.}
The dataset's reference does not specify the period where the data was collected, nor the definition of users or items.
However, it does specify that data were collected from an online game and describes some of the users' contextual and items' content information available in the dataset.
A unique trait of the dataset is that it contains data points collected before and after the deployment of a recommender system.
Another unique trait is that it contains impressions generated for a single user action and a sequence of users' actions.
Consequently, the dataset can be divided into four partitions based on the type of data points and the recommendation task at hand.

\subsection{Public Datasets with Global Impressions}
\label{subsec:public-datasets-with-global-impressions}

We identify ten public datasets containing global impressions. In those datasets, a global impression is a tuple containing a user identifier and an item identifier.
Seven datasets include a label to indicate whether the user-item pair corresponds to an interacted or non-interacted impression.
One dataset includes two binary labels to indicate whether the item was impressed to the user and whether the item received an interaction, respectively.
One dataset includes the number of interactions and impressions for the user-item pair.
One dataset includes the numbers of user-item clicks and user-item purchases for the user-item pair. Hence, a non-interacted impression is a tuple with zero user-item clicks and purchases.

Seven datasets contain a label to indicate an interacted or non-interacted impression.
The \datasetyahoorsixa and \datasetyahoorsixb datasets~\cite{DBLP:conf/wsdm/LiCLW11/datasets/yahoo-r6a,DBLP:conf/kdd/ChuPBMPCZ09/datasets/yahoo-r6a,DBLP:conf/icml/GentileLZ14/datasets/yahoo-r6b,DBLP:conf/sigir/LiKG16/reinforcement-learning/collaborative-filtering-multi-armed-bandits} contain impressions from an online news service.\footnote{\datasetyahoorsixa is accessible upon request at \url{https://webscope.sandbox.yahoo.com/catalog.php?datatype=r&did=49}}\footnote{\datasetyahoorsixb is accessible upon request at \url{https://webscope.sandbox.yahoo.com/catalog.php?datatype=r&did=54}} In the first dataset, the data was collected in the first ten days of May 2009, while in the second it was collected between October \nth{2} and \nth{16} 2011. In both datasets, users are anonymous accounts visiting an online news recommender system, and items are news articles.
Those datasets contain impressions generated by a recommender selecting items at random. Hence, the dataset is useful for evaluating recommenders using reinforcement learning or counterfactual learning.
The \datasetpandor dataset~\cite{DBLP:conf/recsys/SidanaLA18/datasets/PANDOR} contains impressions of online advertisements from a media and news service.\footnote{\datasetpandor is accessible at \url{https://archive.ics.uci.edu/ml/datasets/PANDOR}} Users are anonymous accounts using the service, and items are advertisements.
The dataset contains impressions generated by a top-popular or a similarity-based recommender.
The \datasetinshop and \datasetcrossshop datasets~\cite{DBLP:conf/kdd/ZhuD00C22/deep-learning/combo-fashion-clothes-matching} contain impressions of fashion garments on an e-commerce service. In both datasets, the data was collected over forty days. Users are registered accounts on the service, and items are combinations of bottom and top garments.\footnote{\datasetinshop and \datasetcrossshop are accessible at \url{https://tianchi.aliyun.com/dataset/dataDetail?dataId=131519}.} Each garment has an attribute called \emph{store}; however, the paper does not detail the meaning of such an attribute. Nevertheless, that attribute is used to distinguish the types of items in each dataset: in the \datasetinshop dataset, top and bottom garments share the same store, while in the \datasetcrossshop dataset, they may have different stores.
The \datasetkwaisystem and \datasetkwairandom datasets~\cite{DBLP:conf/kdd/WangMLLZLLJM22/fairness/fair-item-utility-estimation-and-exposure-redistribution} contain impressions of short videos from a social network. Users are accounts registered in the social network, while items are short videos recently published by users.\footnote{\datasetkwaisystem and \datasetkwairandom are accessible at \url{https://github.com/Alice1998/MakeFairnessMoreFair}}
The difference between both datasets is how items were selected: in the \datasetkwairandom dataset, the items were selected randomly, while in the \datasetkwaisystem dataset, the items were selected by a recommender system.

The \datasetalimama dataset~\cite{DBLP:conf/www/ShenWTZLCL22/datasets/alimama-deep-interest-highlight-network} contains two binary labels in the global impression: one indicates whether the user was impressed with the item, and the other indicates whether the user interacted with the item.
The dataset contains impressions of online advertisements from an e-commerce service. The data was collected over eight days.\footnote{\datasetalimama is accessible at \url{https://tianchi.aliyun.com/dataset/dataDetail?dataId=56}} Users are registered accounts, and items are advertisements.
In the dataset, a non-interacted impression has the first label as \emph{true} and the second as \emph{false}, while an interacted impression has both labels as \emph{true}.

The \datasetsearchads dataset~\cite{DBLP:conf/kdd/LeeLTS14/impression-discounting} contains the number of interactions and impressions in the global impression.
The dataset contains impressions of online advertisements shown by a search engine.\footnote{\datasetsearchads is accessible at \url{https://www.kaggle.com/competitions/kddcup2012-track2/data}} Users are individuals using the search engine, and items are advertisements.
The dataset is provided with three partitions: \emph{training}, \emph{validation}, and \emph{testing}. The dataset was available to the participants of the \emph{KDD Cup 2012 - Track 2}: where they were tasked to compute the ratio between the number of interactions and impressions for any user-item pair.\footnote{Details on the KDD Cup 2012 - Track 2 are available at \url{https://www.kaggle.com/competitions/kddcup2012-track2}}

Lastly, the \datasetaliccp dataset~\cite{DBLP:conf/sigir/MaZHWHZG18/datasets/ali-ccp-entire-space-multi-task-model} contains the number of user-item clicks and purchases in a global impression.
The dataset contains impressions of articles from an e-commerce service.\footnote{\datasetaliccp is accessible at \url{https://tianchi.aliyun.com/datalab/dataSet.html?dataId=408}} Users are registered accounts, and items are products and goods. The dataset contains two types of user-item feedback: clicks and purchases.
In the dataset, a non-interacted impression has zero clicks and purchases.
An interacted impression has at least one click or purchase.

\begin{table}[t]
  \centering
  \small
  \caption{Statistics of expired and private datasets with impressions.
    Statistics come from datasets' papers.
    Each paper may define and count interactions and impressions differently than this work.
    Expired datasets (3) are presented first. Then, private datasets (9).
    \textbf{Classification} refers to the accessibility of the dataset.
    \textbf{Year} refers to when the dataset is published.
    \textbf{Users} and \textbf{Items} refer to the number of users and items, respectively.
    \textbf{Impressions} refers to the number of non-interacted user-item impressions.
    \textbf{Interactions} refers to the number of interacted user-item impressions.
    ``-'' indicates the value is not reported.
  }
  \label{tab:expired-datasets:statistics}
  \begin{minipage}{\linewidth}
    \centering
    \begin{tabular}{clcrrrr}
      \toprule
      \textbf{Classification}
        &  \textbf{Dataset}
        &  \textbf{Year}
        &  \textbf{Users}
        &  \textbf{Items}
        &  \textbf{Impressions}
        &  \textbf{Interactions}
      \\
      \midrule
      \multirow{4}[0]{*}{Expired}
        &  \datasetxingsixteen~\cite{DBLP:conf/recsys/AbelBKLP16/recsys-challenge-2016/description-paper,DBLP:conf/recsys/PacukSWWW16/recsys-challenge-2016/second-place}  &  2016
        &  1.37 M
        &  1.36 M
        &  1078.63 M
        &  8.83 M
      \\
      {}
        &  \datasetxingseventeen~\cite{DBLP:conf/recsys/AbelDEK17/recsys-challenge-2017/description-paper}
        &  2017
        &  1.50 M
        &  1.31 M
        &  314.50 M
        &  8.27 M
      \\
      {}
        &  \datasettrivagonineteen~\cite{DBLP:conf/recsys/KneesDMALM19/recsys-challenge-2019/description-paper}
        &  2019
        &  700 K
        &  -
        &  16 M
        &  -
      \\
      {}
        &  \datasetsharechattwentythree\footnote{\url{https://www.recsyschallenge.com/2023/}}
        &  2023
        &  10 M
        &  -
        &  -
        &  -
      \\
      \midrule
      \multirow{9}[0]{*}{Private}
        &  \datasetlinkedinpymk~\cite{DBLP:conf/kdd/LeeLTS14/impression-discounting}
        &  2014
        &  -
        &  -
        &  1800 M
        &  -\\
      {}
        &  \datasetlinkedinendorsement~\cite{DBLP:conf/kdd/LeeLTS14/impression-discounting}
        &  2014
        &  -
        &  -

        &  190 M
        &  -
      \\
      {}
        &  \datasetavazu~\cite{DBLP:conf/sigir/LiKG16/reinforcement-learning/collaborative-filtering-multi-armed-bandits}
        &  2016
        &  -
        &  -
        &  -
        &  -
      \\
      {}
        &  \citet{DBLP:conf/recsys/McInerneyLHHBGM18/reinforcement-learning/multi-armed-bandit-explore-exploit-and-explain}
        &  2018
        &  8.60 K
        &  9.60 K
        &  190.00 K
        &  -
      \\
      {}
        &  \citet{DBLP:conf/www/PeiYCLSJOZ19/reinforcement-learning/economic-ecommerce-recommender-systems}
        &  2019
        &  49 M
        &  200 M
        &  -
        &  670.0 M
      \\
      {}
        &  \datasetlreconeb~\cite{DBLP:conf/cikm/XieZW0L21/reinforcement-learning/teacher-student-rl-network}
        &  2021
        &  36.98 M
        &  -
        &  1092 M
        &  92.80 M
      \\
      {}
        &  \datasetwtsoneb~\cite{DBLP:conf/wsdm/XieZW0L22/adversarial/afe-gan-with-impressions}
        &  2022
        &  37.73 M
        &  13.75 M
        &  1330 M
        &  -
      \\
      {}
        &  \datasetarticlethreem~\cite{DBLP:conf/wsdm/XieZW0L22/adversarial/afe-gan-with-impressions}
        &  2022
        &  37.73 M
        &  7.75 M
        &  332.72 M
        &  -
      \\
      {}
        &  \datasettaobaoindustrial~\cite{DBLP:conf/sigir/ChenWLC0Z22/deep-learning/two-tower-without-impressions-but-exposure-count}
        &  2022
        &  603.89 M
        &  1.76 M
        &  -
        &  251.28 M
      \\
      \bottomrule
    \end{tabular}
  \end{minipage}
\end{table}

\subsection{Expired Datasets}
\label{subsec:datasets-with-impressions:expired-datasets}

Expired datasets were accessible to participants of the ACM RecSys Challenge, a yearly competition where participants are tasked to solve industrial recommendation tasks.
Such challenges run in a limited period, where after such a period, the datasets are not accessible and cannot be used in future research works.
\autoref{tab:expired-datasets:statistics} summarizes statistics of expired datasets, where the statistics values come from each dataset's papers. Four expired datasets exist in the literature, where one dataset contains contextual impressions, and three contain global ones (classification detailed in \autoref{subsec:datasets-with-impressions:public-datasets}).

The \datasettrivagonineteen~\cite{DBLP:conf/recsys/KneesDMALM19/recsys-challenge-2019/description-paper} dataset contains contextual impressions from an online travel service. Users are accounts registered in the service, while items are accommodations. The dataset was released as part of the ACM RecSys Challenge 2019. In the dataset, a contextual impression is a tuple of a user identifier, an item identifier, and an impression as a vector of item identifiers.

Three datasets contain global impressions.
The \datasetxingsixteen~\cite{DBLP:conf/recsys/AbelBKLP16/recsys-challenge-2016/description-paper} dataset contains impressions from a job-oriented social network. Users are accounts registered in the social network, while items are job offers. The dataset was released as part of the ACM RecSys Challenge 2016. In the dataset, a global impression is a tuple containing a user identifier and an item identifier or a tuple containing a user identifier and an impression as a vector of item identifiers.
The \datasetxingseventeen~\cite{DBLP:conf/recsys/AbelDEK17/recsys-challenge-2017/description-paper} dataset contains impressions from the same job-oriented social network. Users are accounts registered in the social network, while items are job offers. The dataset was released as part of the ACM RecSys Challenge 2017. Unlike the 2016 dataset, a global impression in this dataset is a tuple of a user identifier, an item identifier, and a label indicating whether the pair is an interacted or non-interacted impression.
Lastly, The \datasetsharechattwentythree dataset contains impressions from an online advertisement service.\footnote{Details of the dataset are available at: \url{https://www.recsyschallenge.com/2023/}} Users are accounts receiving advertisements, while items are advertisements of mobile applications. The dataset was released as part of the ACM RecSys Challenge 2023. The dataset contains two types of user feedback: clicks and installs. In the dataset, a global impression is a tuple of a user's features, an item's features, a binary label indicating whether the user clicked the item, and a binary label indicating whether the user installed the item. Hence, an interacted impression has at least one label with the value \emph{true}, and a non-interacted impression has both labels with the value \emph{false}.

\section{Evaluation}
\label{sec:evaluation}

Due to the importance of evaluation methodologies in recommender systems, we describe the current trends in evaluating \acrfull{impressionsbased}.
The section discusses three topics on evaluation.
First, the types of evaluations used in the literature, where we compare offline evaluations, user studies, and online evaluations.
Second, the research goals and evaluation methodologies followed when evaluating \gls{impressionsbased}.
Third, the challenges faced when working with impressions.
The evaluation of recommender systems is an active and extensive research area with several open questions.
It is beyond the scope of this section to provide effective answers to those open directions.

\begin{table}[t]
  \centering
  \small
  \caption{
    Classification of reviewed papers according to the type of evaluations they use.
    We group those papers that describe offline and online evaluations of their recommenders into a distinct group.
  }
  \label{tab:discussion:evaluations-counts-and-percentages}
  \begin{minipage}{\linewidth}
    \centering
    \begin{tabular}{lp{9cm}rr}
      \toprule
      \textbf{Classification}
        &  \textbf{Papers References}
        &  \textbf{Count}
        &  \textbf{Percentage}
      \\
      \midrule
      offline \& online
        &  \cite{DBLP:conf/kdd/ZhangZMCZA16/heuristics/glmix-generalized-linear-mixed-models-for-large-scale-response-prediction,DBLP:conf/kdd/BorisyukZK17/heuristics/lijar-job-boosting-by-impressions,DBLP:conf/kdd/LinCSLLJ23/impression-aware/tree-based-progressive-regression-model-for-watch-time-prediction-in-short-video-recommendation,DBLP:conf/cikm/AharonKLSBESSZ19/heuristics/soft-frequency-cap,DBLP:conf/kdd/ZhanPSWWMZJG22/heuristics/impressions-to-compute-statistical-features,DBLP:conf/kdd/ZhuD00C22/deep-learning/combo-fashion-clothes-matching,DBLP:conf/cikm/ZhangCXBHDZ22/impressions/keep-an-industrial-pre-training-framework-for-online-recommendation-via-knowledge-extraction-and-plugging,DBLP:conf/wsdm/RenHZZZ23/impressions/slate-aware-ranking-for-recommendation,DBLP:conf/sigir/ChenWLC0Z22/deep-learning/two-tower-without-impressions-but-exposure-count,DBLP:conf/sigir/WangCLHCYLC22/counterfactual/three-towers-counterfactual-learning-with-impressions,DBLP:conf/recsys/ZhaoHWCNAKSYC19/deep-learning/two-state-reranker/learns-impressions/impressions-features/recommending-what-video-to-watch-next-a-multitask-ranking-system,DBLP:conf/kdd/MaILYCCNB22/ctr-estimation/an-online-multitask-learning-framework-for-google-feed-ads-auction-models,DBLP:conf/cikm/GongFZQDLJG22/impressions/real-time-short-video-recommendation-on-mobile-devices,DBLP:conf/kdd/HuCZWCZ23/impression-aware/boss-a-bilateral-occupational-suitability-aware-recommender-system-for-online-recruitment,DBLP:conf/kdd/WangSXDZ23/impression-aware/bert4ctr-an-efficient-framework-to-combine-pre-trained-language-model-with-non-textual-features-for-ctr-prediction,DBLP:conf/wsdm/XieZW0L22/adversarial/afe-gan-with-impressions,DBLP:conf/kdd/ChenWWLZZLZ023/impression-aware/controllable-multi-objective-re-ranking-with-policy-hypernetworks,DBLP:conf/www/PeiYCLSJOZ19/reinforcement-learning/economic-ecommerce-recommender-systems,DBLP:conf/recsys/McInerneyLHHBGM18/reinforcement-learning/multi-armed-bandit-explore-exploit-and-explain,DBLP:conf/wsdm/GrusonCCMHTC19/reinforcement-learning/impressions-to-stream-for-playlist-recommendation,DBLP:conf/cikm/XieZW0L21/reinforcement-learning/teacher-student-rl-network}
        &  21
        &  48.8\%
      \\
      offline
        &  \cite{DBLP:conf/kdd/LeeLTS14/impression-discounting,DBLP:conf/www/AgarwalCE09/spatio-temporal-models-for-estimating-ctr,DBLP:conf/recsys/WuASB16/user-fatigue/netflix-fatigue-modeling-using-impressions,DBLP:conf/www/MaLS16/user-fatigue/online-news-recommendation-feature-engineering-of-impressions,DBLP:conf/cikm/PerezMaureraFDSSC20/contentwise-impressions,DBLP:conf/ijcai/BiedNPHCCGS23/impression-aware/toward-job-recommendation-for-all,DBLP:journals/tois/0003YW0RLSCR23/impression-aware/on-the-behavior-leakage-from-recommender-system-exposure,DBLP:conf/sigir/GongZ22/deep-learning/session-based-attention-network,DBLP:conf/wsdm/XiLLD0ZT023/impressions/a-birds-eye-view-of-reranking-from-list-level-to-page-level,DBLP:journals/isci/Lu23/impression-aware/knowledge-distillation-enhanced-multitask-framework-for-recommendation,DBLP:conf/sigir/LiKG16/reinforcement-learning/collaborative-filtering-multi-armed-bandits,DBLP:conf/wsdm/GeZYPHHZ22/reinforcement-learning/grouping-items-by-exposure,DBLP:conf/sigir/NayakGM23/impression-aware/news-popularity-beyond-the-click-through-rate-for-personalized-recommendations}
        &  13
        &  30.2\%
      \\
      online
        &  \cite{DBLP:conf/kdd/AgarwalCGHHIKMSSZ14/impression-discounting,DBLP:conf/recsys/CovingtonAS16/deep-learning/two-stage-reranker/learns-impressions/impressions-features/deep-neural-networks-for-youtube-recommendations,DBLP:conf/wsdm/ChenBCJBC19/reinforcement-learning/top-k-off-policy-correction-for-a-reinforce-recommender-system}
        &  3
        &  7.0\%
      \\
      user studies
        &  \cite{DBLP:conf/cscw/ZhaoAHWK17/heuristics/cycling-serpentining,DBLP:conf/recsys/ZhaoWAHK18/impressions-signals/interpreting-user-inaction-in-recsys}
        &  2
        &  4.7\%
      \\
      simulation
        &  \cite{DBLP:conf/wsdm/DeffayetTRR23/impressions/generative-slate-recommendation-with-reinforcement-learning}
        &  1
        &  2.3\%
      \\
      not described
        &  \cite{DBLP:journals/scheduling/BuchbinderFGN14/hard-frequency-cap,DBLP:conf/www/LiuRSKMZLJ17/heuristics/memboost-impressions-as-features,DBLP:conf/sigir/Sun23/impression-aware/take-a-fresh-look-at-recommender-systems-from-an-evaluation-standpoint}
        &  3
        &  7.0\%
      \\
      \bottomrule
    \end{tabular}%
  \end{minipage}
\end{table}

\subsection{Evaluation Types}
\label{subsec:evaluation:evaluation-types}

Evaluation types refer to the methods researchers and practitioners use to measure the recommendation quality of their recommenders.
Four categories of evaluation types exist in the literature of recommender systems~\cite{DBLP:journals/ir/CanamaresCM20/non-impressions/evaluation-offline-evaluation-options-for-recommender-systems,DBLP:journals/csur/ZangerleB23/non-impressions/evaluating-recommender-systems-survey-and-framework}:
\begin{itemize}
  \item \textbf{Simulations:} consist of measuring the recommendation's quality of a recommender using crafted preferences of users. Their main advantage is their low complexity. Their drawback is their low generalization capabilities due to their reliance on artificial user feedback.

  \item \textbf{Offline evaluations:} consist of measuring the recommendation's quality of a recommender using a dataset with impressions. Their advantages are low cost, accessibility, and reproducibility. Their drawback is its low generalization capability due to their reliance on logged and not updated feedback. In \autoref{sec:datasets-with-impressions}, we describe the datasets with impressions used in offline evaluations in the reviewed literature.

  \item \textbf{User studies:} consist of exposing impressions to a selected and reduced group of users in a controlled environment. User studies are more challenging to reproduce and have limited generalization power; however, they are especially useful for collecting explicit user feedback regarding a given set of qualitative or quantitative metrics.

  \item \textbf{Online evaluations:} consist of exposing impressions to users of an online and deployed recommender system. This is the most challenging evaluation because it is costly, time-consuming, and increases business risks. However, they provide the most realistic picture of user preferences toward impressions.
\end{itemize}

As shown in \autoref{tab:discussion:evaluations-counts-and-percentages}, most reviewed papers use offline evaluations to measure the quality of their recommenders.
Moreover, half of the reviewed papers perform offline and online evaluations, two evaluate recommenders via user studies, three exclusively perform online evaluations, and one performs simulations.

\subsection{Research Goals and Evaluation Methodologies}
\label{subsec:evaluation:evaluation-methodologies}

Researchers must ensure the entire methodology, either ad-hoc or based on existing literature, is in line with their research goals, \idest the methods do not conflict, pollutes, or invalidate results.
We classify the research goals of reviewed papers into two categories:

\begin{itemize}
  \item \textbf{Improving the quality of recommendations:} an extension of the traditional research goal in recommender systems applied to \gls{impressionsbased}. The aim is to increase the quality of recommendations by devising novel recommenders using impressions. Consequently, the usual best practices in recommender systems research apply. When using impressions, particular care must be taken to ensure the evaluation methodology is consistent and aligned with the research goal, \eg researchers must not use impressions at test time when evaluating end-to-end or plug-in recommenders.

  \item \textbf{Extracting signals from impressions:} aims to disentangle the user preference on impressions with emphasis on non-interacted impressions. Several papers state~\cite{DBLP:conf/recsys/ZhaoWAHK18/impressions-signals/interpreting-user-inaction-in-recsys,DBLP:conf/recsys/PerezMaureraFDC22/towards-the-evaluation-of-recommender-systems-with-impressions,DBLP:conf/iir/PerezMaureraFDC22/replication-of-impressions} impressions contain complex and mixed signals. For instance, \citet{DBLP:conf/recsys/ZhaoWAHK18/impressions-signals/interpreting-user-inaction-in-recsys} user's inaction on impressions can be attributed to different factors, such as users not being interested in particular items or having already interacted with such items.
\end{itemize}

Reviewed papers use traditional or ad-hoc methodologies depending on the nature of their studies.
Papers extracting signals from impressions use two approaches.
The first approach gathers users' explicit preference for non-interacted impressions with user studies or online evaluations.
\citet{DBLP:conf/recsys/ZhaoWAHK18/impressions-signals/interpreting-user-inaction-in-recsys} describe how to perform a user study to gather users' preferences on non-interacted impressions.
The second approach extracts the preference for impressions using heuristic or machine learning methods in offline or online evaluations.
Several reviewed papers~\cite{DBLP:conf/cikm/AharonKLSBESSZ19/heuristics/soft-frequency-cap,DBLP:conf/kdd/LeeLTS14/impression-discounting} describe recommenders learning the preference of users toward impressions. \citet{DBLP:conf/cikm/AharonKLSBESSZ19/heuristics/soft-frequency-cap} includes an additional \emph{bias} term, called frequency bias, to a traditional matrix factorization recommender, while \citet{DBLP:conf/kdd/LeeLTS14/impression-discounting} defines a weighted factor accounting the user preferences to several features from impressions.
Other papers~\cite{DBLP:conf/recsys/PerezMaureraFDC22/towards-the-evaluation-of-recommender-systems-with-impressions,DBLP:conf/iir/PerezMaureraFDC22/replication-of-impressions} describe how adding an extra hyper-parameter to recommenders aids in identifying whether impressions are positive or negatives signals.

We exemplify a case of a methodology conflicting and confounding the research goal using the ACM RecSys Challenge 2019.
The goal of the competition was to devise \emph{end-to-end} recommenders, \idest a recommender generating learning user preferences and generating impressions.\footnote{The definitions of several types of recommenders are provided in \autoref{subsec:impressions-based-recsys:problem-description}.}
The competition employed a well-known evaluation methodology in recommender systems: it tasked participants to submit impressions containing interacted items at test time and evaluated submissions using the mean reciprocal rank metric.
However, the competition provided the impressions at test time, \idest the impression participants had to submit.
That evaluation methodology is designed to assess the quality of \emph{re-ranking} recommenders instead of \emph{end-to-end} ones, \idest the methodology evaluates recommenders receiving an impression and generating a permutation of it.

\subsection{Challenges}
\label{subsec:evaluation:challenges}

In this section, we identify and describe several challenges researchers and practitioners encounter when handling impressions in recommender systems, either when devising recommendation models, disseminating datasets, or evaluating recommender systems.

\subsubsection{Signals in Impressions} The first challenge is connected to fully utilizing impressions and extracting their signals to learn users' preferences.
In this regard, in \autoref{sec:reviewed-papers}, we describe all the techniques and approaches the literature has employed to use impressions in their recommendation models.
Despite using non-interacted impressions as negative signals is the most popular approach, it may not be the most effective.
That approach is problematic because it does not take into account that user inaction towards good recommendations may be due to factors unrelated to the items' relevance, as found by \citet{DBLP:conf/recsys/ZhaoWAHK18/impressions-signals/interpreting-user-inaction-in-recsys}.
For example, a non-interacted impressed item may be relevant; however, it is superseded by another more relevant impressed item given the users' context, mood, or awareness of all impressed items.
At the same time, users' preferences are not binary or stationary; instead, they depend on many factors, such as the users' context, short and long interest, and location, among other factors~\cite{DBLP:reference/sp/AdomaviciusBTU22/context-aware-recommender-systems-from-foundations-to-recent-developments}.
Thus, a non-interacted impression may be relevant to the user; however, in different contexts or situations.
The literature has not encountered, yet, a set of approaches able to disentangle the signals from impressions; specially, from non-interacted impressions.
Moreover, it is also challenging the effective integration of such signals into existing or newer recommendation models.

\subsubsection{Scalability} Another challenge concerns the scalability of recommenders due to impressions being more abundant than interactions.
As illustrated in \autoref{tab:public-datasets:statistics}, the ratio between the numbers of impressions and interactions ranges from \num{1.82} and \num{143.45} with a median value of \num{18.58} and mean of $\num{26.54} \pm \num{36.30}$ in public datasets.
Notably, in four datasets (\datasetinshop, \datasetkwairandom, \datasetkwaisystem, and \datasetmind), the number of impressions and interactions align closely with their magnitude: millions of records.
In eight datasets (\datasetcw, \datasetfinn, \datasetyahoorsixa, \datasetyahoorsixb, \datasetsearchads, \datasetaliccp, \datasetalimama, \datasetcrossshop), the number of impressions surpasses the number of interactions by one order of magnitude, and in one dataset (\datasetpandor) it is exceeded by two orders of magnitude.
Hence, future works must be attentive to address scalability concerns when using impressions.
Three of our previous papers~\cite{DBLP:conf/recsys/PerezMaureraFDC22/towards-the-evaluation-of-recommender-systems-with-impressions,DBLP:conf/iir/PerezMaureraFDC22/replication-of-impressions} already highlight this challenge, whereby certain recommenders could not be evaluated to scalability issues.

\subsubsection{Public Datasets with Impressions} The dissemination of public datasets with impressions is another challenging task.
The extraction of datasets from real-world recommender systems is already a difficult task, entailing careful considerations encompassing data collection methodologies, data cleansing procedures, privacy safeguards, and other aspects~\cite{DBLP:journals/cacm/GebruMVVWDC21/non-impressions/datasheets-for-datasets}.
Furthermore, disseminating datasets with impressions
introduces novel considerations and exacerbates existing risks.
Specifically, datasets derived from real-world proprietary recommenders entail inherent business risks, as they expose users' interests, system behaviors, and the system's notion of user relevance.
Moreover, additional concerns, such as robust anonymization techniques, must be considered and extended to impressions.

\subsubsection{Incomplete Information} Another challenge arises from the missing information in current public datasets, consequently constraining the efficacy of evaluation methodologies.
As detailed in \autoref{sec:datasets-with-impressions}, eleven public datasets with impressions are available for research purposes.
Among those, eight datasets contain global impressions, where connecting interactions and impressions is not possible.
Three datasets contain contextual impressions, which can connect interactions and impressions; however, two lack all impressions records, and two lack time-related attributes in some or all impressions.
The absence of such information limits future studies and modeling capabilities of users' preferences.
For instance, the research of position biases within datasets with global impressions is not achievable due to those datasets not having position-related attributes.

\subsubsection{Biases within Impressions} The last challenge stems from the biases present in datasets with impressions, which are important to identify in order to adapt methodologies accordingly.
In this context, data bias refers to the disparities between the anticipated and actual statistical distributions within data~\cite{DBLP:journals/tois/0007D0F0023/non-impressions/bias-and-debias-in-recommender-system-a-survey-and-future-directions}. It is worth noting that traditional interaction data and impressions are generated mostly through the same process and combine the biases produced by the recommender system, the user interface and the users themselves. Despite this, impressions may present new unique bias characteristics that deserve to be studied. 
When training with impressions, we can identify two main scenarios. 
First, certain biases may manifest exclusively within impressions, such as new biases related to which non-relevant items are recommended or the position biases~\cite{DBLP:journals/tois/0007D0F0023/non-impressions/bias-and-debias-in-recommender-system-a-survey-and-future-directions} observed in contextual impressions.
Second, impressions likely exhibit biases akin to those found in interactions or other data sources, but possibly to different degrees. 
For instance, exposure biases~\cite{DBLP:journals/tois/0007D0F0023/non-impressions/bias-and-debias-in-recommender-system-a-survey-and-future-directions} are present in impressions as well due to the tendency of recommender systems to include popular items in impressions. 
In order to overcome these challenges, it is also necessary that the community further studies biases in impressions and recommender systems, a direction that we discuss in \autoref{sec:future-directions}.

\section{Open Research Questions \& Future Directions}
\label{sec:future-directions}

In previous sections, we review the state-of-the-art in \acrfull{impressionsbased}.
From such a review, we identify several research questions that remain unanswered.
As impressions are a novel data type in recommender systems, they allow researchers to study different directions and devise more refined evaluation methodologies than those currently used in the literature.
Previous work by \citet{DBLP:conf/recsys/Jeunen19/revisiting-offline-evaluation-for-implicit-feedback-recommender-systems} highlights several open research questions related to impressions.
In this section, we describe such questions, identify research needs, and propose additional research directions for future works.
We focus our discussion on six areas of improvement: recommendation models, datasets with impressions, debiased recommendations and evaluations, impressions signals, and user fatigue.

\subsection{Impressions Signals}
\label{subsec:future-directions:impressions-signals}

In \autoref{sec:impressions-based-recsys}, we propose the signal-centric taxonomy, which classifies recommenders by whether they \emph{assume} or \emph{learn} impressions signals.
When reviewing recommenders, we observe that most papers assume non-interacted impressions are negative signals, \idest users dislike such items.
The literature, however, does not contain strong empirical evidence of non-interacted impressions being negative or positive feedback.
In both \autoref{sec:impressions-based-recsys} and \autoref{sec:evaluation}, we emphasize only three papers~\cite{DBLP:conf/recsys/PerezMaureraFDC22/towards-the-evaluation-of-recommender-systems-with-impressions,DBLP:conf/iir/PerezMaureraFDC22/replication-of-impressions,DBLP:conf/recsys/ZhaoWAHK18/impressions-signals/interpreting-user-inaction-in-recsys} address this topic.
Given such few studies and inconclusive evidence, it remains an open question \emph{how to treat impressions, especially non-interacted impressions}.
Furthermore, it also remains an open question \emph{how to disentangle the signals of users' preferences inside impressions}.

We identify two considerations when addressing such open questions.
First, as happens in several research areas in machine learning, researchers must be aware impressions may be subject to concept drift~\cite{non-impressions/dataset-shift-in-machine-learning,DBLP:journals/pr/Moreno-TorresRACH12/non-impressions/a-unifying-view-on-dataset-shift-in-classification}.
Concept drift represents statistical changes in data points over time, degrading the ability of machine learning models to accurately predict future data points.
Hence, future research works may study detection methods for concept drift using impressions.
Second, as we highlighted in previous discussions, different entities generate impressions, and the signals in impressions may be tied to those entities.
Consequently, future works may need to study whether impressions alone are sufficient to disentangle their signals.

\subsection{User Fatigue}
\label{subsec:future-directions:user-fatigue}

User fatigue is the phenomenon where users dislike the impressions generated by the recommender system regardless of their relevance due to impressions being repetitive or uninteresting.
For example, recommendations of movies similar to those already watched by users may result in positive user experiences.
However, users exhibit fatigue with those movies when they are not rewarded with positive user feedback after many impressions.
At this point, the recommender must vary the recommendations; otherwise, it risks losing users' interest and trust in the recommenders' capabilities.
In the example, impressed items are relevant due to the users' past consumption patterns but not desirable as they do not show interest in them.
It remains an open question \emph{how to identify and model user fatigue using impressions}.
The literature contains five papers~\cite{DBLP:conf/www/MaLS16/user-fatigue/online-news-recommendation-feature-engineering-of-impressions,DBLP:conf/recsys/WuASB16/user-fatigue/netflix-fatigue-modeling-using-impressions,DBLP:conf/kdd/LeeLTS14/impression-discounting,DBLP:journals/scheduling/BuchbinderFGN14/hard-frequency-cap,DBLP:conf/cikm/AharonKLSBESSZ19/heuristics/soft-frequency-cap} addressing this question; however, they use hard-coded rules, ad-hoc fatigue functions, or learn non-personalized functions.

Impressions enable the study of users' fatigue because they contain the necessary information for its study: the items presented to the user and their received back.
In this regard, impressions can be complemented with users' intent and context for a comprehensive study of users' fatigue.
The reviewed literature studies users' fatigue by using the number of interactions and impressions for a given user-item pair.

\subsection{Recommendation Models}
\label{subsec:future-directions:recommendation-models}

Through the lens of the model-centric taxonomy, recommendation models in the literature are varied, where they mostly use deep learning or reinforcement learning to learn users' preferences.
Notably, we observe that the community has not explored many other categories of recommendation models; some of which may be suitable to work with impressions.
Thus, a research area that may be further studied is \emph{the design and development of recommendation models belonging to other categories}.
Although this research area is not unique to \gls{impressionsbased}, addressing it involves the discussion of topics unique to this recommendation paradigm.
For example, redefining the similarity in a k-nearest neighbor recommender to capture the similarities between items based on their impressions.
A good starting point in this area is to consider recommendation models able to encode \emph{side information}, \eg graph-based models~\cite{DBLP:conf/www/CooperLRS14/non-impressions/graph-based-p3-alpha/random-walks-in-recommender-systems-exact-computation-and-simulations,DBLP:conf/recsys/ChristoffelPNB15/non-impressions/graph-based-rp3-beta} or factorization machines~\cite{DBLP:conf/icdm/Rendle10/non-impressions/factorization-machines}, where the impressions are the side information of interactions.
One of our previous works~\cite{DBLP:conf/leri/PerezMaureraFDCC23/incorporating-impressions-to-graph-based-recommenders} shows two approaches to effectively incorporate impressions into graph-based recommenders.
In particular, we redefine the graph and build it using both interacted and non-interacted impressions.

Through the lens of the data-centric taxonomy,  only a handful of works in the literature sample from impressions; however, item sampling is a highly relevant area in recommender systems research.
In our review, we find that the literature has not studied \emph{how different sampling techniques affect the recommendation quality of \gls{impressionsbased}}.
A good starting point in this direction is in differentiating sampling items within the same impressions or globally.
However, more complex types of item sampling can be studied as well.
For instance, impressions can be treated as an additional channel of user feedback.
Thus they can be sampled using the techniques proposed by \citet{DBLP:conf/recsys/LoniPLH16/bayesian-personalized-ranking-with-multi-channel-user-feedback} while considering their entangled signals.
Recently, \citet{DBLP:journals/eaai/JainJ23/sampling-and-noise-filtering-methods-for-recommender-systems-a-literature-review} review and collect into a single document many item sampling techniques used in past recommendation models.

Lastly, through the lens of the signal-centric taxonomy, few recommendation models in the literature learn the signals of impressions.
Instead, most assume non-interacted impressions represent negative signals, while interacted impressions represent positive ones.
This is more noticeable in recommender systems using reinforcement learning.
In those cases, the literature typically assigns a zero reward to non-interacted impressions.
One research direction to pursue is \emph{devising recommendation models able to learn the signals in impressions}.
For reinforcement learning recommenders, this implies modifying the reward function so it does not always yield zero to non-interacted impressions.
Instead, the function may consider other features or factors, \eg users' fatigue or the number of impressions with the same item.
Another research direction without much exploration is the \emph{study of non-interacted impressions as neutral or positive signals}.

\subsection{Biases in Impressions, Debiased Recommendations \& Evaluations}
\label{subsec:future-directions:debiased-recommendations-and-evaluations}

One of the challenges (see \autoref{sec:evaluation}) that the community faces when using impressions is the identification and balancing of biases in impressions.
\citet{DBLP:journals/tois/0007D0F0023/non-impressions/bias-and-debias-in-recommender-system-a-survey-and-future-directions} states that a data bias is a difference between the expected statistical distribution of certain data and their real statistical distribution.
In recommender systems, \citet{DBLP:journals/tois/0007D0F0023/non-impressions/bias-and-debias-in-recommender-system-a-survey-and-future-directions} also argues that biases in interactions may negatively affect the recommendation quality.
In our review, we find that the literature has not deeply studied biases in impressions, has not proposed a characterization of them, and has not studied their effects on recommendations.
Similarly, the literature has not studied debiasing techniques in impressions or by using impressions yet. Since impressions are generated through the same process that generated interactions: the recommender system, the user interface, and the choice made by the user, impressions present a strong connection with interactions but can also be seen as a complementary source of information. In this regard, the biases present in interactions and impressions are likely related, but not identical, and studying one may help to better understand the other.
Broadly speaking, we highlight two directions for further studies.

Concerning the study of existing biases, impressions may enable a more comprehensive study of biases due to the more granular classification of items: never impressed, impressed but not interacted with, and interacted with items.
Additionally, the community can expand the current studies on biases in recommender systems to identify biases in impressions.
One starting point comes from the very definition of impression: a selection of $N$ items from the catalog created by a recommender system, search engine, or any other entity; thus, when the entity generating impressions is biased, then its generated impressions will be biased as well. 
Examples of new types of bias that can be studied are how the recommender system identifies recommended items that are not interacted with (\idest that are non-relevant for the user) and how the contextual impressions change or bias the user assessment of what is relevant and, therefore, their interactions.
Additionally, training a recommender using impressions may amplify or diminish the effects of biases during the feedback loop~\cite{DBLP:journals/tois/0007D0F0023/non-impressions/bias-and-debias-in-recommender-system-a-survey-and-future-directions} that recommender systems go through.
Lastly, the identification of biases in impressions may be challenging due to the incomplete information on the policies of the entities that generated them.

Impressions may also enable the improvement or creation of new debiasing techniques for recommender systems.
Inverse propensity weighting (IPW) is an example of a popular method to correct items' relevance by accounting for their probability of exposure, \idest by computing the items' propensity score~\cite{DBLP:conf/icml/SchnabelSSCJ16/non-impressions/propensity-score-recommendations-as-treatments-debiasing-learning-and-evaluation}.
Generally, \gls{ipw} approaches are computed using interactions; which are an incomplete representation of the users' exposure to items in the catalog.
The use of impressions can mitigate this problem since they are a more comprehensive description of users' exposure, in some cases, a complete one.
It is still an open question whether using impressions for \gls{ipw} (or other debiasing methods) results in effective unbiased estimators or unbiased evaluation of recommenders.
In the case of creation of new debiasing techniques, two recent papers~\cite{DBLP:conf/kdd/ZhangZLWGZG23/debiasing/debiasing-recommendation-by-learning-identifiable-latent-confounders,DBLP:conf/cikm/ZhouHC0W0Y23/debiasing/contrastive-counterfactual-learning-for-causality-aware-interpretable-recommender-systems} propose two distinct approaches to identify, model, and correct exposure biases in recommendations using impressions.
Overall, only very few of the many biases described by \citet{DBLP:journals/tois/0007D0F0023/non-impressions/bias-and-debias-in-recommender-system-a-survey-and-future-directions} have been studied in the context of impressions.
Consequently, using impressions for debiased evaluations still remains a wide and open research direction.

\subsection{Datasets with Impressions}
\label{subsec:future-directions:datasets}

Unlike recommendation models, we do not identify open research questions after reviewing and analyzing existing datasets with impressions.
However, we identify two research needs: one relates to the existing number of public datasets with contextual impressions, and the other relates to the lack of impression's origin in existing datasets.
Publishing those types of datasets allows future works to propose more robust models and further analyze the impact and meaning of impressions on users' preferences.

The first research need is \emph{the publication of public datasets with contextual impressions}.
In \autoref{sec:impressions-based-recsys}, we define \emph{contextual} impressions as those where interacted and non-interacted impressions shown at a given time are recorded, \idest researchers know the position of impressed items and whether users interacted with them or not.
In \autoref{sec:datasets-with-impressions}, we classify datasets with impressions into three categories (\emph{public}, \emph{expired}, and \emph{private} datasets).
There, we highlight the importance of \emph{public} datasets due to their availability and flexible licensing, permitting researchers to use them in future research.
Despite the existence of 13 public datasets, 10 of them contain global impressions, which have limited information when compared to their contextual counterparts.

The second research need is \emph{the publication of datasets, including the origin of impressions}.
As we highlight in \autoref{sec:evaluation}, one research goal is disentangling the signals within impressions.
Under such a goal, existing or ad-hoc evaluation methodologies need labels indicating which system generated impressions.
Those labels characterize whether a given impression comes from a recommender system, a search engine, editorial selections, or other systems.
Without such labels, disentangling the signals is a more challenging task.

\subsection{Recommendation Quality of Impression-Aware Recommender Systems}

Throughout this work, we analyze three dimensions of \gls{impressionsbased}, namely recommendation models, datasets with impressions, and evaluation.
Despite covering a broad selection of dimensions, it is still an open question what is the recommendation quality of the reviewed \gls{impressionsbased} both in the general case and in particular contexts.

As we state at the beginning of \autoref{sec:reviewed-papers}, addressing this direction has its own challenges due to the numerous considerations to make in order to make a fair and representative assessment of the quality of reviewed recommendation models.
Despite the inherent complexity of evaluating recommendation models, this is a topic of increased interest to the community, where research works explore many complementary dimensions.
\citet{DBLP:journals/csur/ZangerleB23/non-impressions/evaluating-recommender-systems-survey-and-framework} in a recent survey describe an evaluation framework for general recommender systems, highlighting the different perspectives to consider. For instance, defining recommendation goals, selection of evaluation methods, selection of metrics, among others.
\citet{DBLP:journals/ir/CanamaresCM20/non-impressions/evaluation-offline-evaluation-options-for-recommender-systems} provides a list of methodological decisions to make when evaluating recommenders when following an \emph{offline evaluation}.
\citet{DBLP:journals/tois/FerrariDacremaBCJ21/a-troubling-analysis} found previous progress claims to be non-reproducible after carefully evaluating recommendation models under the same evaluation methodology.
On a similar note, \citet{DBLP:conf/recsys/ShehzadJ23/non-impressions/everyones-a-winner-on-hyperparameter-tuning-of-recommendation-models} found the hyper-parameter tuning of recommendation models is a necessary step to ensure a fair comparison of recommendation models.

Regarding the evaluation of \gls{impressionsbased}, two of our previous works~\cite{DBLP:conf/iir/PerezMaureraFDC22/replication-of-impressions,DBLP:conf/recsys/PerezMaureraFDC22/towards-the-evaluation-of-recommender-systems-with-impressions} have partially addressed this direction. In those, we assess the recommendation quality \gls{impressionsbased} under a single recommendation task, training a subset of reviewed recommendation models (cycling~\cite{DBLP:conf/cscw/ZhaoAHWK17/heuristics/cycling-serpentining}, \acrlong{hfc}~\cite{DBLP:journals/scheduling/BuchbinderFGN14/hard-frequency-cap}, and \acrlong{idf}~\cite{DBLP:conf/kdd/LeeLTS14/impression-discounting}) on a selection of public datasets with impressions (\datasetcw~\cite{DBLP:conf/cikm/PerezMaureraFDSSC20/contentwise-impressions}, \datasetmind~\cite{DBLP:conf/acl/WuQCWQLLXGWZ20/MIND-a-large-scale-dataset-for-news-recommendation}, and \datasetfinn~\cite{DBLP:journals/datamine/EideLF22/dynamic-slate-recommendation-with-gated-recurrent-units-and-thompson-sampling,DBLP:conf/recsys/EideLFRJV21/finn-no-slates-dataset}) under a single evaluation framework.
Particularly, those studies study the recommendation quality of \emph{plug-in} \gls{impressionsbased} when paired with an already-trained \gls{cf} recommenders.\footnote{See the definition of plug-in recommenders in \autoref{sec:impressions-based-recsys}.}
Despite those studies representing a start in this direction, they are far from representing an exhaustive assessment of the recommendation quality of \gls{impressionsbased}.
Thus, it is still needed comprehensive studies on the recommendation quality of \gls{impressionsbased} on different scenarios and contexts.

\section{Conclusions}
\label{sec:conclusions}

Academic and industrial interest in impressions and their use in recommender systems have steadily increased over the years.
Using impressions as a new data source increases the creation of a novel paradigm for personalized recommendations, termed \acrfull{impressionsbased}.
Recommender systems following this paradigm have the opportunity to model users' preferences more accurately than using interactions alone.
For instance, a recommender may decide whether to recommend a particular item based on the number of impressions it has with a given user~\cite{DBLP:journals/scheduling/BuchbinderFGN14/hard-frequency-cap,DBLP:conf/kdd/LeeLTS14/impression-discounting}.
Several initiatives raise the attention and sustain the use of impressions recommender systems: previous publications, public datasets, and competitions.
To evaluate \gls{impressionsbased}, researchers and practitioners need access to datasets with impressions.
Without public datasets, it is not possible to validate existing works nor compare the effectiveness of recommenders with impressions.

In this work, we systematically review impressions in recommender systems under three perspectives (recommendation models, datasets, and evaluation methodologies).
We term recommenders using impressions as \acrfull{impressionsbased} and define a theoretical framework enclosing them.
Under such a framework, we highlight the similarities and differences between \gls{impressionsbased} and other recommendation paradigms.
In such comparison, we find that although similar paradigms exist, \gls{impressionsbased} are part of a unique type of recommenders.

This work describes a systematic literature review methodology to collect relevant papers in \gls{impressionsbased}.
This methodology consists of discovering papers through academic search engines and selecting only those conference or journal papers published in high-level venues.
Under this methodology, we discovered \numuniqueworks unique papers and selected \numincludedworks to review.

During the work, we propose a classification system composed of three taxonomies for \gls{impressionsbased}, which we term model-centric, data-centric, and signal-centric.
Such taxonomies group recommenders based on how they define their recommendation model, use impressions, and whether they assume or learn a connotation of impressions in users' preferences.
From the review, we highlight several patterns.
First, recommenders have been using more complex paradigms (machine learning, deep learning, and reinforcement learning) since 2016, while the last recommender using simpler paradigms (heuristics and statistics) is from 2017.
Second, most recommenders either learn from impressions or compute features from impressions and then use such features; only a handful of papers in the literature sample from all impressions.
Also, we note the literature does not contain recommenders using factorization machines or graph structures.

Regarding datasets, we describe datasets with impressions and classify them based on their availability to be used in future research works.
In this regard, only one category of datasets (termed \emph{public}) can be used for such purposes.
The literature contains 13 public datasets, where 3 contain \emph{contextual impressions} and the rest \emph{global impressions}.
The former indicates which impression contains a given interaction, while the latter does not contain such information.
We highlight that global impressions are less informative than contextual ones.
For example, it is not possible to adjust presentation and position biases in the data on global datasets, as this information is not present.
When looking at datasets, we note that several papers evaluate their recommenders on private datasets; this is not favorable as such results are not possible to reproduce.
Future research should use public datasets in their experiments.
Moreover, future works may publish public datasets with contextual impressions with labels indicating whether impressions come from recommenders, editorial selections, or other systems.

We present a discussion of current evaluation methodologies of \gls{impressionsbased} in reviewed papers.
In such a discussion, we highlight the importance of sound evaluation methodologies to ensure real progress; we describe research goals with impressions using proper evaluation methodologies.
We also discuss the challenges researchers must consider when evaluating recommenders using impressions, especially the effects of biases, scalability, data collection, and data dissemination.
We identify new evaluation methodologies for future research thanks to impressions and the information they provide.
In particular, we highlight that impressions permit researchers to validate modeling techniques for propensity scores, more effective debiasing techniques, and others.

We close this work by noting several open questions and directions for future works.
In particular, we emphasize the discussion in the three pillars of this work (recommenders, datasets, and evaluation).
We detail papers in the reviewed literature that do not describe certain types of strong recommenders, such as graph-based or factorization machines.
At the same time, we propose novel evaluation methodologies with impressions, accounting for the type of information provided with impressions, \eg to debias evaluations by incorporating layout information on impressed items or measuring propensity scores using impressions.
We close this work with future ideas on topics discussed in a few papers but of high relevance to recommender systems.
For example, to model user fatigue due to repeated impressions or to model biases within recommender systems using impressions.

\bibliographystyle{ACM-Reference-Format} 
\bibliography{main}

\end{document}